\documentclass[12pt,a4paper,aps,preprint,preprintnumbers,superscriptaddress,nofootinbib]{revtex4-1}

\usepackage[utf8]{inputenc}
\usepackage{graphicx}
\usepackage{amssymb}
\usepackage{textcomp}
\usepackage{amsmath}
\usepackage{tabularx}
\usepackage{bm}
\usepackage{times}
\usepackage{color}
\usepackage{slashed}
\usepackage{multirow}
\usepackage{verbatim}
\usepackage{cancel}
\usepackage{subfigure}
\usepackage{ulem}
\usepackage{slashed}

\usepackage[colorlinks=true, pdfstartview=FitV, linkcolor=blue, citecolor=blue, urlcolor=blue]{hyperref}
\allowdisplaybreaks[4]

\def\lsim{\mathrel{\raise.3ex\hbox{$<$\kern-.75em\lower1ex\hbox{$\sim$}}}}
\def\gsim{\mathrel{\raise.3ex\hbox{$>$\kern-.75em\lower1ex\hbox{$\sim$}}}}

\definecolor{orange}{rgb}{1,0.5,0}

\begin{document}

\title{Charged lepton flavor violation in light of the muon magnetic moment anomaly and colliders}

\author{Tong Li}
\email{litong@nankai.edu.cn}
\affiliation{
School of Physics, Nankai University, Tianjin 300071, China
}
\author{Michael A. Schmidt}
\email{m.schmidt@unsw.edu.au}
\affiliation{
Sydney Consortium for Particle Physics and Cosmology, 
School of Physics, The University of New South Wales, Sydney, New South Wales 2052, Australia
}
\author{Chang-Yuan Yao}
\email{yaocy@nankai.edu.cn}
\affiliation{
School of Physics, Nankai University, Tianjin 300071, China
}
\author{Man Yuan}
\email{2120200176@mail.nankai.edu.cn}
\affiliation{
School of Physics, Nankai University, Tianjin 300071, China
}

\preprint{CPPC-2021-06}
\begin{abstract}
Any observation of charged lepton flavor violation (CLFV) implies the existence of new physics beyond the SM in charged lepton sector. CLFV interactions may also contribute to the muon magnetic moment and explain the discrepancy between the SM prediction and the recent muon $g-2$ precision measurement at Fermilab.
We consider the most general SM gauge invariant Lagrangian of $\Delta L=0$ bileptons with CLFV couplings and investigate the interplay of low-energy precision experiments and colliders in light of the muon magnetic moment anomaly. We go beyond previous work by demonstrating the sensitivity of the LHC, the MACE experiment, a proposed muonium-antimuonium conversion experiment, and a muon collider. Currently-available LHC data is already able to probe unexplored parameter space via the CLFV process $pp\to\gamma^*/Z^*\to \ell_1^\pm \ell_1^\pm \ell_2^\mp \ell_2^\mp$.
\end{abstract}

\maketitle

\section{Introduction}
\label{sec:Intro}

The observation of neutrino oscillations and thus non-zero neutrino masses clearly established the existence of lepton flavor violation (LFV) in the neutrino sector.
We also expect that the charged lepton flavor violation (CLFV) occurs in short-distance processes without neutrinos.
The standard convention
defines CLFV as the processes that conserve total lepton number $L\equiv
L_e+L_\mu+L_\tau$ (and baryon number $B$) but violate the global symmetry group
\begin{eqnarray}
{\rm U}(1)_{L_\mu - L_\tau}\times {\rm U}(1)_{L_\mu + L_\tau - 2L_e},
\end{eqnarray}
without involving neutrinos~\cite{Heeck:2016xwg}.
The rates of CLFV processes in the Standard Model (SM) are suppressed by $G_F^2 m_\nu^4\lesssim 10^{-50}$ due to the
unitarity of the leptonic mixing matrix and beyond the sensitivity of any current
or future experiments. Thus, the observation of any CLFV process implies the existence of new physics (NP) beyond the SM.

The muon magnetic dipole moment has a long-standing $\sim 3.7\sigma$ discrepancy between the SM prediction~\cite{Davier:2017zfy,Keshavarzi:2018mgv,Colangelo:2018mtw,Hoferichter:2019gzf,Davier:2019can,Keshavarzi:2019abf,Kurz:2014wya,Chakraborty:2017tqp,
Borsanyi:2017zdw,Blum:2018mom,Giusti:2019xct,Shintani:2019wai,Davies:2019efs,Gerardin:2019rua,Aubin:2019usy,Giusti:2019hkz,Melnikov:2003xd,Masjuan:2017tvw,
Colangelo:2017fiz,Hoferichter:2018kwz,Gerardin:2019vio,Bijnens:2019ghy,Colangelo:2019uex,Pauk:2014rta,Danilkin:2016hnh,Jegerlehner:2017gek,Knecht:2018sci,
Eichmann:2019bqf,Roig:2019reh,Colangelo:2014qya,Blum:2019ugy,Melnikov:2003xd,Masjuan:2017tvw,Colangelo:2017fiz,Hoferichter:2018kwz,Gerardin:2019vio,
Bijnens:2019ghy,Colangelo:2019uex,Pauk:2014rta,Danilkin:2016hnh,Jegerlehner:2017gek,Knecht:2018sci,Eichmann:2019bqf,Roig:2019reh,Blum:2019ugy,
Aoyama:2012wk,Aoyama:2019ryr,Czarnecki:2002nt,Gnendiger:2013pva,Davier:2017zfy,Keshavarzi:2018mgv,Colangelo:2018mtw,Hoferichter:2019gzf,Davier:2019can,
Keshavarzi:2019abf,Kurz:2014wya,Melnikov:2003xd,Masjuan:2017tvw,Colangelo:2017fiz,Hoferichter:2018kwz,Gerardin:2019vio,Bijnens:2019ghy,Colangelo:2019uex,
Pauk:2014rta,Danilkin:2016hnh,Jegerlehner:2017gek,Knecht:2018sci,Eichmann:2019bqf,Roig:2019reh,Blum:2019ugy,Colangelo:2014qya,Aoyama:2012wk,
Aoyama:2019ryr,Czarnecki:2002nt,Gnendiger:2013pva,Davier:2017zfy,Keshavarzi:2018mgv,Colangelo:2018mtw,Hoferichter:2019gzf,Davier:2019can,Keshavarzi:2019abf,
Kurz:2014wya,Melnikov:2003xd,Masjuan:2017tvw,Colangelo:2017fiz,Hoferichter:2018kwz,Gerardin:2019vio,Bijnens:2019ghy,Colangelo:2019uex,Blum:2019ugy,
Colangelo:2014qya,Aoyama:2020ynm} and the Brookhaven experimental measurement~\cite{Bennett:2006fi} over the last decade
\begin{eqnarray}
\Delta a_\mu\equiv a_\mu^{\rm exp}({\rm BNL2006}) - a_\mu^{\rm SM} = (2.74\pm 0.73)\times 10^{-9}.
\end{eqnarray}
Very recently, Fermilab released the Run 1 result of the muon $g-2$ measurement and their combination led to a $4.2\sigma$ tension~\cite{1856534,1856531,FNAL}~\footnote{Note that there is debate about the discrepancy between the $g-2$ experiments and the SM prediction. A recent lattice computation of the SM prediction shows no significant tension with the
FNAL experimental determination~\cite{Borsanyi:2020mff}.}
\begin{eqnarray}
\Delta a_\mu^{\rm NEW}\equiv a_\mu^{\rm exp}({\rm FNAL2021}+{\rm BNL2006}) - a_\mu^{\rm SM} = (2.51\pm 0.59)\times 10^{-9}.
\end{eqnarray}
This new measurement confirms the muon magnetic moment anomaly and may shed light on NP beyond the SM. Many NP solutions have been proposed to explain the result of muon $g-2$ deviation~\cite{Padley:2015uma,Sabatta:2019nfg,Li:2020dbg,Bigaran:2020jil,Yin:2020afe,Baker:2021yli,Bodas:2021fsy,Chen:2021rnl,Yin:2021yqy,Chiang:2021pma,CarcamoHernandez:2021iat,Lee:2021jdr,Crivellin:2021rbq,Endo:2021zal,Iwamoto:2021aaf,Han:2021gfu,Arcadi:2021cwg,Criado:2021qpd,
Zhu:2021vlz,Gu:2021mjd,Wang:2021fkn,VanBeekveld:2021tgn,Nomura:2021oeu,Anselmi:2021chp,Yin:2021mls,Wang:2021bcx,Buen-Abad:2021fwq,Das:2021zea,Abdughani:2021pdc,
Chen:2021jok,Ge:2021cjz,Cadeddu:2021dqx,Brdar:2021pla,Cao:2021tuh,Chakraborti:2021dli,Ibe:2021cvf,Cox:2021gqq,Babu:2021jnu,Han:2021ify,Heinemeyer:2021zpc,
Calibbi:2021qto,Amaral:2021rzw,Bai:2021bau,Baum:2021qzx,Li:2021poy,Zu:2021odn,Keung:2021rps,Ferreira:2021gke,Zhang:2021gun,Ahmed:2021htr,Zhou:2021cfu,Yang:2021duj,
Athron:2021iuf,Chen:2021vzk,Chun:2021dwx,Escribano:2021css,Aboubrahim:2021rwz,Bhattacharya:2021ggm}.

We consider possible explanations of muon $g-2$ in terms of CLFV interactions.
The underlying CLFV interactions may contribute to the calculation of the muon magnetic moment and violate the flavor symmetry twice in loop diagrams.
For a new gauge boson $Z'$ or a new scalar coupled to bilinear leptons with the CLFV coupling denoted by $y$, muon $g-2$ anomaly would constrain the $\mu\mu$ component of the coupling combination $(y^\dagger y)^{\mu\mu}$ and thus the LFV component $y^{\ell\ell'} (\ell\neq \ell')$. Suppose a dominant LFV coupling $y^{\ell\ell'}$, the transition with the broken lepton flavor symmetry $|\Delta(L_\ell-L_{\ell'})|=4$ is present at tree level and the corresponding measurement can reveal the nature of underlying CLFV.
If the lepton flavor conservation of both electron and muon is violated, the probability of muonium-antimuonium conversion $\mu^+ e^-\to \mu^- e^+$ with $|\Delta(L_\mu-L_e)|=4$ is sensitive to the probe of CLFV. After two decades since the search for the muonium-antimuonium conversion at the Paul Scherrer Institut (PSI)~\cite{Willmann:1998gd}, the Muonium-to-Antimuonium Conversion Experiment (MACE)~\cite{MACE:2020} was recently proposed as the next generation experiment to measure this CLFV transition. MACE is expected to improve the sensitivity to the muonium-antimuonium conversion by two orders of magnitude beyond the PSI experiment.

Apart from the implication of low-energy precision experiments for CLFV, the CLFV may well show up at high-energy collider such as the Large Electron-Positron Collider (LEP), the Large Hadron Collider (LHC) or the future $e^+e^-$ colliders. The DELPHI collaboration at LEP interpreted their searches for $e^+e^-\to \ell^+\ell^-$ in terms of four-lepton operators~\cite{Abdallah:2005ph}. Their constraints can be directly applied to underlying bilepton models and CLFV couplings by comparing the effective Lagrangian when the new particle mass is much larger than $\sqrt{s}$. The lepton colliders can directly produce CLFV signature via either on-shell or off-shell bilepton production~\cite{Dev:2017ftk,Li:2018cod,Li:2019xvv}. In this work we go beyond the lepton colliders and emphasize that at hadron colliders the bileptons can also be emitted from one of the opposite-sign leptons in the final states of Drell-Yan process through $\gamma/Z$ exchange. Their leptonic decays next induce the CLFV processes
\begin{eqnarray}
pp\to \gamma^\ast/Z^\ast\to \ell_1^\pm\ell_1^\pm\ell_2^\mp\ell_2^\mp\;,
\end{eqnarray}
which are mediated by the bileptons with $\ell_{1,2}=e,\mu,\tau$ and $\ell_1\neq\ell_2$.
This production scenario violates the lepton flavor symmetry by $|\Delta(L_{\ell_1}-L_{\ell_2})|=4$ as well and is able to probe one single CLFV coupling $y^{\ell_1\ell_2}$ for $\Delta L=0$ bileptons. Note that the $\Delta L=2$ bileptons with single CLFV coupling $y^{\ell_1\ell_2}$ in turn induce lepton flavor conserving processes $pp\to \ell_1^+ \ell_1^- \ell_2^+ \ell_2^-$ and will thus not be considered in this work.

In this work we focus on the CLFV interpretation of the muon magnetic moment anomaly, and investigate the current constraints and the future search potential of MACE and LHC for CLFV bileptons. We also demonstrate the potential of a muon collider to probe the region of parameter space favored by the muon AMM in a two Higgs doublet model.
The CLFV is predicted by many extensions of the SM (see Refs.~\cite{Lindner:2016bgg,Calibbi:2017uvl} for recent reviews), including neutrino
mass models~\cite{Tommasini:1995ii,Cai:2017jrq} as well as the multi-Higgs doublet models~\cite{Branco:2011iw,Iguro:2019sly} or the non-minimal supersymmetric models~\cite{Calibbi:2017uvl,Raidal:2008jk,Han:2020exx}. We construct the most general interactions which couple two charged leptons to either a scalar or a vector bilepton. The interactions are obtained by expanding the most general SM gauge invariant Lagrangian
in terms of explicit leptonic fields. We consider the most relevant constraints from the anomalous magnetic moments (AMMs) of leptons and other constraints from the violation of lepton flavor universality (LFU), electroweak precision observables and previous collider searches at the LEP. We then emphasize the projected sensitivity of future muonium-antimuonium conversion measurement and the LHC to the individual CLFV couplings satisfying the above constraints.
In particular, we demonstrate that currently-available LHC data of $139$ fb$^{-1}$ taken at centre of mass energy $\sqrt{s}=13$ TeV is already able to probe parameter space which is not constrained by other experiments, and we encourage the ATLAS and CMS collaborations to perform a dedicated study of CLFV.

The paper is organized as follows. In Sec.~\ref{sec:Lagrangian} we describe the general Lagrangian for the bileptons with possible CLFV couplings to charged leptons. Then we evaluate the CLFV contribution to the muon magnetic moment anomaly. The constraints from other low-energy experiments are discussed in Sec.~\ref{sec:Const}. In Sec.~\ref{sec:CLFV-LHC} we simulate the CLFV signatures at the LHC and present the projected sensitivity to the CLFV couplings. Finally, in Sec.~\ref{sec:Con} we summarize our conclusions.

\section{Lagrangian for bileptons and leptonic anomalous magnetic moments}
\label{sec:Lagrangian}

In this work we consider the interactions of all $\Delta L=0$ bosonic bileptons with CLFV couplings~\cite{Cuypers:1996ia}. They are obtained by expanding the most general SM gauge invariant Lagrangian in terms of bilinear leptonic fields.
The Lagrangian of $\Delta L=0$ bileptons has four terms
\begin{align}
        \mathcal{L}_{\Delta L=0}&=
        y_1^{ij} V_{1\mu}^0 \bar{L}_i \gamma^\mu P_L L_j
+ y_1^{\prime ij} V_{1\mu}^{\prime 0} \bar{\ell}_i \gamma^\mu P_R \ell_j
+ \left( y_2^{ij} H_{2\alpha} \bar{L}_{i\alpha} P_R \ell_j + h.c. \right)
+ y_3^{ij} \bar{L}_i\gamma^\mu \vec{\sigma}\cdot \vec{V}_{3\mu} L_j
\nonumber\\&
=
\left(
        y_1^{ij} V_{1\mu}^0 \bar{\ell}_i \gamma^\mu P_L \ell_j
+y_1^{ij} V_{1\mu}^0 \bar{\nu}_i \gamma^\mu P_L \nu_j
\right)
+ y_1^{\prime ij} V_{1\mu}^{\prime 0} \bar{\ell}_i \gamma^\mu P_R \ell_j
\nonumber \\&
+ \left( y_2^{ij} H_{2}^+ \bar{\nu}_i P_R \ell_j + y_2^{ij} H_{2}^0 \bar{\ell}_i P_R \ell_j + h.c. \right)
\nonumber \\&
+ \left(y_3^{ij}\sqrt{2}V_{3\mu}^-\bar{\ell}_i\gamma^\mu P_L \nu_j + y_3^{ij}\sqrt{2}V_{3\mu}^+\bar{\nu}_i\gamma^\mu P_L \ell_j - y_3^{ij}V_{3\mu}^0\bar{\ell}_i \gamma^\mu P_L \ell_j\right)\; ,
        \label{eq:delL0}
\end{align}
where $L_i=(\nu_i, \ell_i)$ denotes the left-handed SM lepton doublet with the flavor index $i$.
The subscript of the new bosonic fields, i.e.~1, 2 or 3, manifests their SU$(2)_L$ nature as singlet, doublet or triplet, respectively. The couplings $y_1^{(\prime)}$ and $y_3$ may respectively arise from new gauge interactions with an Abelian LFV $Z^\prime$ or a SU(2)$_L$ triplet gauge boson and $y_2$ naturally appears in two Higgs doublet models with a complex neutral scalar $H_2^0=(h_2 + i a_2)/\sqrt{2}$. The couplings $y_1$, $y_1^\prime$ and $y_3$ are hermitian, while $y_2$ may take any values.
The non-zero elements of the above couplings $y^{(\prime)ij}_{1,2,3}$ can lead to the presence of CLFV processes. See Refs.~\cite{Cuypers:1996ia,Li:2018cod,Li:2019xvv} for a discussion of the CLFV for $\Delta L=2$ bileptons.

\begin{figure}[htb!]
\centering
\includegraphics[width=0.7\textwidth]{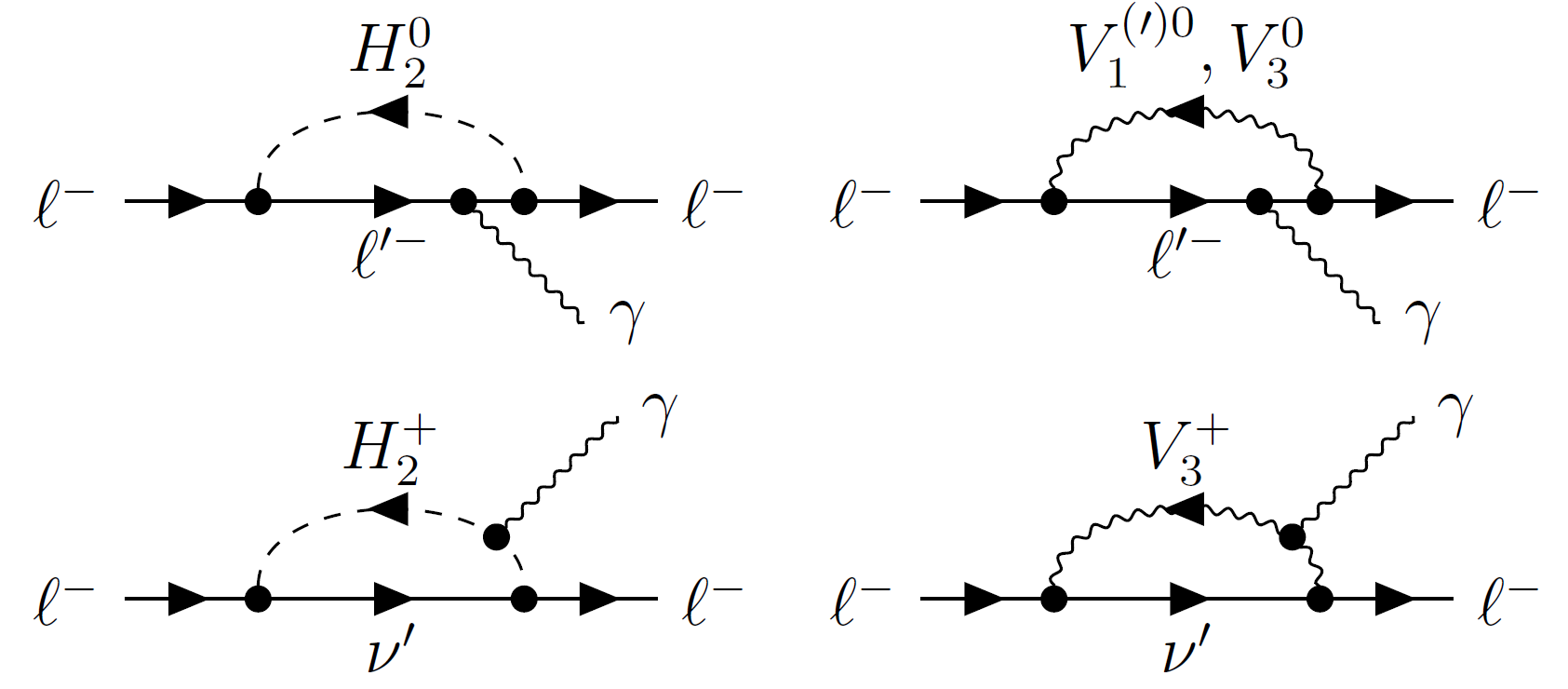}
\caption{
Diagrams contributing to the leptonic anomalous magnetic moments by the Lagrangian in Eq.~(\ref{eq:delL0}).
}
\label{fig:AMM}
\end{figure}

The bileptons contribute to the leptonic anomalous magnetic moments through the one-loop diagrams shown in Fig.~\ref{fig:AMM}.
We follow the general formulas provided by Lavoura in Ref.~\cite{Lavoura:2003xp} and calculate the leading
contributions of the above bileptons to the anomalous magnetic moment of
the lepton $\ell$~\cite{Li:2018cod,Li:2019xvv}.
The general leading order contributions for the $\Delta L=0$ bileptons are given by
\begin{align}
        \Delta a_{\ell}(V_1^{(\prime)0})
        &= \frac{(y_1^{(\prime)\dagger} y_1^{(\prime)})^{\ell\ell}}{12\pi^2} \frac{m_\ell^2}{m_{V_1^{(\prime)0}}^2} \geq 0 \ ,
        \\\nonumber
        \Delta a_\ell(V_3) &= \frac{(y_3^\dagger y_3)^{\ell\ell}}{12\pi^2} \frac{m_\ell^2}{m_{V_3^0}^2} -\frac{5(y_3^\dagger y_3)^{\ell\ell}}{24\pi^2} \frac{m_\ell^2}{m_{V_3^+}^2} \ ,
        \\\nonumber
               \Delta a_\ell(H_2) =&
        -\frac{(y_2^\dagger y_2+y_2y_2^\dagger)^{\ell\ell}}{96\pi^2} \left(\frac{m_\ell^2}{m_{h_2}^2}+ \frac{m_\ell^2}{m_{a_2}^2}  \right) + \frac{ (y_2^\dagger y_2)^{\ell\ell}}{96\pi^2} \frac{m_\ell^2}{m_{H_2^+}^2}
        \\\nonumber &
	+\sum_k \frac{\mathrm{Re}[y_2^{k\ell}y_2^{\ell k}]}{16\pi^2}  \frac{m_\ell}{m_k}
	\left[
		f\left(\tfrac{m_k^2}{m_{h_2}^2}\right)
		-f\left(\tfrac{m_k^2}{m_{a_2}^2}\right)
	\right]\;,
\end{align}
where $k$ denotes the charged lepton flavor in the loop, $f(x) = x (\ln x+\tfrac32)$ and $f(x)$ is negative for $x<e^{-2/3}\simeq0.22$ and thus $m_{h_2,a_2}>2.1 m_k$.
We are particularly interested in the interpretation of the newly measured muon anomalous magnetic moment $\Delta a_\mu$.
The contributions of the bileptons $V_1^{(\prime)0}$
to the anomalous magnetic moment are positive
and thus they are able to explain the discrepancy between the
SM prediction and the measurement of the anomalous magnetic
moment. For the singlet vector bileptons $V_1^{(\prime)0}$ we find the preferred coupling as
\begin{align}
|y_1^{(\prime)\mu\alpha}| &= (0.00516\pm0.00061) \frac{m_{V_1^{(\prime)0}}}{\mathrm{GeV}}
\end{align}
under the assumption of one non-vanishing off-diagonal coupling $y_1^{(\prime)\mu\alpha}$ with $\alpha=e,\tau$.
The contributions from the bileptons $H_2$ and $V_3$ can have either sign. We find for the triplet vector bilepton $V_3$
\begin{align}
|y_3^{\mu\alpha}| &= (0.00516\pm0.00061) \frac{m_{V_3^{(\prime)0}}}{\mathrm{GeV}}
\left(1-\frac{5m_{V_3^0}^2}{2m_{V_3^+}^2}\right)^{-1/2}
\end{align}
with $\alpha=e,\tau$ and thus $V_3$ can only explain the muon AMM for $5m_{V_3^0}^2 < 2m_{V_3^+}^2$.
Finally, the contribution of the electroweak doublet scalar $H_2$ can be
split into three parts $\Delta a_\ell =A_1+A_2+A_3$ with
\begin{align}
	A_1&=-\frac{(y_2^\dagger y_2+y_2y_2^\dagger)^{\ell\ell}}{96\pi^2} \left(\frac{m_\ell^2}{m_{h_2}^2}+ \frac{m_\ell^2}{m_{a_2}^2}  \right)
	\;,
	\\
	A_2 & =+ \frac{ (y_2^\dagger y_2)^{\ell\ell}}{96\pi^2} \frac{m_\ell^2}{m_{H_2^+}^2}
	\;,
	\\
	A_3 &=+\sum_k \frac{\mathrm{Re}[y_2^{k\ell}y_2^{\ell k}]}{16\pi^2}  \frac{m_\ell}{m_k}
	\left[
		f\left(\tfrac{m_k^2}{m_{h_2}^2}\right)
		-f\left(\tfrac{m_k^2}{m_{a_2}^2}\right)
	\right]
	\;.
\end{align}
The first two terms have a definite negative and positive sign, respectively. The sign of the third contribution depends on the neutral scalar masses and Yukawa couplings. A positive contribution to the muon AMM can thus be obtained in various ways, e.g. for $m_{H_2^+}\ll m_{h_2,a_2}$ the second contribution dominates.

The light neutral scalars are of our interest and thus we consider two decoupled cases\footnote{
We show in App.~\ref{app:ST} that the $S$ and $T$ parameter do not pose any constraint on the masses of the $H_2$ components in the limit of vanishing mixing of the CP even scalars.}
	(A) $m_{h_2}\ll m_{a_2}, m_{H_2^+}$ and (B) $m_{a_2}\ll m_{h_2}, m_{H_2^+}$. In these two extremely limiting cases, the contribution to the muon AMM can be rewritten as
\begin{align}
	\Delta a_\mu ({\rm A}) & = -\frac{(y_2^\dagger y_2 + y_2 y_2^\dagger)^{\mu\mu}}{96\pi^2} \frac{m_\mu^2}{m_{h_2}^2} + \sum_k \frac{\mathrm{Re}[y_2^{k\ell} y_2^{\ell k}]}{16\pi^2} \frac{m_\mu}{m_k} f\left(\frac{m_k^2}{m_{h_2}^2}\right)\;,
	\\
	\Delta a_\mu ({\rm B}) & = -\frac{(y_2^\dagger y_2 + y_2 y_2^\dagger)^{\mu\mu}}{96\pi^2} \frac{m_\mu^2}{m_{a_2}^2} - \sum_k \frac{\mathrm{Re}[y_2^{k\ell} y_2^{\ell k}]}{16\pi^2} \frac{m_\mu}{m_k} f\left(\frac{m_k^2}{m_{a_2}^2}\right)\;,
\end{align}
which differ in the sign of the second term. As $f(\frac{m_k^2}{m_{h_2,a_2}^2})$ tends to vanish when $\frac{m_k^2}{m_{h_2,a_2}^2}\ll 1$ and thus the second term is suppressed for small value of $m_k/m_{h_2,a_2}$, the contribution to the muon AMM is generally negative if only the $e-\mu$ couplings are non-zero $y_{2}^{e\mu,\mu e}\neq 0$. The second term however can dominate over the first term if the $\mu-\tau$ coupling is non-zero $y_2^{\mu\tau,\tau\mu}\neq0$, because the second term is enhanced by the $\tau$ lepton mass. If the new neutral scalar is heavier than $2.1 m_\tau$, the function $f$ is negative. Hence, for the masses of interest, a positive contribution to the muon AMM requires antisymmetric (symmetric) Yukawa couplings in case A (B). For real Yukawa couplings, the general contribution of the neutral scalars is finally
\begin{align}
\Delta a_\mu (X=h_2, a_2) & = -\frac{|y_2^{\mu\tau}|^2}{48\pi^2} \frac{m_\mu^2}{m_{X}^2} - \frac{|y_2^{\mu\tau}|^2}{16\pi^2} \frac{m_\mu}{m_\tau} f\left(\frac{m_\tau^2}{m_{X}^2}\right)\;.
\end{align}
Numerically, the Yukawa couplings $y_2^{\mu\tau}$ has to satisfy
\begin{align}
		|y_2^{\mu\tau}| & =
		(0.00145\pm0.00017) \,  \frac{m_X}{\mathrm{GeV} }\, \left(\ln\frac{m_X^2}{m_\tau^2}-\frac32\right)^{-1}
 \left(1 + \frac{m_\mu m_\tau}{3 m_{X}^2 f(m_\tau^2/m_X^2)}\right)^{-1/2}
\end{align}	
with $X=h_2,a_2$ in order to explain the muon anomalous magnetic moment.

The electron anomalous magnetic moment has been measured using a one-electron quantum cyclotron~\cite{Hanneke:2008tm,Hanneke:2010au}.
The SM prediction depends on the electromagnetic fine-structure constant and thus the fine-structure constant is an important input to $\Delta a_e \equiv a_e^{\rm exp} -a_e^{\rm SM}$. The most recent measurements of the fine-structure constant using cesium atoms in a matter-wave interferometer~\cite{Parker:2018vye} and rubidium~\cite{Morel:2020dww} lead to discrepant results which directly imprint on the SM prediction and thus $\Delta a_e$. The fine-structure constant measured using cesium atoms~\cite{Parker:2018vye} yields a $2.4\sigma$ discrepancy
\begin{eqnarray}
	\Delta a_e
	({\rm Berkeley2018})
= (-8.8\pm 3.6)\times 10^{-13}\;,
\end{eqnarray}
while the measurement using rubidium atoms~\cite{Morel:2020dww} results in a different sign with a $1.6\sigma$ discrepancy
\begin{eqnarray}
\Delta a_e
({\rm LKB2020}) = (4.8\pm 3.0)\times 10^{-13}.
\end{eqnarray}
To accommodate both the muon and electron AMMs, we indicate the conservative $3\sigma$ bound for the latest LKB2020 result in the figures below.

\section{Low-energy constraints}
\label{sec:Const}

The relevant low-energy constraints on CLFV have been calculated in Refs.~\cite{Li:2018cod,Li:2019xvv}. They include leptonic anomalous magnetic moments which have been discussed in the previous section, muonium-antimuonium conversion in nuclei, lepton flavor universality, and electroweak precision physics related to the determination of the Fermi constant.
The rare LFV lepton decays do not impose a constraint, suppose there is only one off-diagonal coupling~\cite{Li:2018cod}. We thus will not discuss them here. In Tables~\ref{tab:LFU} and \ref{tab:LEPbounds} we summarize the constraints from lepton flavor universality in lepton decays and the LEP constraints recast from the search for $e^+ e^- \to \ell^+ \ell^-$ at DELPHI~\cite{Abdallah:2005ph}.
We refer the readers to Ref.~\cite{Li:2019xvv} for details. We next only discuss the constraints with new experimental development in more detail, in particular
the proposed muonium-antimuonium experiment MACE~\cite{MACE:2020,Han:2021nod} as well as update the discussion of constraints from electroweak precision physics in light of the Cabibbo angle anomaly~\cite{Belfatto:2019swo,Grossman:2019bzp,Coutinho:2019aiy}.

\begin{table}
\begin{tabular}{|c|c|c|}\hline
		 & $R_{\mu e}$ & $R_{\tau\mu}$ \\\hline
	 \multirow{2}{*}{$V_1^0$}	& $(y_1^{\mu\tau})^2 < 4.9\times 10^{-8} m_{V_1^0}^2$
	& $(y_1^{e\tau})^2 < 5.6\times 10^{-8} m_{V_1^0}^2$ \\
    & $(y_1^{e\tau})^2 < 1.6\times 10^{-7} m_{V_1^0}^2$ & $(y_1^{e\mu})^2 < 1.3\times 10^{-7} m_{V_1^0}^2$
		\\\hline
    \multirow{2}{*}{$H_2$}
    & $|y_2^{\mu\tau}|^2< 6.5\times 10^{-6} m_{H_2^+}^2$ & $|y_2^{e\tau}|^2< 5.8\times 10^{-6} m_{H_2^+}^2$ \\
    & $|y_2^{e\tau}|^2< 3.6\times 10^{-6} m_{H_2^+}^2$ & $|y_2^{e\mu}|^2< 3.8\times 10^{-6} m_{H_2^+}^2$
    \\ \hline
    \multirow{2}{*}{$V_3$}
    & $|y_3^{\mu\tau}|^2< 1.6\times 10^{-6} m_{V_3^+}^2$ & $|y_3^{e\tau}|^2< 1.5\times 10^{-6} m_{V_3^+}^2$ \\
    & $|y_3^{e\tau}|^2< 9.0\times 10^{-7} m_{V_3^+}^2$ & $|y_3^{e\mu}|^2< 9.6\times 10^{-7} m_{V_3^+}^2$
    \\ \hline
\end{tabular}
\caption{The constraints on the CLFV couplings in units of GeV$^{-2}$ from the lepton flavor universality of leptonic $\tau$ decays. The constraints on $V_1^0$ are from the interference with the SM and thus stronger.}
\label{tab:LFU}
\end{table}

\begin{table}[tb]\centering
\resizebox{\linewidth}{!}{\begin{tabular}{|c|c|c|c|}\hline
 & $e^+e^- \to e^+ e^-$ & $e^+e^- \to \mu^+\mu^-$ & $e^+e^- \to \tau^+\tau^-$\\\hline
        $V_{1,3}$ & $|y_{1,3}^{ee}|\leq 6.7\times 10^{-4} m_{V_{1,3}^0}$ & $\sqrt{|y_{1,3}^{ee}y_{1,3}^{\mu\mu} + y_{1,3}^{e\mu}y_{1,3}^{\mu e}|}\leq 4.9\times 10^{-4} m_{V_{1,3}^0}$ & $\sqrt{|y_{1,3}^{ee}y_{1,3}^{\tau\tau}+y_{1,3}^{e\tau}y_{1,3}^{\tau e}|}\leq 4.5\times 10^{-4} m_{V_{1,3}^0}$
\\\hline
$V_1^{\prime0}$ & $|y_1^{\prime ee}|\leq 6.8\times 10^{-4} m_{V_1^{\prime 0}}$ & $\sqrt{|y_1^{\prime ee}y_1^{\prime\mu\mu}+y_1^{\prime e\mu}y_1^{\prime\mu e}|}\leq 5.1\times 10^{-4} m_{V_1^{\prime 0}}$& $\sqrt{|y_1^{\prime ee}y_1^{\prime \tau\tau}+y_1^{\prime e\tau}y_1^{\prime\tau e}|}\leq 4.7\times 10^{-4} m_{V_1^{\prime 0}}$
\\\hline
\end{tabular}}
\caption{LEP limits on couplings for masses well above the center of mass energy $\sqrt{s}\sim 130-207$ GeV.
To make them valid for any masses, one should replace the masses by $(s\cos\theta/2+m^2)^{1/2}$ by averaging over the scattering angle $\langle\cos\theta\rangle\simeq 1/2$.
This limit does not apply for $H_2$ with non-degenerate neutral scalars $m_{h_2}\neq m_{a_2}$.
}
\label{tab:LEPbounds}
\end{table}

\subsection{Muonium-antimuonium oscillation}
The probability of muonium-antimuonium conversion has been firstly calculated in Refs.~\cite{Feinberg:1961zz,Feinberg:1961zza} and in Ref.~\cite{Conlin:2020veq} in effective field theory. For the Lagrangian described in Eq.~(\ref{eq:delL0}), we obtain the muonium-antimuonium conversion probabilities of the vector bileptons as follows~\cite{Li:2019xvv}
\begin{align}
	P(V_1^{(\prime)0})& =\frac{2|y_1^{(\prime)\mu e}|^4}{ \pi^2 a^6 \gamma^2 m_{V_1^{(\prime)0}}^4}  S_{XX}(B)\; ,
			 &
P(V_3)& =\frac{2|y_3^{\mu e}|^4}{ \pi^2 a^6 \gamma^2 m_{V_3^0}^4} S_{XX}(B)\;,
\end{align}
where $a$ denotes the Bohr radius $a=(1/\alpha)(m_e+m_\mu)/(m_e m_\mu)\simeq 1/\alpha m_e$ and the suppression factor is $S_{XX}(0.1 \ {\rm T})=0.36$ for a magnetic field $B=0.1$ T.
The conversion probability of the scalar bilepton is given by
\begin{align}
	P(H_2) & = \frac{1}{\pi^2 a^6 \gamma^2}\left[
		4 |C|^2 + \frac{\left|A-C\right|^2}{1+X^2}
		+\frac{\gamma^2\,\left|A+C\right|^2 }{\gamma^2+b^2 Y^2}
		+\frac{X^2}{1+X^2} \frac{\gamma^2\,\left(|A|^2+|C|^2\right) }{\gamma^2+b^2(1+X^2)}
	\right]\; ,
\end{align}
where $\gamma=G_F^2 m_\mu^5/192\pi^3$ is the muon decay width, $X,Y=\tfrac{\mu_B B}{b}(g_e \pm \frac{m_e}{m_\mu}g_\mu)$ parameterizes the Zeeman effect with $b=1.85\times 10^{-5}$ eV~\cite{Mariam:1982bq,Klempt:1982ge} and $A$ and $C$ are defined as
\begin{align}
	A & \equiv \frac{(y_2^{\mu e})^2+(y_2^{e\mu *})^2}{8} \left(\frac{1}{m_{h_2}^2} - \frac{1}{m_{a_2}^2}\right)\;,
	  &
	C&\equiv \frac{y_2^{\mu e} y_2^{e\mu*}}{4}\left(
	\frac{1}{m_{h_2}^2} + \frac{1}{m_{a_2}^2}\right)\;.
\end{align}
For the two cases of interest with real antisymmetric (symmetric) Yukawa couplings we find
\begin{align}
	P(h_2) & \simeq \frac{|y_2^{e\mu}|^4}{2\pi^2 a^6 \gamma^2 m_{h_2}^4}S_{h_2}(B)\;,
	       &
	P(a_2) & \simeq
	   \frac{\left|y_2^{e\mu}\right|^4}{2\pi^2a^6\gamma^2 m_{a_2}^4} S_{a_2}(B)\;,
\end{align}
which agree with each other at vanishing magnetic field, but slightly differ at finite magnetic field with $S_{h_2}(0.1\,\mathrm{T})=0.86$ and $S_{a_2}(0.1\,\mathrm{T})=0.5$.

The search for muonium-antimuonium conversion at the Paul Scherrer Institut (PSI) placed a constraint on the probability to observe the decay of the muon in antimuonium instead of the decay of the antimuon in muonium with a magnetic field of $B=0.1$ T, that is $P(B=0.1 \ {\rm T}) \leq 8.3\times 10^{-11}$~\cite{Willmann:1998gd}. This bound can be used to obtain the constraints on the CLFV couplings of the bileptons which we summarize in Table~\ref{tab:muonium}. MACE is proposed to improve the sensitivity to the muonium-antimuonium conversion by two orders of magnitude.

\begin{table}[tb!]
\begin{center}
\begin{tabular}{|c|c|}
\hline
 & $\mu^+ e^-\to \mu^- e^+$ \\\hline
$V_1^{(\prime)0}$
& $|y_1^{(\prime)e\mu}|^2< 2.0\times 10^{-7}\, m_{V_1^{(\prime)0}}^2 $\\\hline
\multirow{2}{*}{$H_2$}
& $|y_2^{e\mu}|^2< 2.6 \times 10^{-7}\, m_{h_2}^2$\\
& $|y_2^{e\mu}|^2< 3.4\times 10^{-7}\, m_{a_2}^2$\\\hline
$V_3$
& $|y_3^{e\mu}|^2< 2.0\times 10^{-7}\, m_{V_3^0}^2 $
\\\hline
\end{tabular}
\end{center}
	\caption{Constraints from muonium-antimuonium conversion on the CLFV couplings in units of ${\rm GeV}^{-2}$. Here we assume all the CLFV couplings are real and symmetric. For $H_2$, we provide the limits for the two cases of interest.
}
\label{tab:muonium}
\end{table}

\subsection{Fermi constant and electroweak precision physics}
\label{sec:FermiConstant}

The interactions of leptons with neutrinos lead to new contributions to effective operators with two leptons and two neutrinos, which can be written as
\begin{align}\label{eq:NSI}
	\mathcal{L} & =
	- 2\sqrt{2} G_F[\bar \nu_i\gamma_\mu P_L \nu_j]  [\bar \ell_k\gamma^\mu \left( g_{LL}^{ijkl} P_L + g_{LR}^{ijkl}P_R\right) \ell_l]\;.
\end{align}
In the SM both Wilson coefficients are generated by the exchange of electroweak gauge bosons. They can be expressed in terms of the weak mixing angle $\theta_W$ as
\begin{align}
	g_{LL,SM}^{ijkl} & = \left(-\frac12+\sin^2\theta_W\right) \delta_{ij}\delta_{kl} + \delta_{il}\delta_{jk}
	\qquad\mathrm{and}\qquad
	g_{LR,SM}^{ijkl}  = \sin^2\theta_W \delta_{ij}\delta_{kl}\;.
\end{align}
The contributions of the $\Delta L=0$ bileptons $V_{1,3}$ and $H_2$ to the Wilson coefficients are given by
\begin{align}
\label{gLL}
	g_{LL,NP}^{ijkl}
	& =-\frac{y_1^{ij}y_1^{kl} }{2\sqrt{2}G_F m_{V_1^0}^2}
	-\frac{y_3^{kj}y_3^{il}}{\sqrt{2} G_F m_{V_3^+}^2}\;,
	&
g_{LR,NP}^{ijkl}
& =\frac{y_2^{il}y_2^{jk*}}{4\sqrt{2} G_Fm_{H_2^+}^2}\;.
\end{align}
In particular there is a contribution to muon decay $\mu\to e \nu_\mu \bar \nu_e$, which is generally used to measure the Fermi constant. To leading order it is given by
\begin{equation}\label{eq:muondecay}
	\Gamma(\mu\to e\nu_\mu \bar\nu_e(\gamma)) = \frac{G_F^2 m_\mu^5}{192\pi^3}\, \left(|g_{LL}^{\mu ee \mu}|^2 + |g_{LR}^{\mu ee \mu}|^2\right) \;.
\end{equation}
Thus the Fermi constant extracted in muon decay is given by
\begin{equation}
G_{F,\mu}^2  = G_F^2\left( |1+g_{LL,NP}^{\mu ee\mu}|^2
+ \sum_{\alpha,\beta}{}^\prime |g_{LL,NP}^{\alpha\beta e\mu}|^2
+ \sum_{\alpha,\beta} |g_{LR,NP}^{\alpha\beta e\mu}|^2\right)\; ,
\end{equation}
where the prime on the summation sign indicates that we are not summing over the interfering component with $(\alpha,\beta)=(\mu,e)$. Taking $G_{F,\mu}$ as input, we find to leading order the modification of the Fermi constant in terms of different Wilson coefficients
\begin{equation}
G_{F}  = G_{F,\mu} \left( 1 + \delta G_F \right)\; ,
\qquad \delta G_F \equiv - \mathrm{Re}(g_{LL,NP}^{\mu ee\mu}) -\frac12 \sum_{\alpha,\beta}{}^\prime |g_{LL,NP}^{\alpha\beta e\mu}|^2 - \frac12\sum_{\alpha,\beta} |g_{LR,NP}^{\alpha\beta e\mu}|^2\;.
\end{equation}
This change of the Fermi constant leads to the modifications of other observables. Previously some of us derived constraints on the shift in the Fermi constant~\cite{Li:2019xvv} from the weak mixing angle, the $W$ boson mass, and the unitarity of the CKM matrix. Due to the Cabibbo angle anomaly~\cite{Belfatto:2019swo,Grossman:2019bzp,Coutinho:2019aiy}, we do not impose the constraint from the unitarity of the CKM matrix. Thus the most stringent constraint comes from the weak mixing angle with
\begin{align}
	-0.00056 & < \delta G_F < 0.00062 \;.
\end{align}
This translates into a constraint on the Wilson coefficients
\begin{align}
	-0.00062 < \mathrm{Re}(g_{LL,\rm NP}^{\mu ee\mu}) < 0.00056\;,
	\qquad\qquad
	|g_{LL,\rm NP}^{\alpha\beta e\mu}|,|g_{LR,\rm NP}^{\alpha\beta e\mu}| <0.033\;.
\end{align}
The constraints for the different bileptons are collected in Table~\ref{tab:GF}. Other constraints from non-standard neutrino interactions are less stringent~\cite{Li:2019xvv} and we thus do not discuss them here.

\begin{table}[bt]
\begin{tabular}{|c|c|}\hline
			  & Electroweak precision physics 
 \\\hline
	 $V_1^0$
	 & $(y_1^{e\mu})^2 < 1.3\times10^{-8} m_{V_1^0}^2$
	 \\\hline
		 $H_2$
		 & $|y_2^{e\mu}|^2 < 2.2\times 10^{-6} m_{H_2^+}^2$
		 \\ \hline
		 $V_3$
		 & $|y_3^{e\mu}|^2 < 5.4\times 10^{-7} m_{V_3^+}^2$
		 \\ \hline
\end{tabular}
\caption{Constraints from the electroweak precision physics	on the CLFV couplings in units of GeV$^{-2}$.}
\label{tab:GF}
\end{table}

\section{The search for CLFV at the LHC and lepton colliders}
\label{sec:CLFV-LHC}

\textbf{CLFV at the LHC}:
The bileptons can be emitted from one of the opposite-sign leptons in the Drell-Yan process at a hadron collider.
Their production followed by leptonic decays leads to the CLFV processes with four leptons in final states.
We consider the CLFV processes $pp\to \gamma^\ast/Z^\ast\to \ell_1^\pm\ell_2^\mp X\to\ell_1^\pm\ell_1^\pm\ell_2^\mp\ell_2^\mp$ mediated by the bilepton $X$ with $\ell_{1,2}=e,\mu,\tau$ and $\ell_1\neq\ell_2$.
This production scenario violates the lepton flavor symmetry by $|\Delta(L_{\ell_1}-L_{\ell_2})|=4$ and is induced by one single coupling $y^{\ell_1\ell_2}$ for $\Delta L=0$ bileptons or the coupling product $\lambda^{\ell_1\ell_1}\lambda^{\ell_2\ell_2}$ for $\Delta L=2$ bileptons.
The coupling product $\lambda^{\ell_1\ell_1}\lambda^{\ell_2\ell_2}$ for $\Delta L=2$ bileptons is not directly related to the muon magnetic moment anomaly. For each CLFV signal process, we only study the $\Delta L=0$ bileptons for illustration and assume one dominant single LFV coupling of the bilepton.

The bilepton model files are produced by FeynRules~\cite{Alloul:2013bka} and are interfaced with MadGraph5~\cite{Alwall:2014hca} to generate signal events. The major SM backgrounds are from $\tau^+\tau^-\tau^+\tau^-$, $t\bar{t}t\bar{t}$, $t\bar{t}WW$, $t\bar{t}WZ$, $WWWW$ and $WWWZ$ followed by the parent particles' leptonic decays. For the signal events with tau leptons in final states, the SM backgrounds also include $WZ$+jets with the jets misidentified as tau. The tau leptons are then considered to decay hadronically.
The signal and background events are then passed to Pythia 8~\cite{Sjostrand:2014zea} for parton shower and to Delphes 3~\cite{deFavereau:2013fsa} for detector simulation.
We select the events with exactly two groups of same-sign leptons $\ell_1^\pm \ell_1^\pm$ and $\ell_2^\mp\ell_2^\mp$ satisfying the parton level cuts
\begin{eqnarray}
p_T(\ell)>10~{\rm GeV},~|\eta(\ell)|< 2.5,~\Delta R_{\ell\ell'}>0.4\;,
\end{eqnarray}
where $\ell,\ell'$ are any charged leptons in the events.
Our signal consists of four pure charged leptons and there are always neutrinos produced by the leptonic decay of tau or $W$ in SM background events.
To veto the backgrounds with missing neutrinos, we apply a maximal cut on the transverse missing energy $\slashed{E}_T<20$~GeV.
We also reject the backgrounds with top quarks by requiring no b-jets in the events.
Finally, an invariant mass cut is applied for the four leptons in the high mass region
\begin{eqnarray}
m_{4\ell}>m_X~{\rm for}~m_X> 60~{\rm GeV} \;,
\end{eqnarray}
where $X$ denotes any bileptons mentioned above.
After the above selection cuts, the SM backgrounds are significantly reduced. Note that we do not require the missing energy and the invariant mass cuts when selecting the signal events with tau leptons in order to enhance the signal selection efficiency.

\begin{figure}[phtb!]
\centering
\includegraphics[width=0.32\textwidth]{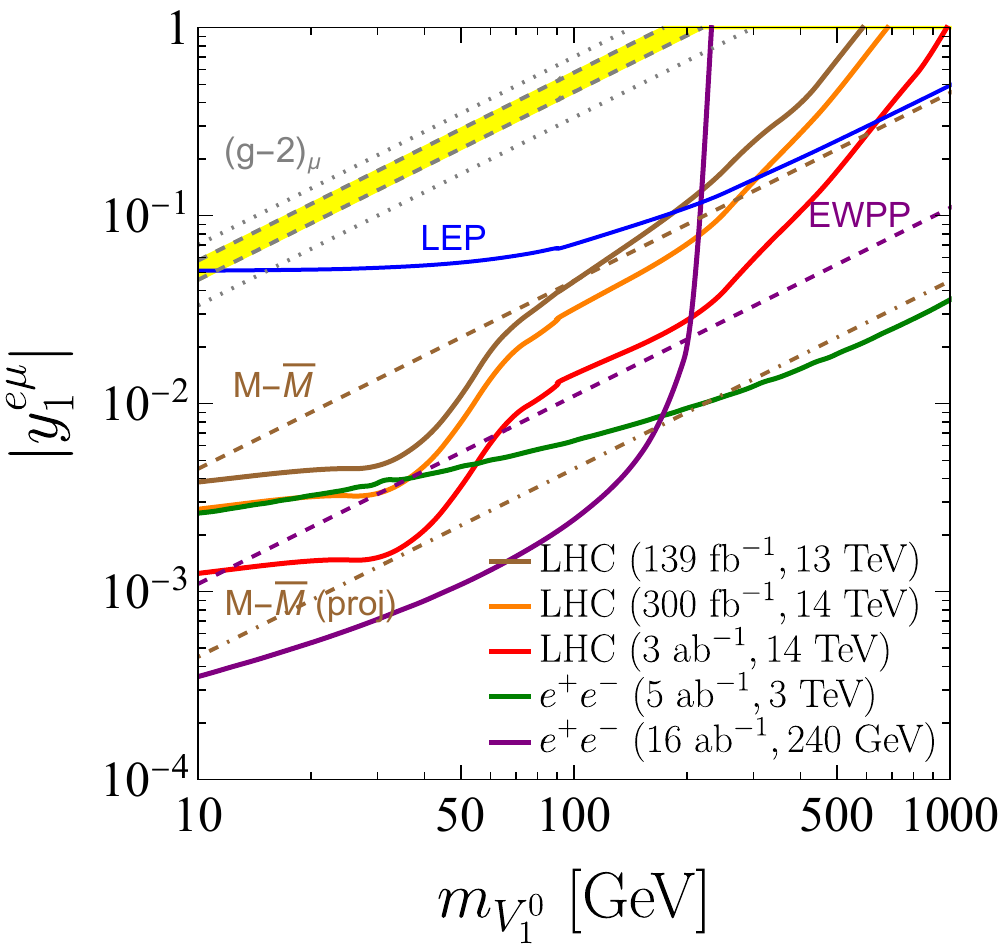}
\includegraphics[width=0.32\textwidth]{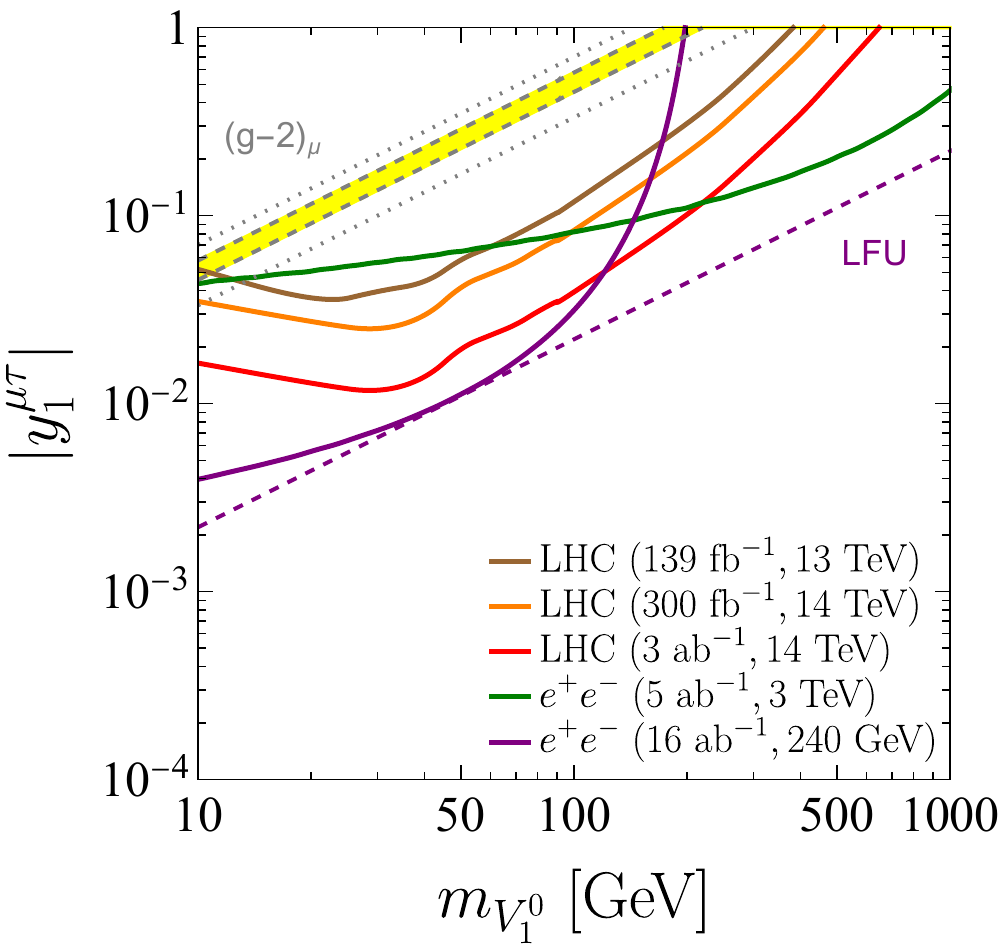}
\includegraphics[width=0.32\textwidth]{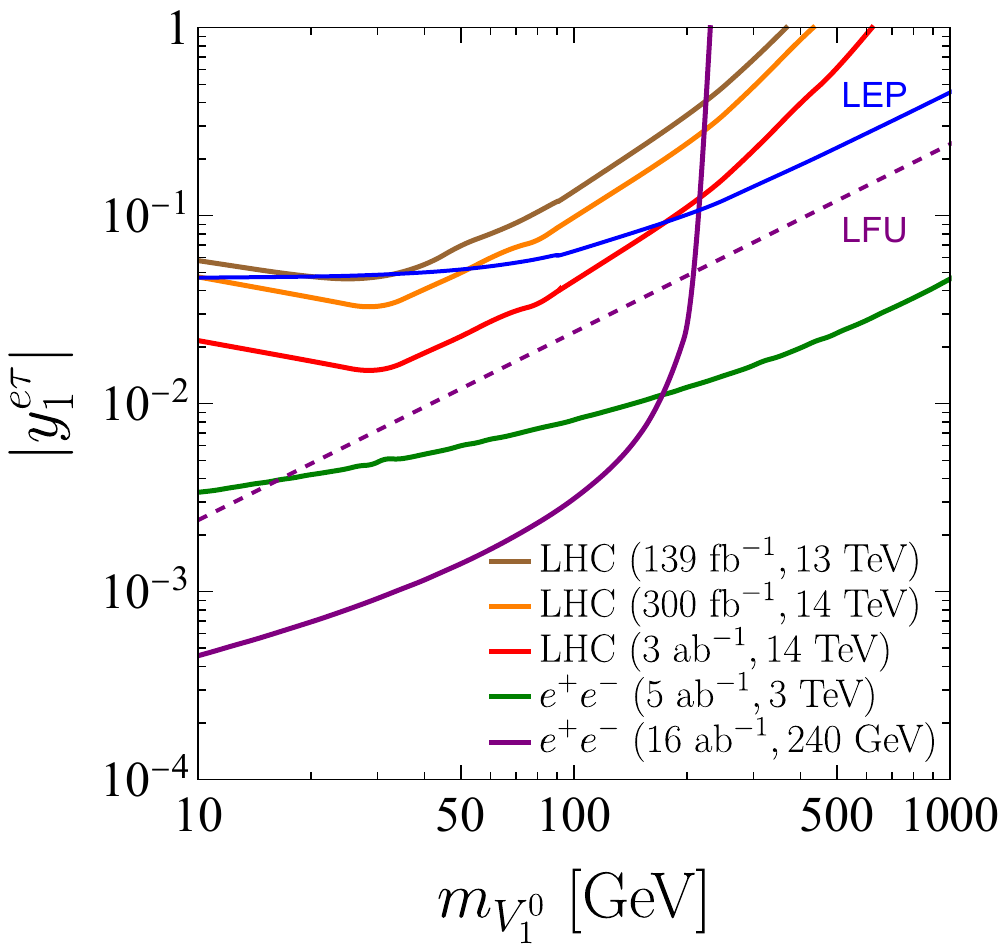}\\
\includegraphics[width=0.32\textwidth]{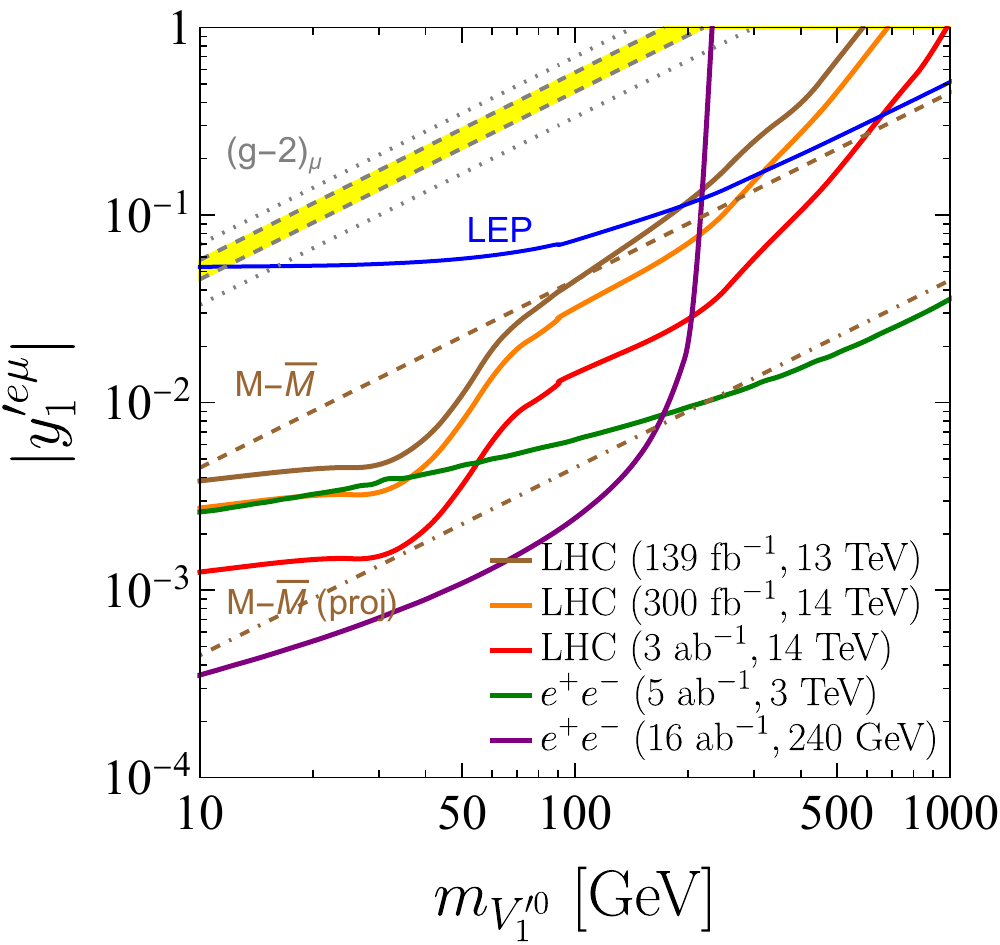}
\includegraphics[width=0.32\textwidth]{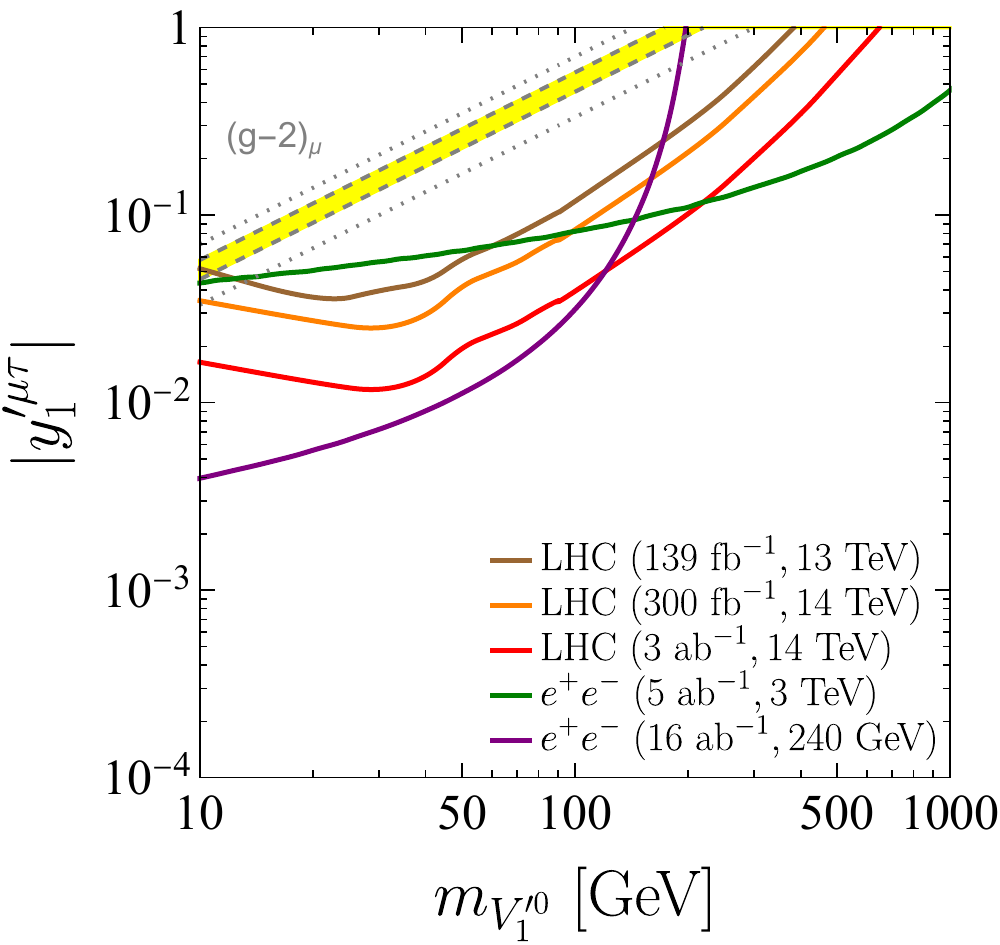}
\includegraphics[width=0.32\textwidth]{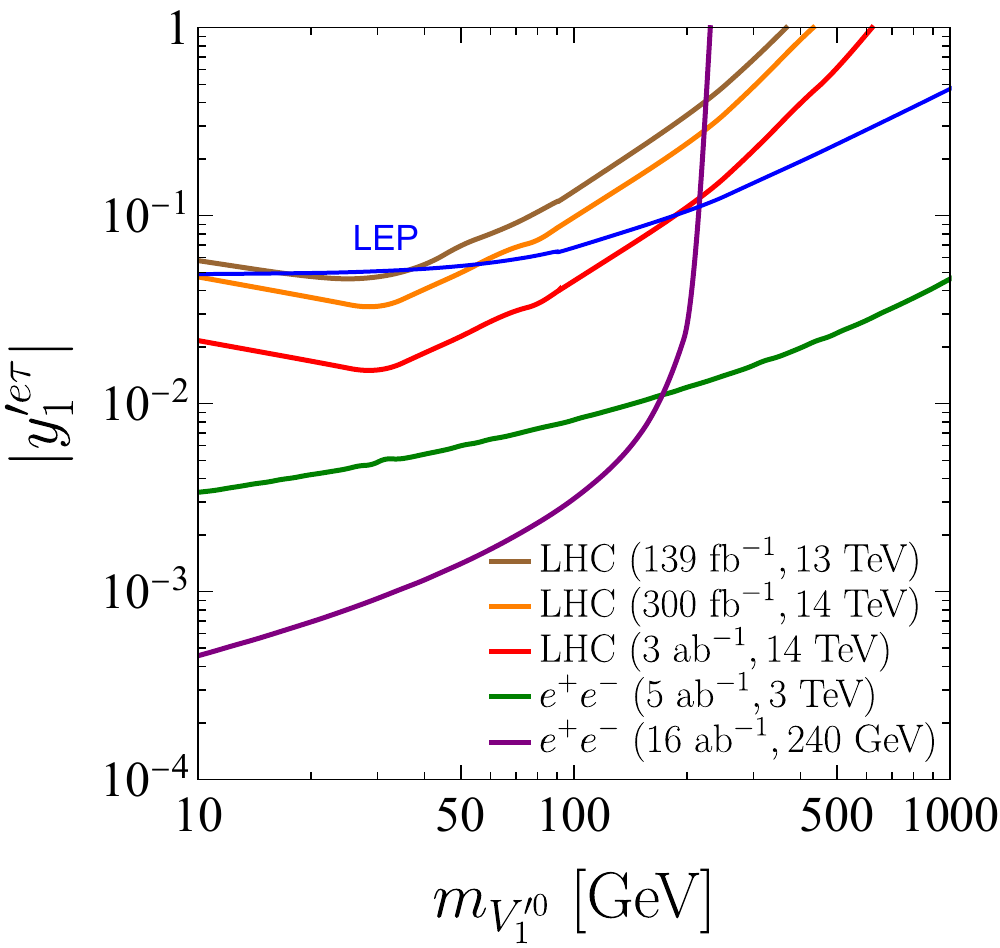}\\
\includegraphics[width=0.32\textwidth]{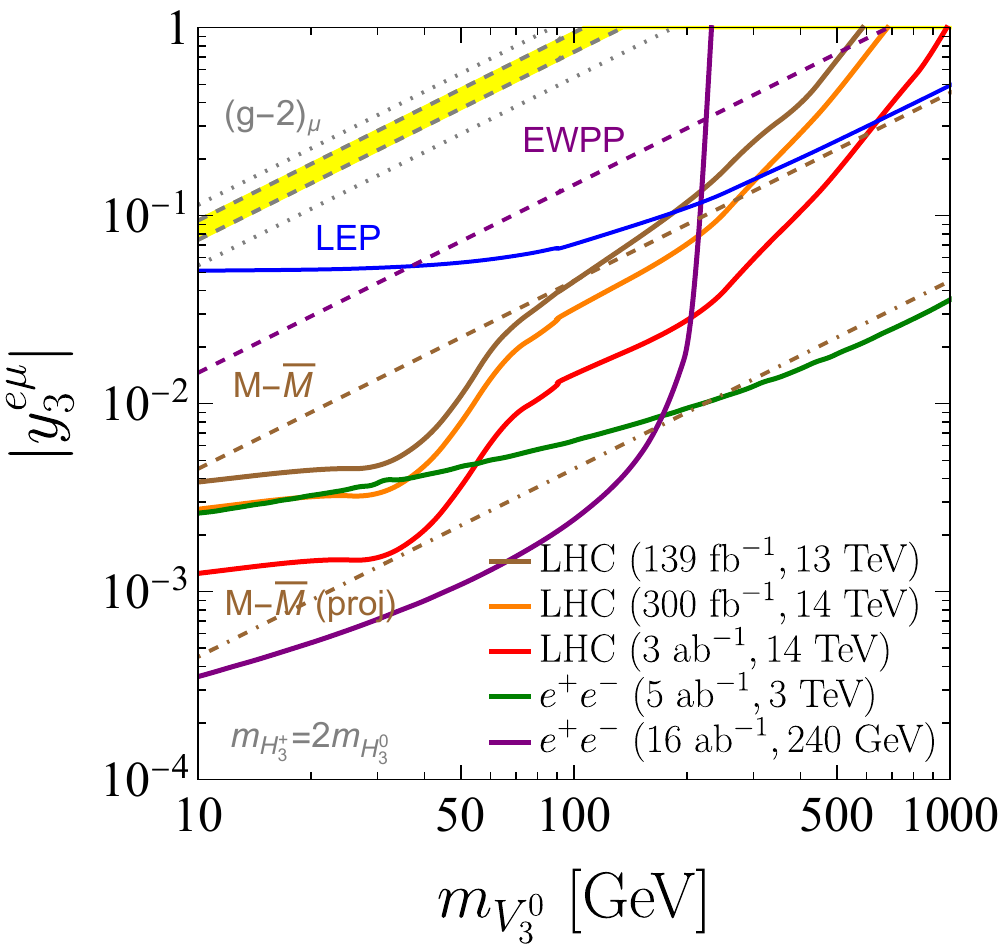}
\includegraphics[width=0.32\textwidth]{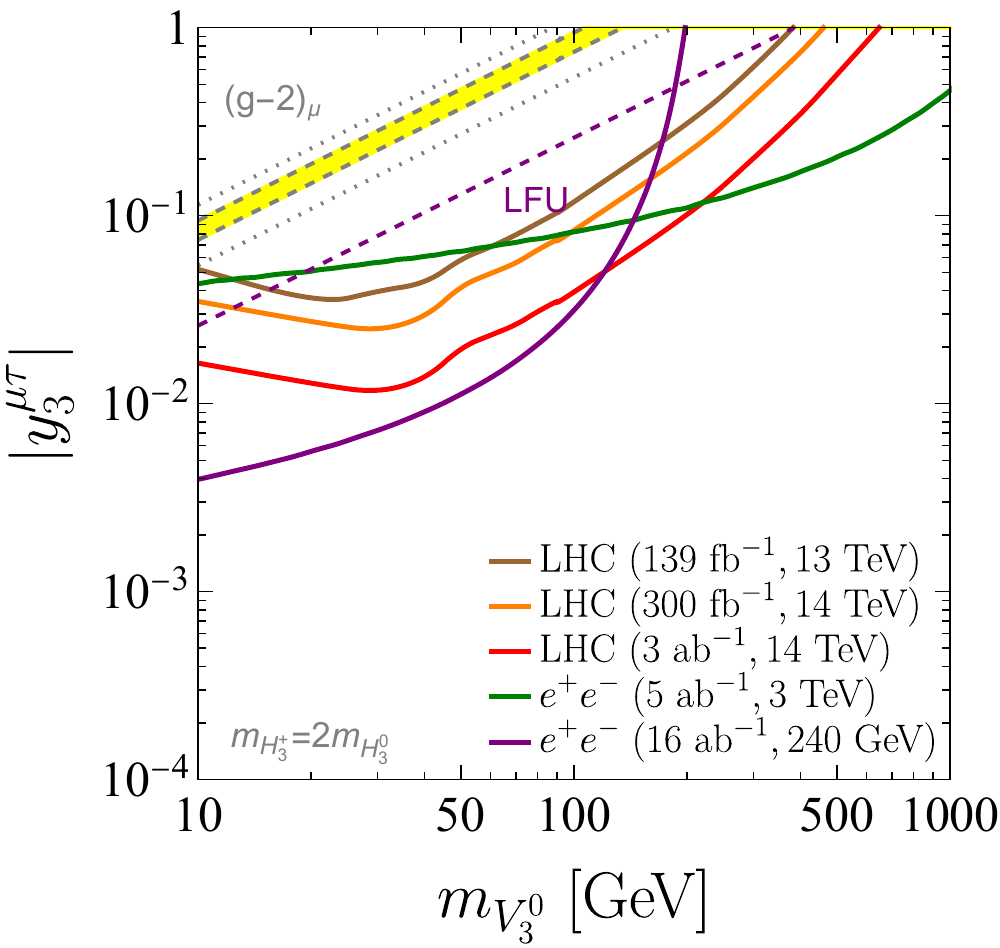}
\includegraphics[width=0.32\textwidth]{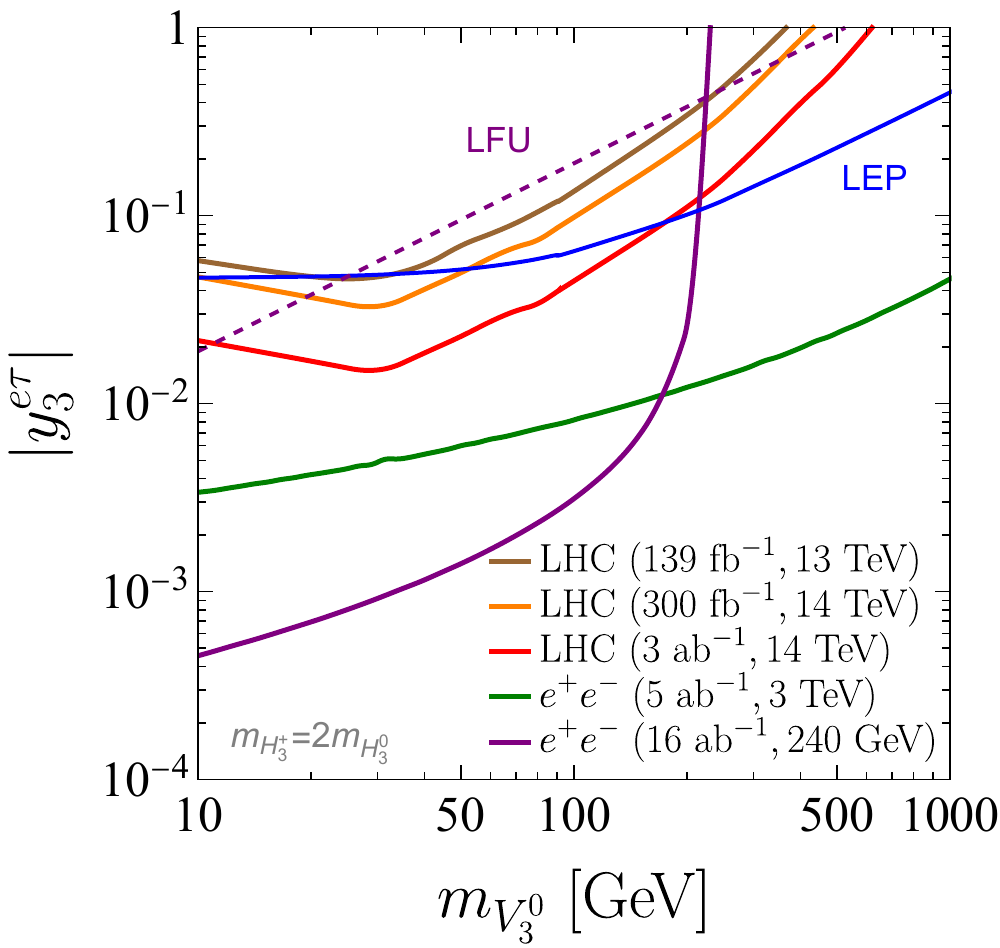}\\
\includegraphics{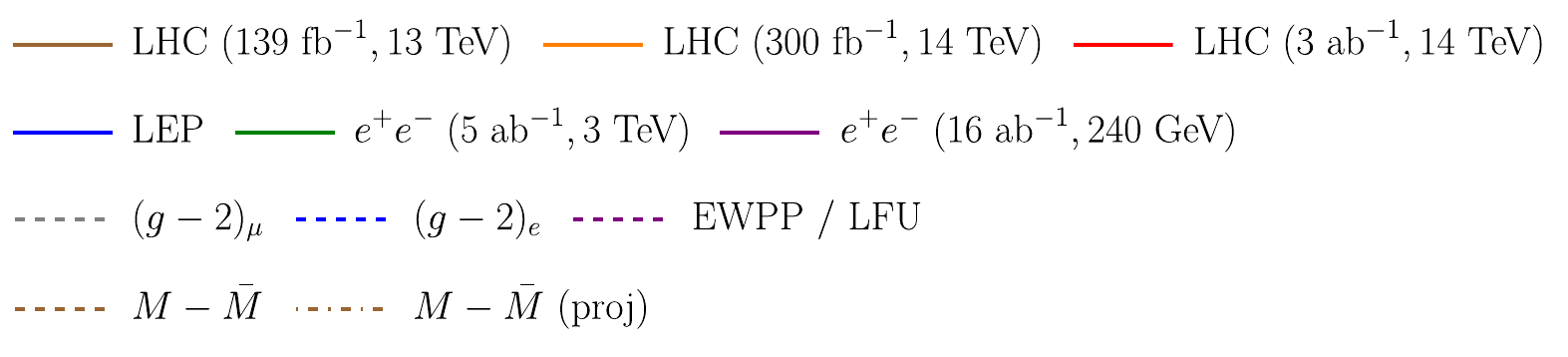}
\caption{
Projected sensitivity of the colliders to the couplings $y^{e\mu}$ (left column), $y^{\mu\tau}$ (middle column) and $y^{e\tau}$ (right column) for $V_1$ (top row), $V_1^{\prime}$ (middle row) and $V_3$ (bottom row). The $1\sigma$ and $3\sigma$ muon AMM favored regions are in the yellow band and dotted band, respectively. For $V_3$ we assume $m_{V_3^+}=2m_{V_3^0}$ to evaluate the muon AMM constraint. Other low-energy constraints are also shown. The brown dashed and dot-dashed lines are for muonium-antimuonium oscillation ($M-\bar{M}$) constraint and MACE projection, respectively. Purple dashed lines are for the constraints from LFU or electroweak precision physics (EWPP). The LEP bound is shown using blue solid line.}
\label{fig:H13}
\end{figure}

\begin{figure}[phtb!]
\centering
\includegraphics[width=0.4\textwidth]{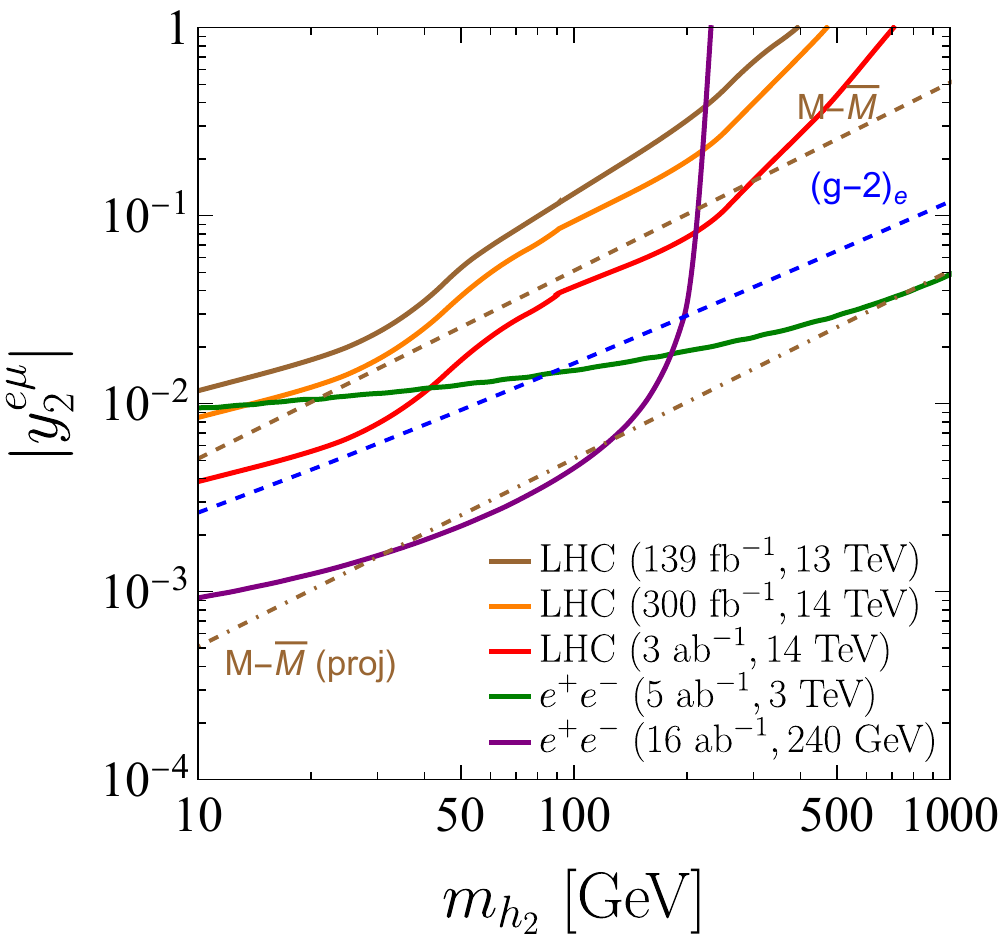}
\includegraphics[width=0.4\textwidth]{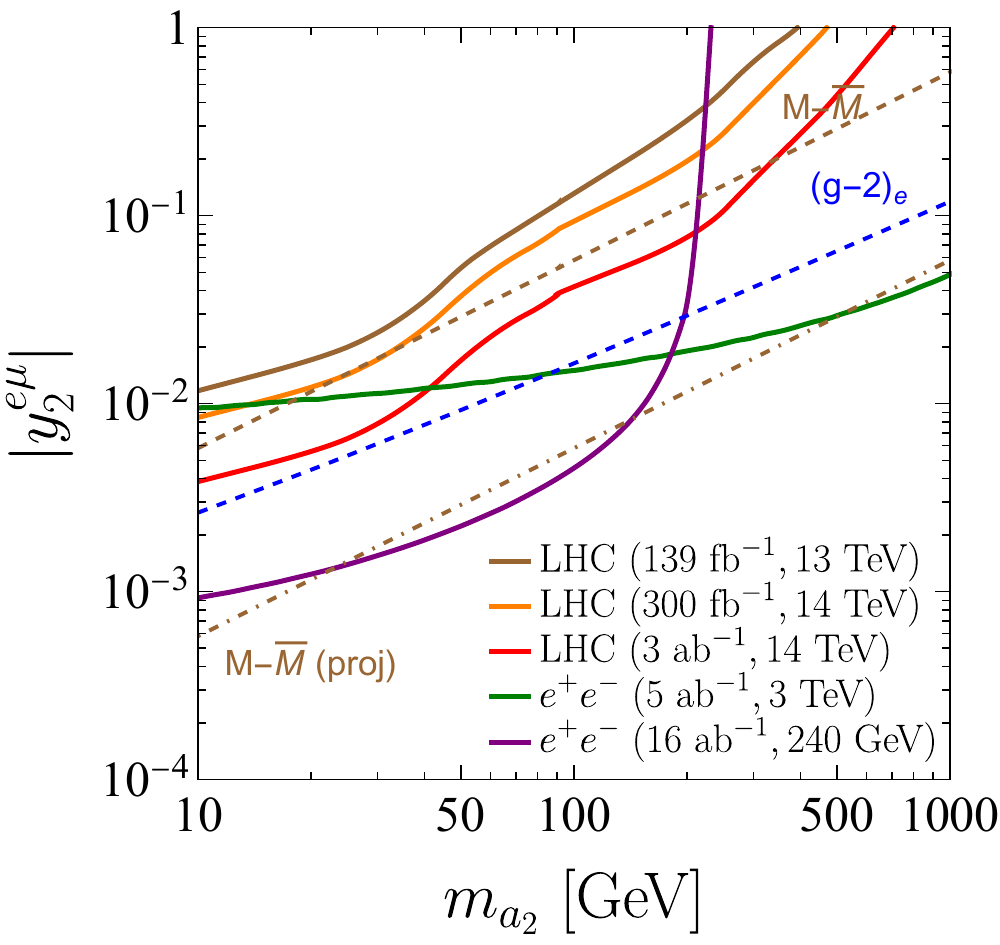}\\
\includegraphics[width=0.4\textwidth]{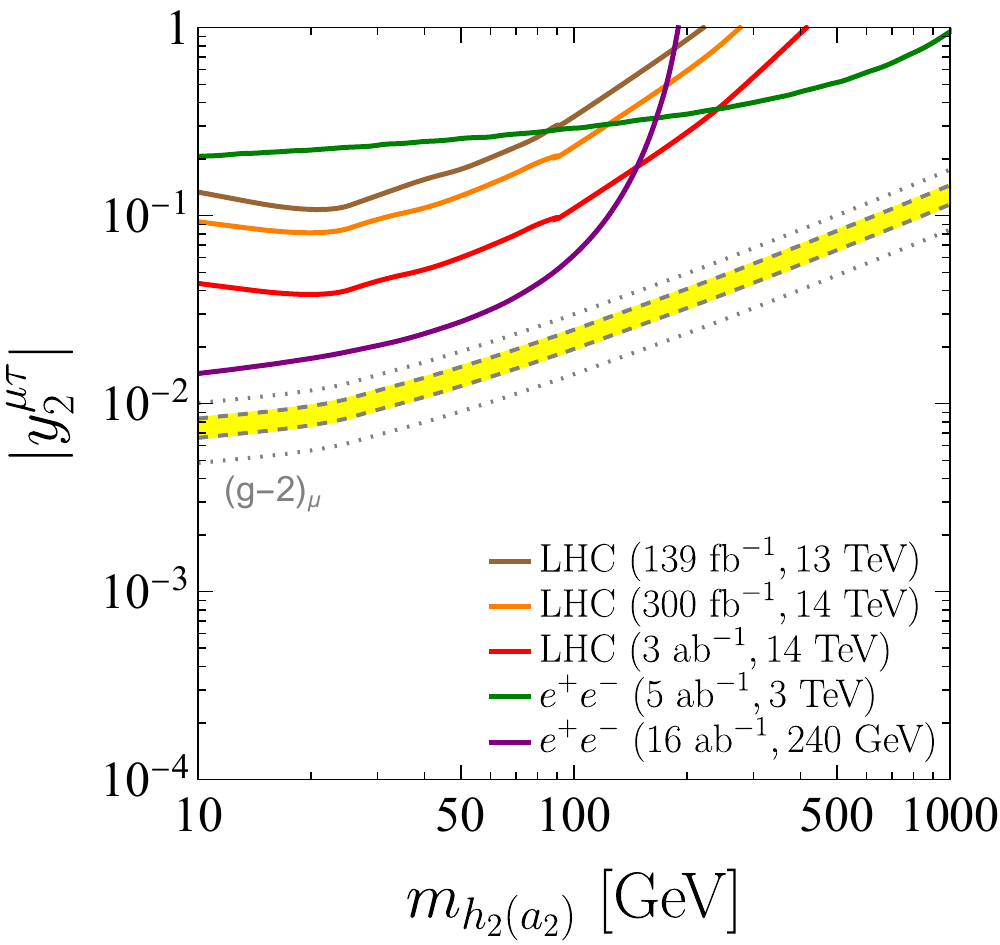}
\includegraphics[width=0.4\textwidth]{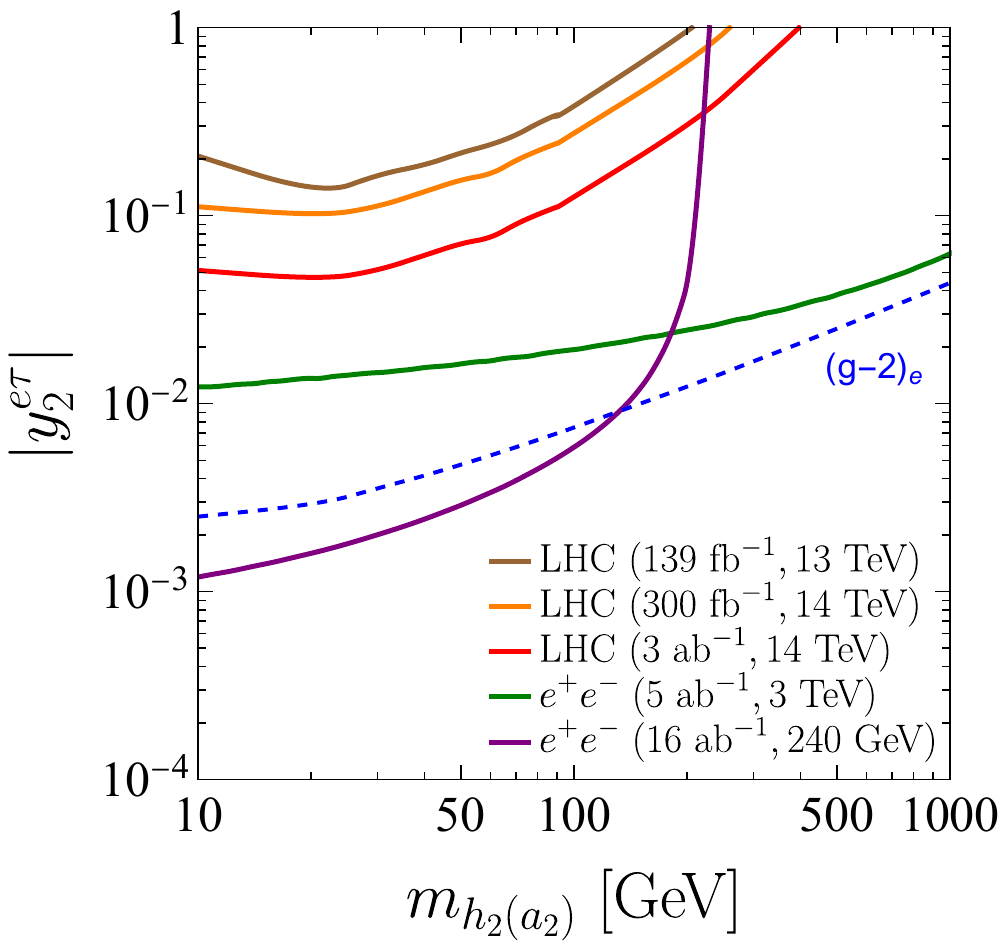}
\caption{
Projected sensitivity of the colliders to the couplings $y_2^{e\mu}$ for $h_2$ (top left), $y_2^{e\mu}$ for $a_2$ (top right), $y_2^{\mu\tau}$ (bottom left) and $y_2^{e\tau}$ (bottom right), as labeled in Fig.~\ref{fig:H13}. }
\label{fig:H2}
\end{figure}

We use the following significance~\cite{Zyla:2020zbs}
\begin{eqnarray}
\sqrt{2\Big((s+b){\rm ln}\Big(1+{s\over b}\Big)-s\Big)}\;,
\end{eqnarray}
where $s$ and $b$ are the signal and background event expectations, respectively.
Figs.~\ref{fig:H13} and \ref{fig:H2} shows the 95\% CL sensitivity to the $\Delta L=0$ coupling for the currently-available data of $139~\mathrm{fb}^{-1}$ taken at a center of mass energy $\sqrt{s}=13$ TeV as a brown solid line and projections for LHC with $\sqrt{s}=14$ TeV and integrated luminosities of $300~{\rm fb}^{-1}$ or $3000~{\rm fb}^{-1}$ using orange and red solid lines, respectively. Current LHC data can already probe unexplored parameter space for the bileptons $V_1^\prime$, $H_2$ and $V_3$.
Note that we do not intend to distinguish the chiral or CP nature of the bileptons couplings.
As a result, the results for vector $V_{1,3}^0$ only coupled to
left-handed leptons are the same as those for $V_1^{\prime 0}$ with only right-handed couplings. For the $H_2$ case, our result is applicable to either CP-even $h_2$ or CP-odd $a_2$ production.
One can see that the cleaner $e^\pm e^\pm \mu^\mp \mu^\mp$ signature make it more sensitive to probe the $y^{e\mu}$ coupling. For the channels with tau leptons in the final states, the sensitivity reach is weakened by the tau tagging rate.

\vspace{2ex}
\textbf{CLFV at $e^+e^-$ colliders}:
The CLFV processes can also happen at $e^+e^-$ colliders with on-shell bileptons in final states, i.e. $e^+e^-\to \ell_1^\pm\ell_2^\mp X$~\cite{Dev:2017ftk,Li:2019xvv,Iguro:2020rby,Endo:2020mev}.
Compared to the above CLFV signatures only produced in the Drell-Yan process at hadron collider, the processes with $e^\pm$ in final states
occur through both s and t channels mediated by $\gamma/Z$ at $e^+e^-$ colliders. The interference between the s and t channels
makes it more sensitive to probe $y^{e\mu}$ and $y^{e\tau}$ couplings. The $\mu^\pm \tau^\mp X$ process only happen in s channel.
We assume 10\% efficiency for the reconstruction of the bileptons and take the significance of $S/\sqrt{S+B}\approx \sqrt{S}$ as 3 for the observation of CLFV.
In Figs.~\ref{fig:H13} and \ref{fig:H2} we present the $e^+e^-$ collider sensitivity to the individual CLFV couplings in Ref.~\cite{Li:2019xvv}, with the proposed center of mass (c.m.) energy and the
integrated luminosity being 16 ab$^{-1}$ at 240 GeV~\cite{FCCee:2017} or 5 ab$^{-1}$ at 3 TeV~\cite{Charles:2018vfv}. The former machine with higher integrated luminosity provides the most sensitive environment in the low mass region and the latter one with larger c.m. energy can probe the high mass region of the bileptons.

The AMM favored parameter space and the constraints from low-energy experiments are
also displayed for the corresponding couplings. The region of coupling $y_{1(3)}^{(\prime)e\mu}$ favored by the muon magnetic moment anomaly is excluded by LEP and muonium-antimuonium oscillation.
The LFU and electroweak precision observables provide strong constraints for $V_1$ and $V_3$, respectively~\footnote{The search of four $e,\mu$ leptons at the LHC could also place constraint on the $e\mu$ couplings~\cite{Aaboud:2019lxo}. As the $e\mu$ couplings favored by muon AMM are already excluded by the low-energy experiments, we do not recast the LHC constraint here.}. The muon AMM favored $y_{1}^{\mu\tau}$ and $y_{3}^{\mu\tau}$ are also excluded by the LFU. The singlet $V_1'$ with right-handed coupling $y_{1}^{\prime\mu\tau}$ is viable in light of muon AMM and can evade the low-energy constraints. The muon AMM preferred $y_{2}^{\mu\tau}$ is beyond the sensitivity of the LHC and future proposed $e^+e^-$ colliders.

Moreover, the electron AMM puts a stringent constraint on the parameter space of $y_{2}^{e\mu}$ and $y_{2}^{e\tau}$. The parameter space consistent with the electron AMM at $3\sigma$ is below the blue dashed line. Part of the parameter space allowed by the electron AMM will be probed at future $e^+e^-$ colliders and the MACE experiment.

\begin{figure}[phtb!]
	\centering
	\includegraphics[width=0.5\textwidth]{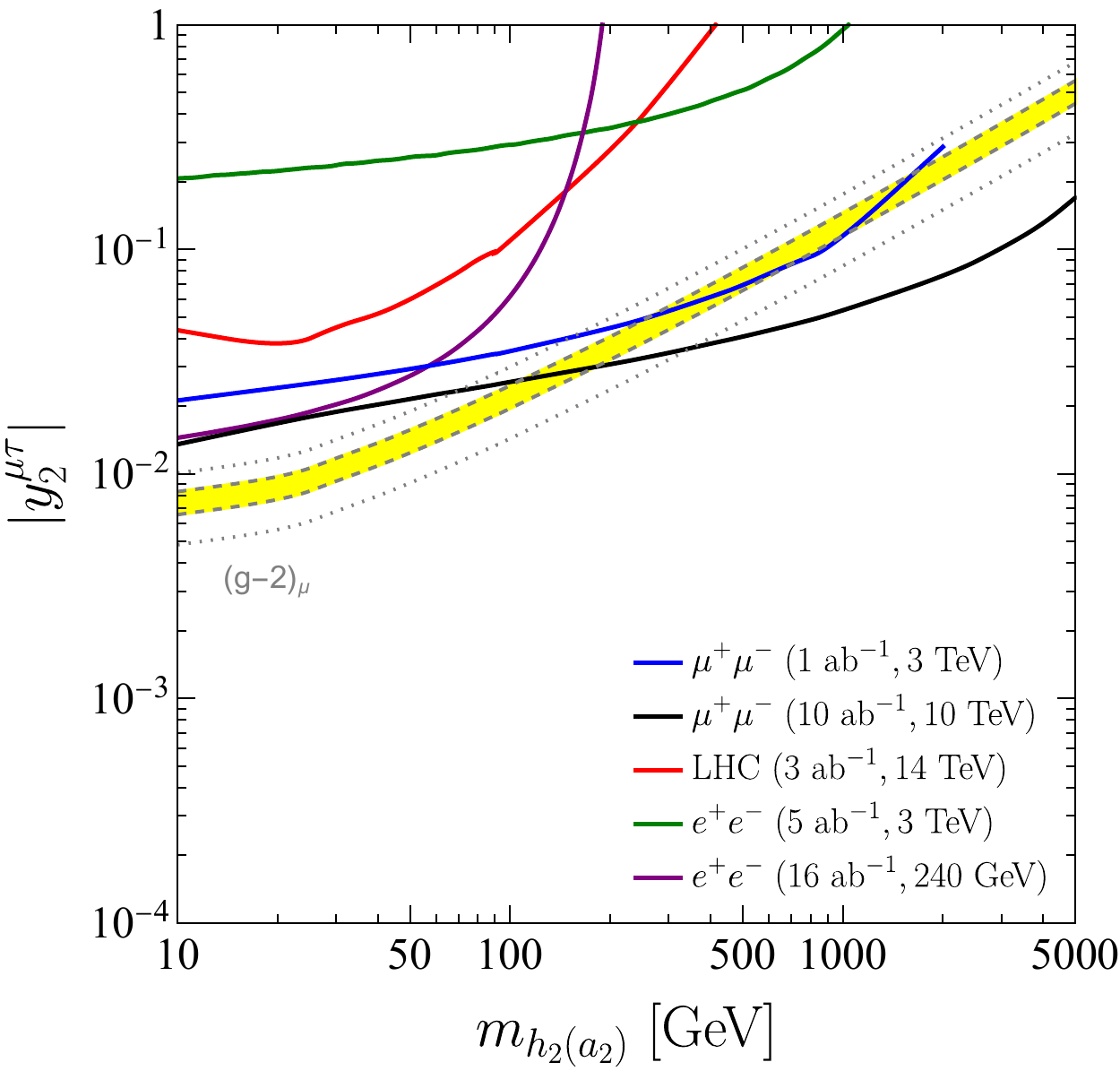}
	\caption{
		Projected sensitivity of two different configurations for a muon collider to the coupling $y_2^{\mu\tau}$ and the region favored by the muon AMM in comparison to the ultimate LHC sensitivity and proposed $e^+e^-$ colliders.}
	\label{fig:muon_collider}
\end{figure}

\vspace{2ex}
\textbf{Sensitivity to $y_2^{\mu\tau}$ at a muon collider:} Recently, due to the cooling technique development of the muon beam, there have been renewed interests for high-energy muon colliders~\cite{Delahaye:2019omf,Han:2020uid,Long:2020wfp,AlAli:2021let}. With its higher center of mass energy and high luminosity, a muon collider has the potential to probe part of the favored region for the muon AMM in the model of a doublet bilepton $H_2$ with a non-vanishing coupling $y_2^{\mu\tau}$. Similar to the $e^+e^-$ collider, a muon collider can also search for the CLFV processes with on-shell bileptons in final states via the annihilation channel $\mu^+\mu^-\to \ell_1^\pm\ell_2^\mp h_2(a_2)$. The corresponding annihilation cross section however falls as $1/s$. Moreover, as the beam energy increases, the photon-photon fusion processes $\gamma\gamma\to \ell_1^\pm\ell_2^\mp h_2(a_2)$ take over the annihilation channels due to the double-logarithmical enhancement from the collinear photon radiation off the high energy muons~\cite{vonWeizsacker:1934nji,Williams:1934ad}. Under the same assumptions as those for the $e^+e^-$ colliders, we then illustrate the sensitivity of two muon collider configurations by combining the $\mu^+\mu^-$ annihilation and $\gamma\gamma$ fusion processes. In Fig.~\ref{fig:muon_collider}, the blue (black) line indicates the sensitivity of a muon collider with center of mass energy of $\sqrt{s}=3~(10)$ TeV and integrated luminosity of 1~(10) ab$^{-1}$.
The high energy and integrated luminosity provides the muon collider with an increased sensitivity reach compared to $e^+e^-$ colliders shown in purple and green lines and the ultimate sensitivity of the LHC (red line). Both muon collider configurations constrain the relevant parameter space of the muon AMM for masses $m_{h_2(a_2)}\gtrsim 100$ GeV, but the low energy configuration looses sensitivity for scalar masses above $2$ TeV.

\section{Conclusion}
\label{sec:Con}

Any CLFV observation implies the existence of new physics beyond the SM in charged lepton sector.
The muon magnetic moment anomaly between the SM prediction and the recent muon $g-2$ precision measurement at Fermilab provides an opportunity to reveal the nature of underlying CLFV. We investigate the current constraints and the future search potential for CLFV inspired by the muon magnetic moment anomaly.

We consider the most general SM invariant Lagrangian of $\Delta L=0$ bileptons coupled to explicit leptonic fields. The muon $g-2$ anomaly constrains the $\mu\mu$ component of the coupling combination $(y^\dagger y)^{\mu\mu}$ and thus the LFV component $y^{\ell\ell'} (\ell\neq \ell')$. A dominant LFV coupling $y^{\ell\ell'}$ induces the transition with $|\Delta(L_\ell-L_{\ell'})|=4$ which can be probed by the proposed muonium-antimuonium conversion experiment MACE, current LHC data, the LHC upgrade and future lepton colliders. We investigate the implication of the muon magnetic moment anomaly and other relevant constraints from low-energy precision experiments.
We then evaluate the projected sensitivity of the future MACE and colliders to the individual CLFV couplings satisfying these constraints.
Our main conclusions are
\begin{itemize}
\item The MACE, the LHC and future $e^+e^-$ colliders are generally sensitive to the relevant CLFV couplings of both vector and scalar bileptons allowed by the low-energy experiments. Current LHC data is able to probe unexplored parameter space.
\item The region of coupling $y_{1(3)}^{(\prime)e\mu}$ favored by the muon magnetic moment anomaly is excluded by LEP and muonium-antimuonium oscillation. The LFU and electroweak precision physics provide strong constraints for $V_1$ and $V_3$, respectively. The muon AMM favored $y_{1}^{\mu\tau}$ and $y_{3}^{\mu\tau}$ are also excluded by the LFU. The singlet $V_1'$ with right-handed coupling $y_{1}^{\prime\mu\tau}$ is viable in light of muon AMM and can evade the low-energy constraints.
\item The muon AMM preferred $y_{2}^{\mu\tau}$ is beyond the sensitivity of future LHC upgrades and $e^+e^-$ colliders, but can be probed by high-energy muon colliders. For $y_{2}^{e\mu}$ and $y_{2}^{e\tau}$, the MACE and future colliders can probe the parameter space allowed by electron AMM.
\end{itemize}

\acknowledgments
TL is supported by the National Natural Science Foundation of China (Grant No. 11975129, 12035008) and ``the Fundamental Research Funds for the Central Universities'', Nankai University (Grants No. 63196013). MS acknowledges support by the Australian Research Council via the Discovery Project DP200101470.
CYY is supported in part by the Grants No.~NSFC-11975130, No.~NSFC-12035008, No.~NSFC-12047533, by the National Key Research and Development Program of China under Grant No.~2017YFA0402200 and the China Post-doctoral Science Foundation under Grant No.~2018M641621.

\appendix
 \section{Oblique corrections}
 \label{app:ST}
One may wonder whether the mass splitting of the bileptons leads to sizable conttribution to the oblique parameters $S$ and $T$. We explicitly show for the electroweak doublet scalar $H_2$ that a large mass splitting is consistent with the current experimental uncertaintites of $S$ and $T$ in the limit of no mixing between the CP even Higgs scalars and vanishing VEV of the second Higgs doublet $H_2$, i.e. $\cos\beta=0$. In this limit we obtain~\cite{Chankowski:1999ta,Chankowski:2000an,Barbieri:2006bg}
\begin{align}
	\Delta T &= A(m_{H_2^+},m_{h_2}) + A(m_{H_2^+},m_{a_2})-A(m_{a_2},m_{h_2})\;, \\
	\Delta S &= F(m_{a_2},m_{h_2}) -F(m_{H_2^+},m_{H_2^+})\;,
\end{align}
with
\begin{align}
	A(m_1,m_2) & = \frac{\sqrt{2}G_F}{16\pi^2 \alpha_{\rm em}} \left[\frac{m_1^2+m_2^2}{2}-\frac{m_1^2m_2^2}{m_1^2-m_2^2}\ln\frac{m_1^2}{m_2^2} \right]\;,
		\\
		F(m_1,m_2) & = \frac{1}{24\pi}\left[-\ln\frac{\Lambda^4}{m_1^2m_2^2} + \frac{4m_1^2m_2^2}{(m_1^2-m_2^2)^2} + \frac{m_1^6+m_2^6 -3 m_1^2m_2^2(m_1^2+m_2^2)}{(m_1^2-m_2^2)^3}\ln \frac{m_1^2}{m_2^2}\right]\;.
\end{align}
In the two limits of $m_{a_2} \simeq m_{H_2^+} \gg m_{h_2}$ and $m_{h_2}\simeq m_{H_2^+}\gg m_{a_2}$, we find $\Delta T=0$ and $\Delta S=-5/72\pi \simeq -0.02$, which is well below the experimental uncertainty of the oblique parameters~\cite{Zyla:2020zbs}.

\bibliography{refs}

\begin{thebibliography}{148}%
\makeatletter
\providecommand \@ifxundefined [1]{%
 \@ifx{#1\undefined}
}%
\providecommand \@ifnum [1]{%
 \ifnum #1\expandafter \@firstoftwo
 \else \expandafter \@secondoftwo
 \fi
}%
\providecommand \@ifx [1]{%
 \ifx #1\expandafter \@firstoftwo
 \else \expandafter \@secondoftwo
 \fi
}%
\providecommand \natexlab [1]{#1}%
\providecommand \enquote  [1]{``#1''}%
\providecommand \bibnamefont  [1]{#1}%
\providecommand \bibfnamefont [1]{#1}%
\providecommand \citenamefont [1]{#1}%
\providecommand \href@noop [0]{\@secondoftwo}%
\providecommand \href [0]{\begingroup \@sanitize@url \@href}%
\providecommand \@href[1]{\@@startlink{#1}\@@href}%
\providecommand \@@href[1]{\endgroup#1\@@endlink}%
\providecommand \@sanitize@url [0]{\catcode `\\12\catcode `\$12\catcode
  `\&12\catcode `\#12\catcode `\^12\catcode `\_12\catcode `\%12\relax}%
\providecommand \@@startlink[1]{}%
\providecommand \@@endlink[0]{}%
\providecommand \url  [0]{\begingroup\@sanitize@url \@url }%
\providecommand \@url [1]{\endgroup\@href {#1}{\urlprefix }}%
\providecommand \urlprefix  [0]{URL }%
\providecommand \Eprint [0]{\href }%
\providecommand \doibase [0]{http://dx.doi.org/}%
\providecommand \selectlanguage [0]{\@gobble}%
\providecommand \bibinfo  [0]{\@secondoftwo}%
\providecommand \bibfield  [0]{\@secondoftwo}%
\providecommand \translation [1]{[#1]}%
\providecommand \BibitemOpen [0]{}%
\providecommand \bibitemStop [0]{}%
\providecommand \bibitemNoStop [0]{.\EOS\space}%
\providecommand \EOS [0]{\spacefactor3000\relax}%
\providecommand \BibitemShut  [1]{\csname bibitem#1\endcsname}%
\let\auto@bib@innerbib\@empty
\bibitem [{\citenamefont {Heeck}(2017)}]{Heeck:2016xwg}%
  \BibitemOpen
  \bibfield  {author} {\bibinfo {author} {\bibfnamefont {J.}~\bibnamefont
  {Heeck}},\ }\href {\doibase 10.1103/PhysRevD.95.015022} {\bibfield  {journal}
  {\bibinfo  {journal} {Phys. Rev. D}\ }\textbf {\bibinfo {volume} {95}},\
  \bibinfo {pages} {015022} (\bibinfo {year} {2017})},\ \Eprint
  {http://arxiv.org/abs/1610.07623} {arXiv:1610.07623 [hep-ph]} \BibitemShut
  {NoStop}%
\bibitem [{\citenamefont {Davier}\ \emph {et~al.}(2017)\citenamefont {Davier},
  \citenamefont {Hoecker}, \citenamefont {Malaescu},\ and\ \citenamefont
  {Zhang}}]{Davier:2017zfy}%
  \BibitemOpen
  \bibfield  {author} {\bibinfo {author} {\bibfnamefont {M.}~\bibnamefont
  {Davier}}, \bibinfo {author} {\bibfnamefont {A.}~\bibnamefont {Hoecker}},
  \bibinfo {author} {\bibfnamefont {B.}~\bibnamefont {Malaescu}}, \ and\
  \bibinfo {author} {\bibfnamefont {Z.}~\bibnamefont {Zhang}},\ }\href
  {\doibase 10.1140/epjc/s10052-017-5161-6} {\bibfield  {journal} {\bibinfo
  {journal} {Eur. Phys. J.}\ }\textbf {\bibinfo {volume} {C77}},\ \bibinfo
  {pages} {827} (\bibinfo {year} {2017})},\ \Eprint
  {http://arxiv.org/abs/1706.09436} {arXiv:1706.09436 [hep-ph]} \BibitemShut
  {NoStop}%
\bibitem [{\citenamefont {Keshavarzi}\ \emph {et~al.}(2018)\citenamefont
  {Keshavarzi}, \citenamefont {Nomura},\ and\ \citenamefont
  {Teubner}}]{Keshavarzi:2018mgv}%
  \BibitemOpen
  \bibfield  {author} {\bibinfo {author} {\bibfnamefont {A.}~\bibnamefont
  {Keshavarzi}}, \bibinfo {author} {\bibfnamefont {D.}~\bibnamefont {Nomura}},
  \ and\ \bibinfo {author} {\bibfnamefont {T.}~\bibnamefont {Teubner}},\ }\href
  {\doibase 10.1103/PhysRevD.97.114025} {\bibfield  {journal} {\bibinfo
  {journal} {Phys. Rev. D}\ }\textbf {\bibinfo {volume} {97}},\ \bibinfo
  {pages} {114025} (\bibinfo {year} {2018})},\ \Eprint
  {http://arxiv.org/abs/1802.02995} {arXiv:1802.02995 [hep-ph]} \BibitemShut
  {NoStop}%
\bibitem [{\citenamefont {Colangelo}\ \emph {et~al.}(2019)\citenamefont
  {Colangelo}, \citenamefont {Hoferichter},\ and\ \citenamefont
  {Stoffer}}]{Colangelo:2018mtw}%
  \BibitemOpen
  \bibfield  {author} {\bibinfo {author} {\bibfnamefont {G.}~\bibnamefont
  {Colangelo}}, \bibinfo {author} {\bibfnamefont {M.}~\bibnamefont
  {Hoferichter}}, \ and\ \bibinfo {author} {\bibfnamefont {P.}~\bibnamefont
  {Stoffer}},\ }\href {\doibase 10.1007/JHEP02(2019)006} {\bibfield  {journal}
  {\bibinfo  {journal} {JHEP}\ }\textbf {\bibinfo {volume} {02}},\ \bibinfo
  {pages} {006} (\bibinfo {year} {2019})},\ \Eprint
  {http://arxiv.org/abs/1810.00007} {arXiv:1810.00007 [hep-ph]} \BibitemShut
  {NoStop}%
\bibitem [{\citenamefont {Hoferichter}\ \emph {et~al.}(2019)\citenamefont
  {Hoferichter}, \citenamefont {Hoid},\ and\ \citenamefont
  {Kubis}}]{Hoferichter:2019gzf}%
  \BibitemOpen
  \bibfield  {author} {\bibinfo {author} {\bibfnamefont {M.}~\bibnamefont
  {Hoferichter}}, \bibinfo {author} {\bibfnamefont {B.-L.}\ \bibnamefont
  {Hoid}}, \ and\ \bibinfo {author} {\bibfnamefont {B.}~\bibnamefont {Kubis}},\
  }\href {\doibase 10.1007/JHEP08(2019)137} {\bibfield  {journal} {\bibinfo
  {journal} {JHEP}\ }\textbf {\bibinfo {volume} {08}},\ \bibinfo {pages} {137}
  (\bibinfo {year} {2019})},\ \Eprint {http://arxiv.org/abs/1907.01556}
  {arXiv:1907.01556 [hep-ph]} \BibitemShut {NoStop}%
\bibitem [{\citenamefont {Davier}\ \emph {et~al.}(2020)\citenamefont {Davier},
  \citenamefont {Hoecker}, \citenamefont {Malaescu},\ and\ \citenamefont
  {Zhang}}]{Davier:2019can}%
  \BibitemOpen
  \bibfield  {author} {\bibinfo {author} {\bibfnamefont {M.}~\bibnamefont
  {Davier}}, \bibinfo {author} {\bibfnamefont {A.}~\bibnamefont {Hoecker}},
  \bibinfo {author} {\bibfnamefont {B.}~\bibnamefont {Malaescu}}, \ and\
  \bibinfo {author} {\bibfnamefont {Z.}~\bibnamefont {Zhang}},\ }\href
  {\doibase 10.1140/epjc/s10052-020-7792-2} {\bibfield  {journal} {\bibinfo
  {journal} {Eur. Phys. J.}\ }\textbf {\bibinfo {volume} {C80}},\ \bibinfo
  {pages} {241} (\bibinfo {year} {2020})},\ \bibinfo {note} {[Erratum: Eur.
  Phys. J. {\bf C80}, 410 (2020)]},\ \Eprint {http://arxiv.org/abs/1908.00921}
  {arXiv:1908.00921 [hep-ph]} \BibitemShut {NoStop}%
\bibitem [{\citenamefont {Keshavarzi}\ \emph {et~al.}(2020)\citenamefont
  {Keshavarzi}, \citenamefont {Nomura},\ and\ \citenamefont
  {Teubner}}]{Keshavarzi:2019abf}%
  \BibitemOpen
  \bibfield  {author} {\bibinfo {author} {\bibfnamefont {A.}~\bibnamefont
  {Keshavarzi}}, \bibinfo {author} {\bibfnamefont {D.}~\bibnamefont {Nomura}},
  \ and\ \bibinfo {author} {\bibfnamefont {T.}~\bibnamefont {Teubner}},\ }\href
  {\doibase 10.1103/PhysRevD.101.014029} {\bibfield  {journal} {\bibinfo
  {journal} {Phys. Rev.}\ }\textbf {\bibinfo {volume} {D101}},\ \bibinfo
  {pages} {014029} (\bibinfo {year} {2020})},\ \Eprint
  {http://arxiv.org/abs/1911.00367} {arXiv:1911.00367 [hep-ph]} \BibitemShut
  {NoStop}%
\bibitem [{\citenamefont {Kurz}\ \emph {et~al.}(2014)\citenamefont {Kurz},
  \citenamefont {Liu}, \citenamefont {Marquard},\ and\ \citenamefont
  {Steinhauser}}]{Kurz:2014wya}%
  \BibitemOpen
  \bibfield  {author} {\bibinfo {author} {\bibfnamefont {A.}~\bibnamefont
  {Kurz}}, \bibinfo {author} {\bibfnamefont {T.}~\bibnamefont {Liu}}, \bibinfo
  {author} {\bibfnamefont {P.}~\bibnamefont {Marquard}}, \ and\ \bibinfo
  {author} {\bibfnamefont {M.}~\bibnamefont {Steinhauser}},\ }\href {\doibase
  10.1016/j.physletb.2014.05.043} {\bibfield  {journal} {\bibinfo  {journal}
  {Phys. Lett.}\ }\textbf {\bibinfo {volume} {B734}},\ \bibinfo {pages} {144}
  (\bibinfo {year} {2014})},\ \Eprint {http://arxiv.org/abs/1403.6400}
  {arXiv:1403.6400 [hep-ph]} \BibitemShut {NoStop}%
\bibitem [{\citenamefont {Chakraborty}\ \emph {et~al.}(2018)\citenamefont
  {Chakraborty} \emph {et~al.}}]{Chakraborty:2017tqp}%
  \BibitemOpen
  \bibfield  {author} {\bibinfo {author} {\bibfnamefont {B.}~\bibnamefont
  {Chakraborty}} \emph {et~al.} (\bibinfo {collaboration} {Fermilab Lattice,
  LATTICE-HPQCD, MILC}),\ }\href {\doibase 10.1103/PhysRevLett.120.152001}
  {\bibfield  {journal} {\bibinfo  {journal} {Phys. Rev. Lett.}\ }\textbf
  {\bibinfo {volume} {120}},\ \bibinfo {pages} {152001} (\bibinfo {year}
  {2018})},\ \Eprint {http://arxiv.org/abs/1710.11212} {arXiv:1710.11212
  [hep-lat]} \BibitemShut {NoStop}%
\bibitem [{\citenamefont {Borsanyi}\ \emph {et~al.}(2018)\citenamefont
  {Borsanyi} \emph {et~al.}}]{Borsanyi:2017zdw}%
  \BibitemOpen
  \bibfield  {author} {\bibinfo {author} {\bibfnamefont {S.}~\bibnamefont
  {Borsanyi}} \emph {et~al.} (\bibinfo {collaboration}
  {Budapest-Marseille-Wuppertal}),\ }\href {\doibase
  10.1103/PhysRevLett.121.022002} {\bibfield  {journal} {\bibinfo  {journal}
  {Phys. Rev. Lett.}\ }\textbf {\bibinfo {volume} {121}},\ \bibinfo {pages}
  {022002} (\bibinfo {year} {2018})},\ \Eprint
  {http://arxiv.org/abs/1711.04980} {arXiv:1711.04980 [hep-lat]} \BibitemShut
  {NoStop}%
\bibitem [{\citenamefont {Blum}\ \emph {et~al.}(2018)\citenamefont {Blum},
  \citenamefont {Boyle}, \citenamefont {G\"ulpers}, \citenamefont {Izubuchi},
  \citenamefont {Jin}, \citenamefont {Jung}, \citenamefont {J\"uttner},
  \citenamefont {Lehner}, \citenamefont {Portelli},\ and\ \citenamefont
  {Tsang}}]{Blum:2018mom}%
  \BibitemOpen
  \bibfield  {author} {\bibinfo {author} {\bibfnamefont {T.}~\bibnamefont
  {Blum}}, \bibinfo {author} {\bibfnamefont {P.~A.}\ \bibnamefont {Boyle}},
  \bibinfo {author} {\bibfnamefont {V.}~\bibnamefont {G\"ulpers}}, \bibinfo
  {author} {\bibfnamefont {T.}~\bibnamefont {Izubuchi}}, \bibinfo {author}
  {\bibfnamefont {L.}~\bibnamefont {Jin}}, \bibinfo {author} {\bibfnamefont
  {C.}~\bibnamefont {Jung}}, \bibinfo {author} {\bibfnamefont {A.}~\bibnamefont
  {J\"uttner}}, \bibinfo {author} {\bibfnamefont {C.}~\bibnamefont {Lehner}},
  \bibinfo {author} {\bibfnamefont {A.}~\bibnamefont {Portelli}}, \ and\
  \bibinfo {author} {\bibfnamefont {J.~T.}\ \bibnamefont {Tsang}} (\bibinfo
  {collaboration} {RBC, UKQCD}),\ }\href {\doibase
  10.1103/PhysRevLett.121.022003} {\bibfield  {journal} {\bibinfo  {journal}
  {Phys. Rev. Lett.}\ }\textbf {\bibinfo {volume} {121}},\ \bibinfo {pages}
  {022003} (\bibinfo {year} {2018})},\ \Eprint
  {http://arxiv.org/abs/1801.07224} {arXiv:1801.07224 [hep-lat]} \BibitemShut
  {NoStop}%
\bibitem [{\citenamefont {Giusti}\ \emph {et~al.}(2019)\citenamefont {Giusti},
  \citenamefont {Lubicz}, \citenamefont {Martinelli}, \citenamefont
  {Sanfilippo},\ and\ \citenamefont {Simula}}]{Giusti:2019xct}%
  \BibitemOpen
  \bibfield  {author} {\bibinfo {author} {\bibfnamefont {D.}~\bibnamefont
  {Giusti}}, \bibinfo {author} {\bibfnamefont {V.}~\bibnamefont {Lubicz}},
  \bibinfo {author} {\bibfnamefont {G.}~\bibnamefont {Martinelli}}, \bibinfo
  {author} {\bibfnamefont {F.}~\bibnamefont {Sanfilippo}}, \ and\ \bibinfo
  {author} {\bibfnamefont {S.}~\bibnamefont {Simula}} (\bibinfo {collaboration}
  {ETM}),\ }\href {\doibase 10.1103/PhysRevD.99.114502} {\bibfield  {journal}
  {\bibinfo  {journal} {Phys. Rev.}\ }\textbf {\bibinfo {volume} {D99}},\
  \bibinfo {pages} {114502} (\bibinfo {year} {2019})},\ \Eprint
  {http://arxiv.org/abs/1901.10462} {arXiv:1901.10462 [hep-lat]} \BibitemShut
  {NoStop}%
\bibitem [{\citenamefont {Shintani}\ and\ \citenamefont
  {Kuramashi}(2019)}]{Shintani:2019wai}%
  \BibitemOpen
  \bibfield  {author} {\bibinfo {author} {\bibfnamefont {E.}~\bibnamefont
  {Shintani}}\ and\ \bibinfo {author} {\bibfnamefont {Y.}~\bibnamefont
  {Kuramashi}},\ }\href {\doibase 10.1103/PhysRevD.100.034517} {\bibfield
  {journal} {\bibinfo  {journal} {Phys. Rev.}\ }\textbf {\bibinfo {volume}
  {D100}},\ \bibinfo {pages} {034517} (\bibinfo {year} {2019})},\ \Eprint
  {http://arxiv.org/abs/1902.00885} {arXiv:1902.00885 [hep-lat]} \BibitemShut
  {NoStop}%
\bibitem [{\citenamefont {Davies}\ \emph {et~al.}(2020)\citenamefont {Davies}
  \emph {et~al.}}]{Davies:2019efs}%
  \BibitemOpen
  \bibfield  {author} {\bibinfo {author} {\bibfnamefont {C.~T.~H.}\
  \bibnamefont {Davies}} \emph {et~al.} (\bibinfo {collaboration} {Fermilab
  Lattice, LATTICE-HPQCD, MILC}),\ }\href {\doibase
  10.1103/PhysRevD.101.034512} {\bibfield  {journal} {\bibinfo  {journal}
  {Phys. Rev.}\ }\textbf {\bibinfo {volume} {D101}},\ \bibinfo {pages} {034512}
  (\bibinfo {year} {2020})},\ \Eprint {http://arxiv.org/abs/1902.04223}
  {arXiv:1902.04223 [hep-lat]} \BibitemShut {NoStop}%
\bibitem [{\citenamefont {G\'erardin}\ \emph {et~al.}(2019)\citenamefont
  {G\'erardin}, \citenamefont {C\`e}, \citenamefont {von Hippel}, \citenamefont
  {H{\"o}rz}, \citenamefont {Meyer}, \citenamefont {Mohler}, \citenamefont
  {Ottnad}, \citenamefont {Wilhelm},\ and\ \citenamefont
  {Wittig}}]{Gerardin:2019rua}%
  \BibitemOpen
  \bibfield  {author} {\bibinfo {author} {\bibfnamefont {A.}~\bibnamefont
  {G\'erardin}}, \bibinfo {author} {\bibfnamefont {M.}~\bibnamefont {C\`e}},
  \bibinfo {author} {\bibfnamefont {G.}~\bibnamefont {von Hippel}}, \bibinfo
  {author} {\bibfnamefont {B.}~\bibnamefont {H{\"o}rz}}, \bibinfo {author}
  {\bibfnamefont {H.~B.}\ \bibnamefont {Meyer}}, \bibinfo {author}
  {\bibfnamefont {D.}~\bibnamefont {Mohler}}, \bibinfo {author} {\bibfnamefont
  {K.}~\bibnamefont {Ottnad}}, \bibinfo {author} {\bibfnamefont
  {J.}~\bibnamefont {Wilhelm}}, \ and\ \bibinfo {author} {\bibfnamefont
  {H.}~\bibnamefont {Wittig}},\ }\href {\doibase 10.1103/PhysRevD.100.014510}
  {\bibfield  {journal} {\bibinfo  {journal} {Phys. Rev.}\ }\textbf {\bibinfo
  {volume} {D100}},\ \bibinfo {pages} {014510} (\bibinfo {year} {2019})},\
  \Eprint {http://arxiv.org/abs/1904.03120} {arXiv:1904.03120 [hep-lat]}
  \BibitemShut {NoStop}%
\bibitem [{\citenamefont {Aubin}\ \emph {et~al.}(2020)\citenamefont {Aubin},
  \citenamefont {Blum}, \citenamefont {Tu}, \citenamefont {Golterman},
  \citenamefont {Jung},\ and\ \citenamefont {Peris}}]{Aubin:2019usy}%
  \BibitemOpen
  \bibfield  {author} {\bibinfo {author} {\bibfnamefont {C.}~\bibnamefont
  {Aubin}}, \bibinfo {author} {\bibfnamefont {T.}~\bibnamefont {Blum}},
  \bibinfo {author} {\bibfnamefont {C.}~\bibnamefont {Tu}}, \bibinfo {author}
  {\bibfnamefont {M.}~\bibnamefont {Golterman}}, \bibinfo {author}
  {\bibfnamefont {C.}~\bibnamefont {Jung}}, \ and\ \bibinfo {author}
  {\bibfnamefont {S.}~\bibnamefont {Peris}},\ }\href {\doibase
  10.1103/PhysRevD.101.014503} {\bibfield  {journal} {\bibinfo  {journal}
  {Phys. Rev.}\ }\textbf {\bibinfo {volume} {D101}},\ \bibinfo {pages} {014503}
  (\bibinfo {year} {2020})},\ \Eprint {http://arxiv.org/abs/1905.09307}
  {arXiv:1905.09307 [hep-lat]} \BibitemShut {NoStop}%
\bibitem [{\citenamefont {Giusti}\ and\ \citenamefont
  {Simula}(2019)}]{Giusti:2019hkz}%
  \BibitemOpen
  \bibfield  {author} {\bibinfo {author} {\bibfnamefont {D.}~\bibnamefont
  {Giusti}}\ and\ \bibinfo {author} {\bibfnamefont {S.}~\bibnamefont
  {Simula}},\ }\href {\doibase 10.22323/1.363.0104} {\bibfield  {journal}
  {\bibinfo  {journal} {PoS}\ }\textbf {\bibinfo {volume} {LATTICE2019}},\
  \bibinfo {pages} {104} (\bibinfo {year} {2019})},\ \Eprint
  {http://arxiv.org/abs/1910.03874} {arXiv:1910.03874 [hep-lat]} \BibitemShut
  {NoStop}%
\bibitem [{\citenamefont {Melnikov}\ and\ \citenamefont
  {Vainshtein}(2004)}]{Melnikov:2003xd}%
  \BibitemOpen
  \bibfield  {author} {\bibinfo {author} {\bibfnamefont {K.}~\bibnamefont
  {Melnikov}}\ and\ \bibinfo {author} {\bibfnamefont {A.}~\bibnamefont
  {Vainshtein}},\ }\href {\doibase 10.1103/PhysRevD.70.113006} {\bibfield
  {journal} {\bibinfo  {journal} {Phys. Rev.}\ }\textbf {\bibinfo {volume}
  {D70}},\ \bibinfo {pages} {113006} (\bibinfo {year} {2004})},\ \Eprint
  {http://arxiv.org/abs/hep-ph/0312226} {arXiv:hep-ph/0312226 [hep-ph]}
  \BibitemShut {NoStop}%
\bibitem [{\citenamefont {Masjuan}\ and\ \citenamefont
  {S{\'a}nchez-Puertas}(2017)}]{Masjuan:2017tvw}%
  \BibitemOpen
  \bibfield  {author} {\bibinfo {author} {\bibfnamefont {P.}~\bibnamefont
  {Masjuan}}\ and\ \bibinfo {author} {\bibfnamefont {P.}~\bibnamefont
  {S{\'a}nchez-Puertas}},\ }\href {\doibase 10.1103/PhysRevD.95.054026}
  {\bibfield  {journal} {\bibinfo  {journal} {Phys. Rev.}\ }\textbf {\bibinfo
  {volume} {D95}},\ \bibinfo {pages} {054026} (\bibinfo {year} {2017})},\
  \Eprint {http://arxiv.org/abs/1701.05829} {arXiv:1701.05829 [hep-ph]}
  \BibitemShut {NoStop}%
\bibitem [{\citenamefont {Colangelo}\ \emph {et~al.}(2017)\citenamefont
  {Colangelo}, \citenamefont {Hoferichter}, \citenamefont {Procura},\ and\
  \citenamefont {Stoffer}}]{Colangelo:2017fiz}%
  \BibitemOpen
  \bibfield  {author} {\bibinfo {author} {\bibfnamefont {G.}~\bibnamefont
  {Colangelo}}, \bibinfo {author} {\bibfnamefont {M.}~\bibnamefont
  {Hoferichter}}, \bibinfo {author} {\bibfnamefont {M.}~\bibnamefont
  {Procura}}, \ and\ \bibinfo {author} {\bibfnamefont {P.}~\bibnamefont
  {Stoffer}},\ }\href {\doibase 10.1007/JHEP04(2017)161} {\bibfield  {journal}
  {\bibinfo  {journal} {JHEP}\ }\textbf {\bibinfo {volume} {04}},\ \bibinfo
  {pages} {161} (\bibinfo {year} {2017})},\ \Eprint
  {http://arxiv.org/abs/1702.07347} {arXiv:1702.07347 [hep-ph]} \BibitemShut
  {NoStop}%
\bibitem [{\citenamefont {Hoferichter}\ \emph {et~al.}(2018)\citenamefont
  {Hoferichter}, \citenamefont {Hoid}, \citenamefont {Kubis}, \citenamefont
  {Leupold},\ and\ \citenamefont {Schneider}}]{Hoferichter:2018kwz}%
  \BibitemOpen
  \bibfield  {author} {\bibinfo {author} {\bibfnamefont {M.}~\bibnamefont
  {Hoferichter}}, \bibinfo {author} {\bibfnamefont {B.-L.}\ \bibnamefont
  {Hoid}}, \bibinfo {author} {\bibfnamefont {B.}~\bibnamefont {Kubis}},
  \bibinfo {author} {\bibfnamefont {S.}~\bibnamefont {Leupold}}, \ and\
  \bibinfo {author} {\bibfnamefont {S.~P.}\ \bibnamefont {Schneider}},\ }\href
  {\doibase 10.1007/JHEP10(2018)141} {\bibfield  {journal} {\bibinfo  {journal}
  {JHEP}\ }\textbf {\bibinfo {volume} {10}},\ \bibinfo {pages} {141} (\bibinfo
  {year} {2018})},\ \Eprint {http://arxiv.org/abs/1808.04823} {arXiv:1808.04823
  [hep-ph]} \BibitemShut {NoStop}%
\bibitem [{\citenamefont {G{\'e}rardin}\ \emph {et~al.}(2019)\citenamefont
  {G{\'e}rardin}, \citenamefont {Meyer},\ and\ \citenamefont
  {Nyffeler}}]{Gerardin:2019vio}%
  \BibitemOpen
  \bibfield  {author} {\bibinfo {author} {\bibfnamefont {A.}~\bibnamefont
  {G{\'e}rardin}}, \bibinfo {author} {\bibfnamefont {H.~B.}\ \bibnamefont
  {Meyer}}, \ and\ \bibinfo {author} {\bibfnamefont {A.}~\bibnamefont
  {Nyffeler}},\ }\href {\doibase 10.1103/PhysRevD.100.034520} {\bibfield
  {journal} {\bibinfo  {journal} {Phys. Rev.}\ }\textbf {\bibinfo {volume}
  {D100}},\ \bibinfo {pages} {034520} (\bibinfo {year} {2019})},\ \Eprint
  {http://arxiv.org/abs/1903.09471} {arXiv:1903.09471 [hep-lat]} \BibitemShut
  {NoStop}%
\bibitem [{\citenamefont {Bijnens}\ \emph {et~al.}(2019)\citenamefont
  {Bijnens}, \citenamefont {Hermansson-Truedsson},\ and\ \citenamefont
  {Rodr{\'i}guez-S{\'a}nchez}}]{Bijnens:2019ghy}%
  \BibitemOpen
  \bibfield  {author} {\bibinfo {author} {\bibfnamefont {J.}~\bibnamefont
  {Bijnens}}, \bibinfo {author} {\bibfnamefont {N.}~\bibnamefont
  {Hermansson-Truedsson}}, \ and\ \bibinfo {author} {\bibfnamefont
  {A.}~\bibnamefont {Rodr{\'i}guez-S{\'a}nchez}},\ }\href {\doibase
  10.1016/j.physletb.2019.134994} {\bibfield  {journal} {\bibinfo  {journal}
  {Phys. Lett.}\ }\textbf {\bibinfo {volume} {B798}},\ \bibinfo {pages}
  {134994} (\bibinfo {year} {2019})},\ \Eprint
  {http://arxiv.org/abs/1908.03331} {arXiv:1908.03331 [hep-ph]} \BibitemShut
  {NoStop}%
\bibitem [{\citenamefont {Colangelo}\ \emph {et~al.}(2020)\citenamefont
  {Colangelo}, \citenamefont {Hagelstein}, \citenamefont {Hoferichter},
  \citenamefont {Laub},\ and\ \citenamefont {Stoffer}}]{Colangelo:2019uex}%
  \BibitemOpen
  \bibfield  {author} {\bibinfo {author} {\bibfnamefont {G.}~\bibnamefont
  {Colangelo}}, \bibinfo {author} {\bibfnamefont {F.}~\bibnamefont
  {Hagelstein}}, \bibinfo {author} {\bibfnamefont {M.}~\bibnamefont
  {Hoferichter}}, \bibinfo {author} {\bibfnamefont {L.}~\bibnamefont {Laub}}, \
  and\ \bibinfo {author} {\bibfnamefont {P.}~\bibnamefont {Stoffer}},\ }\href
  {\doibase 10.1007/JHEP03(2020)101} {\bibfield  {journal} {\bibinfo  {journal}
  {JHEP}\ }\textbf {\bibinfo {volume} {03}},\ \bibinfo {pages} {101} (\bibinfo
  {year} {2020})},\ \Eprint {http://arxiv.org/abs/1910.13432} {arXiv:1910.13432
  [hep-ph]} \BibitemShut {NoStop}%
\bibitem [{\citenamefont {Pauk}\ and\ \citenamefont
  {Vanderhaeghen}(2014)}]{Pauk:2014rta}%
  \BibitemOpen
  \bibfield  {author} {\bibinfo {author} {\bibfnamefont {V.}~\bibnamefont
  {Pauk}}\ and\ \bibinfo {author} {\bibfnamefont {M.}~\bibnamefont
  {Vanderhaeghen}},\ }\href {\doibase 10.1140/epjc/s10052-014-3008-y}
  {\bibfield  {journal} {\bibinfo  {journal} {Eur. Phys. J.}\ }\textbf
  {\bibinfo {volume} {C74}},\ \bibinfo {pages} {3008} (\bibinfo {year}
  {2014})},\ \Eprint {http://arxiv.org/abs/1401.0832} {arXiv:1401.0832
  [hep-ph]} \BibitemShut {NoStop}%
\bibitem [{\citenamefont {Danilkin}\ and\ \citenamefont
  {Vanderhaeghen}(2017)}]{Danilkin:2016hnh}%
  \BibitemOpen
  \bibfield  {author} {\bibinfo {author} {\bibfnamefont {I.}~\bibnamefont
  {Danilkin}}\ and\ \bibinfo {author} {\bibfnamefont {M.}~\bibnamefont
  {Vanderhaeghen}},\ }\href {\doibase 10.1103/PhysRevD.95.014019} {\bibfield
  {journal} {\bibinfo  {journal} {Phys. Rev.}\ }\textbf {\bibinfo {volume}
  {D95}},\ \bibinfo {pages} {014019} (\bibinfo {year} {2017})},\ \Eprint
  {http://arxiv.org/abs/1611.04646} {arXiv:1611.04646 [hep-ph]} \BibitemShut
  {NoStop}%
\bibitem [{\citenamefont {Jegerlehner}(2017)}]{Jegerlehner:2017gek}%
  \BibitemOpen
  \bibfield  {author} {\bibinfo {author} {\bibfnamefont {F.}~\bibnamefont
  {Jegerlehner}},\ }\href {\doibase 10.1007/978-3-319-63577-4} {\bibfield
  {journal} {\bibinfo  {journal} {Springer Tracts Mod. Phys.}\ }\textbf
  {\bibinfo {volume} {274}},\ \bibinfo {pages} {1} (\bibinfo {year}
  {2017})}\BibitemShut {NoStop}%
\bibitem [{\citenamefont {Knecht}\ \emph {et~al.}(2018)\citenamefont {Knecht},
  \citenamefont {Narison}, \citenamefont {Rabemananjara},\ and\ \citenamefont
  {Rabetiarivony}}]{Knecht:2018sci}%
  \BibitemOpen
  \bibfield  {author} {\bibinfo {author} {\bibfnamefont {M.}~\bibnamefont
  {Knecht}}, \bibinfo {author} {\bibfnamefont {S.}~\bibnamefont {Narison}},
  \bibinfo {author} {\bibfnamefont {A.}~\bibnamefont {Rabemananjara}}, \ and\
  \bibinfo {author} {\bibfnamefont {D.}~\bibnamefont {Rabetiarivony}},\ }\href
  {\doibase 10.1016/j.physletb.2018.10.048} {\bibfield  {journal} {\bibinfo
  {journal} {Phys. Lett.}\ }\textbf {\bibinfo {volume} {B787}},\ \bibinfo
  {pages} {111} (\bibinfo {year} {2018})},\ \Eprint
  {http://arxiv.org/abs/1808.03848} {arXiv:1808.03848 [hep-ph]} \BibitemShut
  {NoStop}%
\bibitem [{\citenamefont {Eichmann}\ \emph {et~al.}(2020)\citenamefont
  {Eichmann}, \citenamefont {Fischer},\ and\ \citenamefont
  {Williams}}]{Eichmann:2019bqf}%
  \BibitemOpen
  \bibfield  {author} {\bibinfo {author} {\bibfnamefont {G.}~\bibnamefont
  {Eichmann}}, \bibinfo {author} {\bibfnamefont {C.~S.}\ \bibnamefont
  {Fischer}}, \ and\ \bibinfo {author} {\bibfnamefont {R.}~\bibnamefont
  {Williams}},\ }\href {\doibase 10.1103/PhysRevD.101.054015} {\bibfield
  {journal} {\bibinfo  {journal} {Phys. Rev.}\ }\textbf {\bibinfo {volume}
  {D101}},\ \bibinfo {pages} {054015} (\bibinfo {year} {2020})},\ \Eprint
  {http://arxiv.org/abs/1910.06795} {arXiv:1910.06795 [hep-ph]} \BibitemShut
  {NoStop}%
\bibitem [{\citenamefont {Roig}\ and\ \citenamefont
  {S{\'a}nchez-Puertas}(2020)}]{Roig:2019reh}%
  \BibitemOpen
  \bibfield  {author} {\bibinfo {author} {\bibfnamefont {P.}~\bibnamefont
  {Roig}}\ and\ \bibinfo {author} {\bibfnamefont {P.}~\bibnamefont
  {S{\'a}nchez-Puertas}},\ }\href {\doibase 10.1103/PhysRevD.101.074019}
  {\bibfield  {journal} {\bibinfo  {journal} {Phys. Rev.}\ }\textbf {\bibinfo
  {volume} {D101}},\ \bibinfo {pages} {074019} (\bibinfo {year} {2020})},\
  \Eprint {http://arxiv.org/abs/1910.02881} {arXiv:1910.02881 [hep-ph]}
  \BibitemShut {NoStop}%
\bibitem [{\citenamefont {Colangelo}\ \emph {et~al.}(2014)\citenamefont
  {Colangelo}, \citenamefont {Hoferichter}, \citenamefont {Nyffeler},
  \citenamefont {Passera},\ and\ \citenamefont {Stoffer}}]{Colangelo:2014qya}%
  \BibitemOpen
  \bibfield  {author} {\bibinfo {author} {\bibfnamefont {G.}~\bibnamefont
  {Colangelo}}, \bibinfo {author} {\bibfnamefont {M.}~\bibnamefont
  {Hoferichter}}, \bibinfo {author} {\bibfnamefont {A.}~\bibnamefont
  {Nyffeler}}, \bibinfo {author} {\bibfnamefont {M.}~\bibnamefont {Passera}}, \
  and\ \bibinfo {author} {\bibfnamefont {P.}~\bibnamefont {Stoffer}},\ }\href
  {\doibase 10.1016/j.physletb.2014.06.012} {\bibfield  {journal} {\bibinfo
  {journal} {Phys. Lett.}\ }\textbf {\bibinfo {volume} {B735}},\ \bibinfo
  {pages} {90} (\bibinfo {year} {2014})},\ \Eprint
  {http://arxiv.org/abs/1403.7512} {arXiv:1403.7512 [hep-ph]} \BibitemShut
  {NoStop}%
\bibitem [{\citenamefont {Blum}\ \emph {et~al.}(2020)\citenamefont {Blum},
  \citenamefont {Christ}, \citenamefont {Hayakawa}, \citenamefont {Izubuchi},
  \citenamefont {Jin}, \citenamefont {Jung},\ and\ \citenamefont
  {Lehner}}]{Blum:2019ugy}%
  \BibitemOpen
  \bibfield  {author} {\bibinfo {author} {\bibfnamefont {T.}~\bibnamefont
  {Blum}}, \bibinfo {author} {\bibfnamefont {N.}~\bibnamefont {Christ}},
  \bibinfo {author} {\bibfnamefont {M.}~\bibnamefont {Hayakawa}}, \bibinfo
  {author} {\bibfnamefont {T.}~\bibnamefont {Izubuchi}}, \bibinfo {author}
  {\bibfnamefont {L.}~\bibnamefont {Jin}}, \bibinfo {author} {\bibfnamefont
  {C.}~\bibnamefont {Jung}}, \ and\ \bibinfo {author} {\bibfnamefont
  {C.}~\bibnamefont {Lehner}},\ }\href {\doibase
  10.1103/PhysRevLett.124.132002} {\bibfield  {journal} {\bibinfo  {journal}
  {Phys. Rev. Lett.}\ }\textbf {\bibinfo {volume} {124}},\ \bibinfo {pages}
  {132002} (\bibinfo {year} {2020})},\ \Eprint
  {http://arxiv.org/abs/1911.08123} {arXiv:1911.08123 [hep-lat]} \BibitemShut
  {NoStop}%
\bibitem [{\citenamefont {Aoyama}\ \emph {et~al.}(2012)\citenamefont {Aoyama},
  \citenamefont {Hayakawa}, \citenamefont {Kinoshita},\ and\ \citenamefont
  {Nio}}]{Aoyama:2012wk}%
  \BibitemOpen
  \bibfield  {author} {\bibinfo {author} {\bibfnamefont {T.}~\bibnamefont
  {Aoyama}}, \bibinfo {author} {\bibfnamefont {M.}~\bibnamefont {Hayakawa}},
  \bibinfo {author} {\bibfnamefont {T.}~\bibnamefont {Kinoshita}}, \ and\
  \bibinfo {author} {\bibfnamefont {M.}~\bibnamefont {Nio}},\ }\href {\doibase
  10.1103/PhysRevLett.109.111808} {\bibfield  {journal} {\bibinfo  {journal}
  {Phys. Rev. Lett.}\ }\textbf {\bibinfo {volume} {109}},\ \bibinfo {pages}
  {111808} (\bibinfo {year} {2012})},\ \Eprint {http://arxiv.org/abs/1205.5370}
  {arXiv:1205.5370 [hep-ph]} \BibitemShut {NoStop}%
\bibitem [{\citenamefont {Aoyama}\ \emph {et~al.}(2019)\citenamefont {Aoyama},
  \citenamefont {Kinoshita},\ and\ \citenamefont {Nio}}]{Aoyama:2019ryr}%
  \BibitemOpen
  \bibfield  {author} {\bibinfo {author} {\bibfnamefont {T.}~\bibnamefont
  {Aoyama}}, \bibinfo {author} {\bibfnamefont {T.}~\bibnamefont {Kinoshita}}, \
  and\ \bibinfo {author} {\bibfnamefont {M.}~\bibnamefont {Nio}},\ }\href
  {\doibase 10.3390/atoms7010028} {\bibfield  {journal} {\bibinfo  {journal}
  {Atoms}\ }\textbf {\bibinfo {volume} {7}},\ \bibinfo {pages} {28} (\bibinfo
  {year} {2019})}\BibitemShut {NoStop}%
\bibitem [{\citenamefont {Czarnecki}\ \emph {et~al.}(2003)\citenamefont
  {Czarnecki}, \citenamefont {Marciano},\ and\ \citenamefont
  {Vainshtein}}]{Czarnecki:2002nt}%
  \BibitemOpen
  \bibfield  {author} {\bibinfo {author} {\bibfnamefont {A.}~\bibnamefont
  {Czarnecki}}, \bibinfo {author} {\bibfnamefont {W.~J.}\ \bibnamefont
  {Marciano}}, \ and\ \bibinfo {author} {\bibfnamefont {A.}~\bibnamefont
  {Vainshtein}},\ }\href {\doibase 10.1103/PhysRevD.67.073006} {\bibfield
  {journal} {\bibinfo  {journal} {Phys. Rev. D}\ }\textbf {\bibinfo {volume}
  {67}},\ \bibinfo {pages} {073006} (\bibinfo {year} {2003})},\ \bibinfo {note}
  {[Erratum: Phys.Rev.D 73, 119901 (2006)]},\ \Eprint
  {http://arxiv.org/abs/hep-ph/0212229} {arXiv:hep-ph/0212229} \BibitemShut
  {NoStop}%
\bibitem [{\citenamefont {Gnendiger}\ \emph {et~al.}(2013)\citenamefont
  {Gnendiger}, \citenamefont {St\"ockinger},\ and\ \citenamefont
  {St\"ockinger-Kim}}]{Gnendiger:2013pva}%
  \BibitemOpen
  \bibfield  {author} {\bibinfo {author} {\bibfnamefont {C.}~\bibnamefont
  {Gnendiger}}, \bibinfo {author} {\bibfnamefont {D.}~\bibnamefont
  {St\"ockinger}}, \ and\ \bibinfo {author} {\bibfnamefont {H.}~\bibnamefont
  {St\"ockinger-Kim}},\ }\href {\doibase 10.1103/PhysRevD.88.053005} {\bibfield
   {journal} {\bibinfo  {journal} {Phys. Rev. D}\ }\textbf {\bibinfo {volume}
  {88}},\ \bibinfo {pages} {053005} (\bibinfo {year} {2013})},\ \Eprint
  {http://arxiv.org/abs/1306.5546} {arXiv:1306.5546 [hep-ph]} \BibitemShut
  {NoStop}%
\bibitem [{\citenamefont {Aoyama}\ \emph {et~al.}(2020)\citenamefont {Aoyama}
  \emph {et~al.}}]{Aoyama:2020ynm}%
  \BibitemOpen
  \bibfield  {author} {\bibinfo {author} {\bibfnamefont {T.}~\bibnamefont
  {Aoyama}} \emph {et~al.},\ }\href {\doibase 10.1016/j.physrep.2020.07.006}
  {\bibfield  {journal} {\bibinfo  {journal} {Phys. Rept.}\ }\textbf {\bibinfo
  {volume} {887}},\ \bibinfo {pages} {1} (\bibinfo {year} {2020})},\ \Eprint
  {http://arxiv.org/abs/2006.04822} {arXiv:2006.04822 [hep-ph]} \BibitemShut
  {NoStop}%
\bibitem [{\citenamefont {Bennett}\ \emph {et~al.}(2006)\citenamefont {Bennett}
  \emph {et~al.}}]{Bennett:2006fi}%
  \BibitemOpen
  \bibfield  {author} {\bibinfo {author} {\bibfnamefont {G.~W.}\ \bibnamefont
  {Bennett}} \emph {et~al.} (\bibinfo {collaboration} {Muon g-2}),\ }\href
  {\doibase 10.1103/PhysRevD.73.072003} {\bibfield  {journal} {\bibinfo
  {journal} {Phys. Rev. D}\ }\textbf {\bibinfo {volume} {73}},\ \bibinfo
  {pages} {072003} (\bibinfo {year} {2006})},\ \Eprint
  {http://arxiv.org/abs/hep-ex/0602035} {arXiv:hep-ex/0602035} \BibitemShut
  {NoStop}%
\bibitem [{\citenamefont {Albahri}\ \emph
  {et~al.}(2021{\natexlab{a}})\citenamefont {Albahri} \emph
  {et~al.}}]{1856534}%
  \BibitemOpen
  \bibfield  {author} {\bibinfo {author} {\bibfnamefont {T.}~\bibnamefont
  {Albahri}} \emph {et~al.},\ }\href {\doibase 10.1103/PhysRevA.103.042208}
  {\bibfield  {journal} {\bibinfo  {journal} {Phys. Rev. A}\ }\textbf {\bibinfo
  {volume} {103}},\ \bibinfo {pages} {042208} (\bibinfo {year}
  {2021}{\natexlab{a}})},\ \Eprint {http://arxiv.org/abs/2104.03201}
  {arXiv:2104.03201 [hep-ex]} \BibitemShut {NoStop}%
\bibitem [{\citenamefont {Albahri}\ \emph
  {et~al.}(2021{\natexlab{b}})\citenamefont {Albahri} \emph
  {et~al.}}]{1856531}%
  \BibitemOpen
  \bibfield  {author} {\bibinfo {author} {\bibfnamefont {T.}~\bibnamefont
  {Albahri}} \emph {et~al.},\ }\href {\doibase 10.1103/PhysRevD.103.072002}
  {\bibfield  {journal} {\bibinfo  {journal} {Phys. Rev. D}\ }\textbf {\bibinfo
  {volume} {103}},\ \bibinfo {pages} {072002} (\bibinfo {year}
  {2021}{\natexlab{b}})},\ \Eprint {http://arxiv.org/abs/2104.03247}
  {arXiv:2104.03247 [hep-ex]} \BibitemShut {NoStop}%
\bibitem [{\citenamefont {Abi}\ \emph {et~al.}(2021)\citenamefont {Abi} \emph
  {et~al.}}]{FNAL}%
  \BibitemOpen
  \bibfield  {author} {\bibinfo {author} {\bibfnamefont {B.}~\bibnamefont
  {Abi}} \emph {et~al.} (\bibinfo {collaboration} {Muon g-2 Collaboration}),\
  }\href {\doibase 10.1103/PhysRevLett.126.141801} {\bibfield  {journal}
  {\bibinfo  {journal} {Phys. Rev. Lett.}\ }\textbf {\bibinfo {volume} {126}},\
  \bibinfo {pages} {141801} (\bibinfo {year} {2021})}\BibitemShut {NoStop}%
\bibitem [{\citenamefont {Borsanyi}\ \emph {et~al.}(2021)\citenamefont
  {Borsanyi} \emph {et~al.}}]{Borsanyi:2020mff}%
  \BibitemOpen
  \bibfield  {author} {\bibinfo {author} {\bibfnamefont {S.}~\bibnamefont
  {Borsanyi}} \emph {et~al.},\ }\href {\doibase 10.1038/s41586-021-03418-1}
  {\bibfield  {journal} {\bibinfo  {journal} {Nature}\ }\textbf {\bibinfo
  {volume} {593}},\ \bibinfo {pages} {51} (\bibinfo {year} {2021})},\ \Eprint
  {http://arxiv.org/abs/2002.12347} {arXiv:2002.12347 [hep-lat]} \BibitemShut
  {NoStop}%
\bibitem [{\citenamefont {Padley}\ \emph {et~al.}(2015)\citenamefont {Padley},
  \citenamefont {Sinha},\ and\ \citenamefont {Wang}}]{Padley:2015uma}%
  \BibitemOpen
  \bibfield  {author} {\bibinfo {author} {\bibfnamefont {B.~P.}\ \bibnamefont
  {Padley}}, \bibinfo {author} {\bibfnamefont {K.}~\bibnamefont {Sinha}}, \
  and\ \bibinfo {author} {\bibfnamefont {K.}~\bibnamefont {Wang}},\ }\href
  {\doibase 10.1103/PhysRevD.92.055025} {\bibfield  {journal} {\bibinfo
  {journal} {Phys. Rev. D}\ }\textbf {\bibinfo {volume} {92}},\ \bibinfo
  {pages} {055025} (\bibinfo {year} {2015})},\ \Eprint
  {http://arxiv.org/abs/1505.05877} {arXiv:1505.05877 [hep-ph]} \BibitemShut
  {NoStop}%
\bibitem [{\citenamefont {Sabatta}\ \emph {et~al.}(2020)\citenamefont
  {Sabatta}, \citenamefont {Cornell}, \citenamefont {Goyal}, \citenamefont
  {Kumar}, \citenamefont {Mellado},\ and\ \citenamefont
  {Ruan}}]{Sabatta:2019nfg}%
  \BibitemOpen
  \bibfield  {author} {\bibinfo {author} {\bibfnamefont {D.}~\bibnamefont
  {Sabatta}}, \bibinfo {author} {\bibfnamefont {A.~S.}\ \bibnamefont
  {Cornell}}, \bibinfo {author} {\bibfnamefont {A.}~\bibnamefont {Goyal}},
  \bibinfo {author} {\bibfnamefont {M.}~\bibnamefont {Kumar}}, \bibinfo
  {author} {\bibfnamefont {B.}~\bibnamefont {Mellado}}, \ and\ \bibinfo
  {author} {\bibfnamefont {X.}~\bibnamefont {Ruan}},\ }\href {\doibase
  10.1088/1674-1137/44/6/063103} {\bibfield  {journal} {\bibinfo  {journal}
  {Chin. Phys. C}\ }\textbf {\bibinfo {volume} {44}},\ \bibinfo {pages}
  {063103} (\bibinfo {year} {2020})},\ \Eprint
  {http://arxiv.org/abs/1909.03969} {arXiv:1909.03969 [hep-ph]} \BibitemShut
  {NoStop}%
\bibitem [{\citenamefont {Li}\ \emph {et~al.}(2021{\natexlab{a}})\citenamefont
  {Li}, \citenamefont {Li}, \citenamefont {Li}, \citenamefont {Yang},\ and\
  \citenamefont {Zhang}}]{Li:2020dbg}%
  \BibitemOpen
  \bibfield  {author} {\bibinfo {author} {\bibfnamefont {S.-P.}\ \bibnamefont
  {Li}}, \bibinfo {author} {\bibfnamefont {X.-Q.}\ \bibnamefont {Li}}, \bibinfo
  {author} {\bibfnamefont {Y.-Y.}\ \bibnamefont {Li}}, \bibinfo {author}
  {\bibfnamefont {Y.-D.}\ \bibnamefont {Yang}}, \ and\ \bibinfo {author}
  {\bibfnamefont {X.}~\bibnamefont {Zhang}},\ }\href {\doibase
  10.1007/JHEP01(2021)034} {\bibfield  {journal} {\bibinfo  {journal} {JHEP}\
  }\textbf {\bibinfo {volume} {01}},\ \bibinfo {pages} {034} (\bibinfo {year}
  {2021}{\natexlab{a}})},\ \Eprint {http://arxiv.org/abs/2010.02799}
  {arXiv:2010.02799 [hep-ph]} \BibitemShut {NoStop}%
\bibitem [{\citenamefont {Bigaran}\ and\ \citenamefont
  {Volkas}(2020)}]{Bigaran:2020jil}%
  \BibitemOpen
  \bibfield  {author} {\bibinfo {author} {\bibfnamefont {I.}~\bibnamefont
  {Bigaran}}\ and\ \bibinfo {author} {\bibfnamefont {R.~R.}\ \bibnamefont
  {Volkas}},\ }\href {\doibase 10.1103/PhysRevD.102.075037} {\bibfield
  {journal} {\bibinfo  {journal} {Phys. Rev. D}\ }\textbf {\bibinfo {volume}
  {102}},\ \bibinfo {pages} {075037} (\bibinfo {year} {2020})},\ \Eprint
  {http://arxiv.org/abs/2002.12544} {arXiv:2002.12544 [hep-ph]} \BibitemShut
  {NoStop}%
\bibitem [{\citenamefont {Yin}\ and\ \citenamefont
  {Yamaguchi}(2020)}]{Yin:2020afe}%
  \BibitemOpen
  \bibfield  {author} {\bibinfo {author} {\bibfnamefont {W.}~\bibnamefont
  {Yin}}\ and\ \bibinfo {author} {\bibfnamefont {M.}~\bibnamefont
  {Yamaguchi}},\ }\href@noop {} {\  (\bibinfo {year} {2020})},\ \Eprint
  {http://arxiv.org/abs/2012.03928} {arXiv:2012.03928 [hep-ph]} \BibitemShut
  {NoStop}%
\bibitem [{\citenamefont {Baker}\ \emph {et~al.}(2021)\citenamefont {Baker},
  \citenamefont {Cox},\ and\ \citenamefont {Volkas}}]{Baker:2021yli}%
  \BibitemOpen
  \bibfield  {author} {\bibinfo {author} {\bibfnamefont {M.~J.}\ \bibnamefont
  {Baker}}, \bibinfo {author} {\bibfnamefont {P.}~\bibnamefont {Cox}}, \ and\
  \bibinfo {author} {\bibfnamefont {R.~R.}\ \bibnamefont {Volkas}},\
  }\href@noop {} {\  (\bibinfo {year} {2021})},\ \Eprint
  {http://arxiv.org/abs/2103.13401} {arXiv:2103.13401 [hep-ph]} \BibitemShut
  {NoStop}%
\bibitem [{\citenamefont {Bodas}\ \emph {et~al.}(2021)\citenamefont {Bodas},
  \citenamefont {Coy},\ and\ \citenamefont {King}}]{Bodas:2021fsy}%
  \BibitemOpen
  \bibfield  {author} {\bibinfo {author} {\bibfnamefont {A.}~\bibnamefont
  {Bodas}}, \bibinfo {author} {\bibfnamefont {R.}~\bibnamefont {Coy}}, \ and\
  \bibinfo {author} {\bibfnamefont {S.~J.~D.}\ \bibnamefont {King}},\
  }\href@noop {} {\  (\bibinfo {year} {2021})},\ \Eprint
  {http://arxiv.org/abs/2102.07781} {arXiv:2102.07781 [hep-ph]} \BibitemShut
  {NoStop}%
\bibitem [{\citenamefont {Chen}\ \emph
  {et~al.}(2021{\natexlab{a}})\citenamefont {Chen}, \citenamefont {Wang},\ and\
  \citenamefont {Yao}}]{Chen:2021rnl}%
  \BibitemOpen
  \bibfield  {author} {\bibinfo {author} {\bibfnamefont {N.}~\bibnamefont
  {Chen}}, \bibinfo {author} {\bibfnamefont {B.}~\bibnamefont {Wang}}, \ and\
  \bibinfo {author} {\bibfnamefont {C.-Y.}\ \bibnamefont {Yao}},\ }\href@noop
  {} {\  (\bibinfo {year} {2021}{\natexlab{a}})},\ \Eprint
  {http://arxiv.org/abs/2102.05619} {arXiv:2102.05619 [hep-ph]} \BibitemShut
  {NoStop}%
\bibitem [{\citenamefont {Yin}\ and\ \citenamefont {Yin}(2021)}]{Yin:2021yqy}%
  \BibitemOpen
  \bibfield  {author} {\bibinfo {author} {\bibfnamefont {W.}~\bibnamefont
  {Yin}}\ and\ \bibinfo {author} {\bibfnamefont {W.}~\bibnamefont {Yin}},\
  }\href@noop {} {\  (\bibinfo {year} {2021})},\ \Eprint
  {http://arxiv.org/abs/2103.14234} {arXiv:2103.14234 [hep-ph]} \BibitemShut
  {NoStop}%
\bibitem [{\citenamefont {Chiang}\ and\ \citenamefont
  {Yagyu}(2021)}]{Chiang:2021pma}%
  \BibitemOpen
  \bibfield  {author} {\bibinfo {author} {\bibfnamefont {C.-W.}\ \bibnamefont
  {Chiang}}\ and\ \bibinfo {author} {\bibfnamefont {K.}~\bibnamefont {Yagyu}},\
  }\href@noop {} {\  (\bibinfo {year} {2021})},\ \Eprint
  {http://arxiv.org/abs/2104.00890} {arXiv:2104.00890 [hep-ph]} \BibitemShut
  {NoStop}%
\bibitem [{\citenamefont {C\'arcamo~Hern\'andez}\ \emph
  {et~al.}(2021)\citenamefont {C\'arcamo~Hern\'andez}, \citenamefont
  {Espinoza}, \citenamefont {Carlos G\'omez-Izquierdo},\ and\ \citenamefont
  {Mondrag\'on}}]{CarcamoHernandez:2021iat}%
  \BibitemOpen
  \bibfield  {author} {\bibinfo {author} {\bibfnamefont {A.~E.}\ \bibnamefont
  {C\'arcamo~Hern\'andez}}, \bibinfo {author} {\bibfnamefont {C.}~\bibnamefont
  {Espinoza}}, \bibinfo {author} {\bibfnamefont {J.}~\bibnamefont {Carlos
  G\'omez-Izquierdo}}, \ and\ \bibinfo {author} {\bibfnamefont
  {M.}~\bibnamefont {Mondrag\'on}},\ }\href@noop {} {\  (\bibinfo {year}
  {2021})},\ \Eprint {http://arxiv.org/abs/2104.02730} {arXiv:2104.02730
  [hep-ph]} \BibitemShut {NoStop}%
\bibitem [{\citenamefont {Lee}(2021)}]{Lee:2021jdr}%
  \BibitemOpen
  \bibfield  {author} {\bibinfo {author} {\bibfnamefont {H.~M.}\ \bibnamefont
  {Lee}},\ }\href@noop {} {\  (\bibinfo {year} {2021})},\ \Eprint
  {http://arxiv.org/abs/2104.02982} {arXiv:2104.02982 [hep-ph]} \BibitemShut
  {NoStop}%
\bibitem [{\citenamefont {Crivellin}\ and\ \citenamefont
  {Hoferichter}(2021)}]{Crivellin:2021rbq}%
  \BibitemOpen
  \bibfield  {author} {\bibinfo {author} {\bibfnamefont {A.}~\bibnamefont
  {Crivellin}}\ and\ \bibinfo {author} {\bibfnamefont {M.}~\bibnamefont
  {Hoferichter}},\ }\href@noop {} {\  (\bibinfo {year} {2021})},\ \Eprint
  {http://arxiv.org/abs/2104.03202} {arXiv:2104.03202 [hep-ph]} \BibitemShut
  {NoStop}%
\bibitem [{\citenamefont {Endo}\ \emph {et~al.}(2021)\citenamefont {Endo},
  \citenamefont {Hamaguchi}, \citenamefont {Iwamoto},\ and\ \citenamefont
  {Kitahara}}]{Endo:2021zal}%
  \BibitemOpen
  \bibfield  {author} {\bibinfo {author} {\bibfnamefont {M.}~\bibnamefont
  {Endo}}, \bibinfo {author} {\bibfnamefont {K.}~\bibnamefont {Hamaguchi}},
  \bibinfo {author} {\bibfnamefont {S.}~\bibnamefont {Iwamoto}}, \ and\
  \bibinfo {author} {\bibfnamefont {T.}~\bibnamefont {Kitahara}},\ }\href@noop
  {} {\  (\bibinfo {year} {2021})},\ \Eprint {http://arxiv.org/abs/2104.03217}
  {arXiv:2104.03217 [hep-ph]} \BibitemShut {NoStop}%
\bibitem [{\citenamefont {Iwamoto}\ \emph {et~al.}(2021)\citenamefont
  {Iwamoto}, \citenamefont {Yanagida},\ and\ \citenamefont
  {Yokozaki}}]{Iwamoto:2021aaf}%
  \BibitemOpen
  \bibfield  {author} {\bibinfo {author} {\bibfnamefont {S.}~\bibnamefont
  {Iwamoto}}, \bibinfo {author} {\bibfnamefont {T.~T.}\ \bibnamefont
  {Yanagida}}, \ and\ \bibinfo {author} {\bibfnamefont {N.}~\bibnamefont
  {Yokozaki}},\ }\href@noop {} {\  (\bibinfo {year} {2021})},\ \Eprint
  {http://arxiv.org/abs/2104.03223} {arXiv:2104.03223 [hep-ph]} \BibitemShut
  {NoStop}%
\bibitem [{\citenamefont {Han}\ \emph {et~al.}(2021{\natexlab{a}})\citenamefont
  {Han}, \citenamefont {Li}, \citenamefont {Wang}, \citenamefont {Wang},\ and\
  \citenamefont {Zhang}}]{Han:2021gfu}%
  \BibitemOpen
  \bibfield  {author} {\bibinfo {author} {\bibfnamefont {X.-F.}\ \bibnamefont
  {Han}}, \bibinfo {author} {\bibfnamefont {T.}~\bibnamefont {Li}}, \bibinfo
  {author} {\bibfnamefont {H.-X.}\ \bibnamefont {Wang}}, \bibinfo {author}
  {\bibfnamefont {L.}~\bibnamefont {Wang}}, \ and\ \bibinfo {author}
  {\bibfnamefont {Y.}~\bibnamefont {Zhang}},\ }\href@noop {} {\  (\bibinfo
  {year} {2021}{\natexlab{a}})},\ \Eprint {http://arxiv.org/abs/2104.03227}
  {arXiv:2104.03227 [hep-ph]} \BibitemShut {NoStop}%
\bibitem [{\citenamefont {Arcadi}\ \emph {et~al.}(2021)\citenamefont {Arcadi},
  \citenamefont {Calibbi}, \citenamefont {Fedele},\ and\ \citenamefont
  {Mescia}}]{Arcadi:2021cwg}%
  \BibitemOpen
  \bibfield  {author} {\bibinfo {author} {\bibfnamefont {G.}~\bibnamefont
  {Arcadi}}, \bibinfo {author} {\bibfnamefont {L.}~\bibnamefont {Calibbi}},
  \bibinfo {author} {\bibfnamefont {M.}~\bibnamefont {Fedele}}, \ and\ \bibinfo
  {author} {\bibfnamefont {F.}~\bibnamefont {Mescia}},\ }\href@noop {} {\
  (\bibinfo {year} {2021})},\ \Eprint {http://arxiv.org/abs/2104.03228}
  {arXiv:2104.03228 [hep-ph]} \BibitemShut {NoStop}%
\bibitem [{\citenamefont {Criado}\ \emph {et~al.}(2021)\citenamefont {Criado},
  \citenamefont {Djouadi}, \citenamefont {Koivunen}, \citenamefont
  {M\"u\"ursepp}, \citenamefont {Raidal},\ and\ \citenamefont
  {Veerm\"ae}}]{Criado:2021qpd}%
  \BibitemOpen
  \bibfield  {author} {\bibinfo {author} {\bibfnamefont {J.~C.}\ \bibnamefont
  {Criado}}, \bibinfo {author} {\bibfnamefont {A.}~\bibnamefont {Djouadi}},
  \bibinfo {author} {\bibfnamefont {N.}~\bibnamefont {Koivunen}}, \bibinfo
  {author} {\bibfnamefont {K.}~\bibnamefont {M\"u\"ursepp}}, \bibinfo {author}
  {\bibfnamefont {M.}~\bibnamefont {Raidal}}, \ and\ \bibinfo {author}
  {\bibfnamefont {H.}~\bibnamefont {Veerm\"ae}},\ }\href@noop {} {\  (\bibinfo
  {year} {2021})},\ \Eprint {http://arxiv.org/abs/2104.03231} {arXiv:2104.03231
  [hep-ph]} \BibitemShut {NoStop}%
\bibitem [{\citenamefont {Zhu}\ and\ \citenamefont {Liu}(2021)}]{Zhu:2021vlz}%
  \BibitemOpen
  \bibfield  {author} {\bibinfo {author} {\bibfnamefont {B.}~\bibnamefont
  {Zhu}}\ and\ \bibinfo {author} {\bibfnamefont {X.}~\bibnamefont {Liu}},\
  }\href@noop {} {\  (\bibinfo {year} {2021})},\ \Eprint
  {http://arxiv.org/abs/2104.03238} {arXiv:2104.03238 [hep-ph]} \BibitemShut
  {NoStop}%
\bibitem [{\citenamefont {Gu}\ \emph {et~al.}(2021)\citenamefont {Gu},
  \citenamefont {Liu}, \citenamefont {Su},\ and\ \citenamefont
  {Wang}}]{Gu:2021mjd}%
  \BibitemOpen
  \bibfield  {author} {\bibinfo {author} {\bibfnamefont {Y.}~\bibnamefont
  {Gu}}, \bibinfo {author} {\bibfnamefont {N.}~\bibnamefont {Liu}}, \bibinfo
  {author} {\bibfnamefont {L.}~\bibnamefont {Su}}, \ and\ \bibinfo {author}
  {\bibfnamefont {D.}~\bibnamefont {Wang}},\ }\href@noop {} {\  (\bibinfo
  {year} {2021})},\ \Eprint {http://arxiv.org/abs/2104.03239} {arXiv:2104.03239
  [hep-ph]} \BibitemShut {NoStop}%
\bibitem [{\citenamefont {Wang}\ \emph
  {et~al.}(2021{\natexlab{a}})\citenamefont {Wang}, \citenamefont {Wang},\ and\
  \citenamefont {Zhang}}]{Wang:2021fkn}%
  \BibitemOpen
  \bibfield  {author} {\bibinfo {author} {\bibfnamefont {H.-X.}\ \bibnamefont
  {Wang}}, \bibinfo {author} {\bibfnamefont {L.}~\bibnamefont {Wang}}, \ and\
  \bibinfo {author} {\bibfnamefont {Y.}~\bibnamefont {Zhang}},\ }\href@noop {}
  {\  (\bibinfo {year} {2021}{\natexlab{a}})},\ \Eprint
  {http://arxiv.org/abs/2104.03242} {arXiv:2104.03242 [hep-ph]} \BibitemShut
  {NoStop}%
\bibitem [{\citenamefont {Van~Beekveld}\ \emph {et~al.}(2021)\citenamefont
  {Van~Beekveld}, \citenamefont {Beenakker}, \citenamefont {Schutten},\ and\
  \citenamefont {De~Wit}}]{VanBeekveld:2021tgn}%
  \BibitemOpen
  \bibfield  {author} {\bibinfo {author} {\bibfnamefont {M.}~\bibnamefont
  {Van~Beekveld}}, \bibinfo {author} {\bibfnamefont {W.}~\bibnamefont
  {Beenakker}}, \bibinfo {author} {\bibfnamefont {M.}~\bibnamefont {Schutten}},
  \ and\ \bibinfo {author} {\bibfnamefont {J.}~\bibnamefont {De~Wit}},\
  }\href@noop {} {\  (\bibinfo {year} {2021})},\ \Eprint
  {http://arxiv.org/abs/2104.03245} {arXiv:2104.03245 [hep-ph]} \BibitemShut
  {NoStop}%
\bibitem [{\citenamefont {Nomura}\ and\ \citenamefont
  {Okada}(2021)}]{Nomura:2021oeu}%
  \BibitemOpen
  \bibfield  {author} {\bibinfo {author} {\bibfnamefont {T.}~\bibnamefont
  {Nomura}}\ and\ \bibinfo {author} {\bibfnamefont {H.}~\bibnamefont {Okada}},\
  }\href@noop {} {\  (\bibinfo {year} {2021})},\ \Eprint
  {http://arxiv.org/abs/2104.03248} {arXiv:2104.03248 [hep-ph]} \BibitemShut
  {NoStop}%
\bibitem [{\citenamefont {Anselmi}\ \emph {et~al.}(2021)\citenamefont
  {Anselmi}, \citenamefont {Kannike}, \citenamefont {Marzo}, \citenamefont
  {Marzola}, \citenamefont {Melis}, \citenamefont {M\"u\"ursepp}, \citenamefont
  {Piva},\ and\ \citenamefont {Raidal}}]{Anselmi:2021chp}%
  \BibitemOpen
  \bibfield  {author} {\bibinfo {author} {\bibfnamefont {D.}~\bibnamefont
  {Anselmi}}, \bibinfo {author} {\bibfnamefont {K.}~\bibnamefont {Kannike}},
  \bibinfo {author} {\bibfnamefont {C.}~\bibnamefont {Marzo}}, \bibinfo
  {author} {\bibfnamefont {L.}~\bibnamefont {Marzola}}, \bibinfo {author}
  {\bibfnamefont {A.}~\bibnamefont {Melis}}, \bibinfo {author} {\bibfnamefont
  {K.}~\bibnamefont {M\"u\"ursepp}}, \bibinfo {author} {\bibfnamefont
  {M.}~\bibnamefont {Piva}}, \ and\ \bibinfo {author} {\bibfnamefont
  {M.}~\bibnamefont {Raidal}},\ }\href@noop {} {\  (\bibinfo {year} {2021})},\
  \Eprint {http://arxiv.org/abs/2104.03249} {arXiv:2104.03249 [hep-ph]}
  \BibitemShut {NoStop}%
\bibitem [{\citenamefont {Yin}(2021)}]{Yin:2021mls}%
  \BibitemOpen
  \bibfield  {author} {\bibinfo {author} {\bibfnamefont {W.}~\bibnamefont
  {Yin}},\ }\href@noop {} {\  (\bibinfo {year} {2021})},\ \Eprint
  {http://arxiv.org/abs/2104.03259} {arXiv:2104.03259 [hep-ph]} \BibitemShut
  {NoStop}%
\bibitem [{\citenamefont {Wang}\ \emph
  {et~al.}(2021{\natexlab{b}})\citenamefont {Wang}, \citenamefont {Wu},
  \citenamefont {Xiao}, \citenamefont {Yang},\ and\ \citenamefont
  {Zhang}}]{Wang:2021bcx}%
  \BibitemOpen
  \bibfield  {author} {\bibinfo {author} {\bibfnamefont {F.}~\bibnamefont
  {Wang}}, \bibinfo {author} {\bibfnamefont {L.}~\bibnamefont {Wu}}, \bibinfo
  {author} {\bibfnamefont {Y.}~\bibnamefont {Xiao}}, \bibinfo {author}
  {\bibfnamefont {J.~M.}\ \bibnamefont {Yang}}, \ and\ \bibinfo {author}
  {\bibfnamefont {Y.}~\bibnamefont {Zhang}},\ }\href@noop {} {\  (\bibinfo
  {year} {2021}{\natexlab{b}})},\ \Eprint {http://arxiv.org/abs/2104.03262}
  {arXiv:2104.03262 [hep-ph]} \BibitemShut {NoStop}%
\bibitem [{\citenamefont {Buen-Abad}\ \emph {et~al.}(2021)\citenamefont
  {Buen-Abad}, \citenamefont {Fan}, \citenamefont {Reece},\ and\ \citenamefont
  {Sun}}]{Buen-Abad:2021fwq}%
  \BibitemOpen
  \bibfield  {author} {\bibinfo {author} {\bibfnamefont {M.~A.}\ \bibnamefont
  {Buen-Abad}}, \bibinfo {author} {\bibfnamefont {J.}~\bibnamefont {Fan}},
  \bibinfo {author} {\bibfnamefont {M.}~\bibnamefont {Reece}}, \ and\ \bibinfo
  {author} {\bibfnamefont {C.}~\bibnamefont {Sun}},\ }\href@noop {} {\
  (\bibinfo {year} {2021})},\ \Eprint {http://arxiv.org/abs/2104.03267}
  {arXiv:2104.03267 [hep-ph]} \BibitemShut {NoStop}%
\bibitem [{\citenamefont {Das}\ \emph {et~al.}(2021)\citenamefont {Das},
  \citenamefont {Kumar~Das},\ and\ \citenamefont {Khan}}]{Das:2021zea}%
  \BibitemOpen
  \bibfield  {author} {\bibinfo {author} {\bibfnamefont {P.}~\bibnamefont
  {Das}}, \bibinfo {author} {\bibfnamefont {M.}~\bibnamefont {Kumar~Das}}, \
  and\ \bibinfo {author} {\bibfnamefont {N.}~\bibnamefont {Khan}},\ }\href@noop
  {} {\  (\bibinfo {year} {2021})},\ \Eprint {http://arxiv.org/abs/2104.03271}
  {arXiv:2104.03271 [hep-ph]} \BibitemShut {NoStop}%
\bibitem [{\citenamefont {Abdughani}\ \emph {et~al.}(2021)\citenamefont
  {Abdughani}, \citenamefont {Fan}, \citenamefont {Feng}, \citenamefont
  {Sming~Tsai}, \citenamefont {Wu},\ and\ \citenamefont
  {Yuan}}]{Abdughani:2021pdc}%
  \BibitemOpen
  \bibfield  {author} {\bibinfo {author} {\bibfnamefont {M.}~\bibnamefont
  {Abdughani}}, \bibinfo {author} {\bibfnamefont {Y.-Z.}\ \bibnamefont {Fan}},
  \bibinfo {author} {\bibfnamefont {L.}~\bibnamefont {Feng}}, \bibinfo {author}
  {\bibfnamefont {Y.-L.}\ \bibnamefont {Sming~Tsai}}, \bibinfo {author}
  {\bibfnamefont {L.}~\bibnamefont {Wu}}, \ and\ \bibinfo {author}
  {\bibfnamefont {Q.}~\bibnamefont {Yuan}},\ }\href@noop {} {\  (\bibinfo
  {year} {2021})},\ \Eprint {http://arxiv.org/abs/2104.03274} {arXiv:2104.03274
  [hep-ph]} \BibitemShut {NoStop}%
\bibitem [{\citenamefont {Chen}\ \emph
  {et~al.}(2021{\natexlab{b}})\citenamefont {Chen}, \citenamefont {Chiang},\
  and\ \citenamefont {Nomura}}]{Chen:2021jok}%
  \BibitemOpen
  \bibfield  {author} {\bibinfo {author} {\bibfnamefont {C.-H.}\ \bibnamefont
  {Chen}}, \bibinfo {author} {\bibfnamefont {C.-W.}\ \bibnamefont {Chiang}}, \
  and\ \bibinfo {author} {\bibfnamefont {T.}~\bibnamefont {Nomura}},\
  }\href@noop {} {\  (\bibinfo {year} {2021}{\natexlab{b}})},\ \Eprint
  {http://arxiv.org/abs/2104.03275} {arXiv:2104.03275 [hep-ph]} \BibitemShut
  {NoStop}%
\bibitem [{\citenamefont {Ge}\ \emph {et~al.}(2021)\citenamefont {Ge},
  \citenamefont {Ma},\ and\ \citenamefont {Pasquini}}]{Ge:2021cjz}%
  \BibitemOpen
  \bibfield  {author} {\bibinfo {author} {\bibfnamefont {S.-F.}\ \bibnamefont
  {Ge}}, \bibinfo {author} {\bibfnamefont {X.-D.}\ \bibnamefont {Ma}}, \ and\
  \bibinfo {author} {\bibfnamefont {P.}~\bibnamefont {Pasquini}},\ }\href@noop
  {} {\  (\bibinfo {year} {2021})},\ \Eprint {http://arxiv.org/abs/2104.03276}
  {arXiv:2104.03276 [hep-ph]} \BibitemShut {NoStop}%
\bibitem [{\citenamefont {Cadeddu}\ \emph {et~al.}(2021)\citenamefont
  {Cadeddu}, \citenamefont {Cargioli}, \citenamefont {Dordei}, \citenamefont
  {Giunti},\ and\ \citenamefont {Picciau}}]{Cadeddu:2021dqx}%
  \BibitemOpen
  \bibfield  {author} {\bibinfo {author} {\bibfnamefont {M.}~\bibnamefont
  {Cadeddu}}, \bibinfo {author} {\bibfnamefont {N.}~\bibnamefont {Cargioli}},
  \bibinfo {author} {\bibfnamefont {F.}~\bibnamefont {Dordei}}, \bibinfo
  {author} {\bibfnamefont {C.}~\bibnamefont {Giunti}}, \ and\ \bibinfo {author}
  {\bibfnamefont {E.}~\bibnamefont {Picciau}},\ }\href@noop {} {\  (\bibinfo
  {year} {2021})},\ \Eprint {http://arxiv.org/abs/2104.03280} {arXiv:2104.03280
  [hep-ph]} \BibitemShut {NoStop}%
\bibitem [{\citenamefont {Brdar}\ \emph {et~al.}(2021)\citenamefont {Brdar},
  \citenamefont {Jana}, \citenamefont {Kubo},\ and\ \citenamefont
  {Lindner}}]{Brdar:2021pla}%
  \BibitemOpen
  \bibfield  {author} {\bibinfo {author} {\bibfnamefont {V.}~\bibnamefont
  {Brdar}}, \bibinfo {author} {\bibfnamefont {S.}~\bibnamefont {Jana}},
  \bibinfo {author} {\bibfnamefont {J.}~\bibnamefont {Kubo}}, \ and\ \bibinfo
  {author} {\bibfnamefont {M.}~\bibnamefont {Lindner}},\ }\href@noop {} {\
  (\bibinfo {year} {2021})},\ \Eprint {http://arxiv.org/abs/2104.03282}
  {arXiv:2104.03282 [hep-ph]} \BibitemShut {NoStop}%
\bibitem [{\citenamefont {Cao}\ \emph {et~al.}(2021)\citenamefont {Cao},
  \citenamefont {Lian}, \citenamefont {Pan}, \citenamefont {Zhang},\ and\
  \citenamefont {Zhu}}]{Cao:2021tuh}%
  \BibitemOpen
  \bibfield  {author} {\bibinfo {author} {\bibfnamefont {J.}~\bibnamefont
  {Cao}}, \bibinfo {author} {\bibfnamefont {J.}~\bibnamefont {Lian}}, \bibinfo
  {author} {\bibfnamefont {Y.}~\bibnamefont {Pan}}, \bibinfo {author}
  {\bibfnamefont {D.}~\bibnamefont {Zhang}}, \ and\ \bibinfo {author}
  {\bibfnamefont {P.}~\bibnamefont {Zhu}},\ }\href@noop {} {\  (\bibinfo {year}
  {2021})},\ \Eprint {http://arxiv.org/abs/2104.03284} {arXiv:2104.03284
  [hep-ph]} \BibitemShut {NoStop}%
\bibitem [{\citenamefont {Chakraborti}\ \emph {et~al.}(2021)\citenamefont
  {Chakraborti}, \citenamefont {Heinemeyer},\ and\ \citenamefont
  {Saha}}]{Chakraborti:2021dli}%
  \BibitemOpen
  \bibfield  {author} {\bibinfo {author} {\bibfnamefont {M.}~\bibnamefont
  {Chakraborti}}, \bibinfo {author} {\bibfnamefont {S.}~\bibnamefont
  {Heinemeyer}}, \ and\ \bibinfo {author} {\bibfnamefont {I.}~\bibnamefont
  {Saha}},\ }\href@noop {} {\  (\bibinfo {year} {2021})},\ \Eprint
  {http://arxiv.org/abs/2104.03287} {arXiv:2104.03287 [hep-ph]} \BibitemShut
  {NoStop}%
\bibitem [{\citenamefont {Ibe}\ \emph {et~al.}(2021)\citenamefont {Ibe},
  \citenamefont {Kobayashi}, \citenamefont {Nakayama},\ and\ \citenamefont
  {Shirai}}]{Ibe:2021cvf}%
  \BibitemOpen
  \bibfield  {author} {\bibinfo {author} {\bibfnamefont {M.}~\bibnamefont
  {Ibe}}, \bibinfo {author} {\bibfnamefont {S.}~\bibnamefont {Kobayashi}},
  \bibinfo {author} {\bibfnamefont {Y.}~\bibnamefont {Nakayama}}, \ and\
  \bibinfo {author} {\bibfnamefont {S.}~\bibnamefont {Shirai}},\ }\href@noop {}
  {\  (\bibinfo {year} {2021})},\ \Eprint {http://arxiv.org/abs/2104.03289}
  {arXiv:2104.03289 [hep-ph]} \BibitemShut {NoStop}%
\bibitem [{\citenamefont {Cox}\ \emph {et~al.}(2021)\citenamefont {Cox},
  \citenamefont {Han},\ and\ \citenamefont {Yanagida}}]{Cox:2021gqq}%
  \BibitemOpen
  \bibfield  {author} {\bibinfo {author} {\bibfnamefont {P.}~\bibnamefont
  {Cox}}, \bibinfo {author} {\bibfnamefont {C.}~\bibnamefont {Han}}, \ and\
  \bibinfo {author} {\bibfnamefont {T.~T.}\ \bibnamefont {Yanagida}},\
  }\href@noop {} {\  (\bibinfo {year} {2021})},\ \Eprint
  {http://arxiv.org/abs/2104.03290} {arXiv:2104.03290 [hep-ph]} \BibitemShut
  {NoStop}%
\bibitem [{\citenamefont {Babu}\ \emph {et~al.}(2021)\citenamefont {Babu},
  \citenamefont {Jana}, \citenamefont {Lindner},\ and\ \citenamefont
  {K}}]{Babu:2021jnu}%
  \BibitemOpen
  \bibfield  {author} {\bibinfo {author} {\bibfnamefont {K.~S.}\ \bibnamefont
  {Babu}}, \bibinfo {author} {\bibfnamefont {S.}~\bibnamefont {Jana}}, \bibinfo
  {author} {\bibfnamefont {M.}~\bibnamefont {Lindner}}, \ and\ \bibinfo
  {author} {\bibfnamefont {V.~P.}\ \bibnamefont {K}},\ }\href@noop {} {\
  (\bibinfo {year} {2021})},\ \Eprint {http://arxiv.org/abs/2104.03291}
  {arXiv:2104.03291 [hep-ph]} \BibitemShut {NoStop}%
\bibitem [{\citenamefont {Han}(2021)}]{Han:2021ify}%
  \BibitemOpen
  \bibfield  {author} {\bibinfo {author} {\bibfnamefont {C.}~\bibnamefont
  {Han}},\ }\href@noop {} {\  (\bibinfo {year} {2021})},\ \Eprint
  {http://arxiv.org/abs/2104.03292} {arXiv:2104.03292 [hep-ph]} \BibitemShut
  {NoStop}%
\bibitem [{\citenamefont {Heinemeyer}\ \emph {et~al.}(2021)\citenamefont
  {Heinemeyer}, \citenamefont {Kpatcha}, \citenamefont {Lara}, \citenamefont
  {L\'opez-Fogliani}, \citenamefont {Mu\~noz},\ and\ \citenamefont
  {Nagata}}]{Heinemeyer:2021zpc}%
  \BibitemOpen
  \bibfield  {author} {\bibinfo {author} {\bibfnamefont {S.}~\bibnamefont
  {Heinemeyer}}, \bibinfo {author} {\bibfnamefont {E.}~\bibnamefont {Kpatcha}},
  \bibinfo {author} {\bibfnamefont {I.~n.}\ \bibnamefont {Lara}}, \bibinfo
  {author} {\bibfnamefont {D.~E.}\ \bibnamefont {L\'opez-Fogliani}}, \bibinfo
  {author} {\bibfnamefont {C.}~\bibnamefont {Mu\~noz}}, \ and\ \bibinfo
  {author} {\bibfnamefont {N.}~\bibnamefont {Nagata}},\ }\href@noop {} {\
  (\bibinfo {year} {2021})},\ \Eprint {http://arxiv.org/abs/2104.03294}
  {arXiv:2104.03294 [hep-ph]} \BibitemShut {NoStop}%
\bibitem [{\citenamefont {Calibbi}\ \emph {et~al.}(2021)\citenamefont
  {Calibbi}, \citenamefont {L\'opez-Ib\'a\~nez}, \citenamefont {Melis},\ and\
  \citenamefont {Vives}}]{Calibbi:2021qto}%
  \BibitemOpen
  \bibfield  {author} {\bibinfo {author} {\bibfnamefont {L.}~\bibnamefont
  {Calibbi}}, \bibinfo {author} {\bibfnamefont {M.~L.}\ \bibnamefont
  {L\'opez-Ib\'a\~nez}}, \bibinfo {author} {\bibfnamefont {A.}~\bibnamefont
  {Melis}}, \ and\ \bibinfo {author} {\bibfnamefont {O.}~\bibnamefont
  {Vives}},\ }\href@noop {} {\  (\bibinfo {year} {2021})},\ \Eprint
  {http://arxiv.org/abs/2104.03296} {arXiv:2104.03296 [hep-ph]} \BibitemShut
  {NoStop}%
\bibitem [{\citenamefont {Amaral}\ \emph {et~al.}(2021)\citenamefont {Amaral},
  \citenamefont {Cerde\~no}, \citenamefont {Cheek},\ and\ \citenamefont
  {Foldenauer}}]{Amaral:2021rzw}%
  \BibitemOpen
  \bibfield  {author} {\bibinfo {author} {\bibfnamefont {D.~W.~P.}\
  \bibnamefont {Amaral}}, \bibinfo {author} {\bibfnamefont {D.~G.}\
  \bibnamefont {Cerde\~no}}, \bibinfo {author} {\bibfnamefont {A.}~\bibnamefont
  {Cheek}}, \ and\ \bibinfo {author} {\bibfnamefont {P.}~\bibnamefont
  {Foldenauer}},\ }\href@noop {} {\  (\bibinfo {year} {2021})},\ \Eprint
  {http://arxiv.org/abs/2104.03297} {arXiv:2104.03297 [hep-ph]} \BibitemShut
  {NoStop}%
\bibitem [{\citenamefont {Bai}\ and\ \citenamefont
  {Berger}(2021)}]{Bai:2021bau}%
  \BibitemOpen
  \bibfield  {author} {\bibinfo {author} {\bibfnamefont {Y.}~\bibnamefont
  {Bai}}\ and\ \bibinfo {author} {\bibfnamefont {J.}~\bibnamefont {Berger}},\
  }\href@noop {} {\  (\bibinfo {year} {2021})},\ \Eprint
  {http://arxiv.org/abs/2104.03301} {arXiv:2104.03301 [hep-ph]} \BibitemShut
  {NoStop}%
\bibitem [{\citenamefont {Baum}\ \emph {et~al.}(2021)\citenamefont {Baum},
  \citenamefont {Carena}, \citenamefont {Shah},\ and\ \citenamefont
  {Wagner}}]{Baum:2021qzx}%
  \BibitemOpen
  \bibfield  {author} {\bibinfo {author} {\bibfnamefont {S.}~\bibnamefont
  {Baum}}, \bibinfo {author} {\bibfnamefont {M.}~\bibnamefont {Carena}},
  \bibinfo {author} {\bibfnamefont {N.~R.}\ \bibnamefont {Shah}}, \ and\
  \bibinfo {author} {\bibfnamefont {C.~E.~M.}\ \bibnamefont {Wagner}},\
  }\href@noop {} {\  (\bibinfo {year} {2021})},\ \Eprint
  {http://arxiv.org/abs/2104.03302} {arXiv:2104.03302 [hep-ph]} \BibitemShut
  {NoStop}%
\bibitem [{\citenamefont {Li}\ \emph {et~al.}(2021{\natexlab{b}})\citenamefont
  {Li}, \citenamefont {Pei},\ and\ \citenamefont {Zhang}}]{Li:2021poy}%
  \BibitemOpen
  \bibfield  {author} {\bibinfo {author} {\bibfnamefont {T.}~\bibnamefont
  {Li}}, \bibinfo {author} {\bibfnamefont {J.}~\bibnamefont {Pei}}, \ and\
  \bibinfo {author} {\bibfnamefont {W.}~\bibnamefont {Zhang}},\ }\href@noop {}
  {\  (\bibinfo {year} {2021}{\natexlab{b}})},\ \Eprint
  {http://arxiv.org/abs/2104.03334} {arXiv:2104.03334 [hep-ph]} \BibitemShut
  {NoStop}%
\bibitem [{\citenamefont {Zu}\ \emph {et~al.}(2021)\citenamefont {Zu},
  \citenamefont {Pan}, \citenamefont {Feng}, \citenamefont {Yuan},\ and\
  \citenamefont {Fan}}]{Zu:2021odn}%
  \BibitemOpen
  \bibfield  {author} {\bibinfo {author} {\bibfnamefont {L.}~\bibnamefont
  {Zu}}, \bibinfo {author} {\bibfnamefont {X.}~\bibnamefont {Pan}}, \bibinfo
  {author} {\bibfnamefont {L.}~\bibnamefont {Feng}}, \bibinfo {author}
  {\bibfnamefont {Q.}~\bibnamefont {Yuan}}, \ and\ \bibinfo {author}
  {\bibfnamefont {Y.-Z.}\ \bibnamefont {Fan}},\ }\href@noop {} {\  (\bibinfo
  {year} {2021})},\ \Eprint {http://arxiv.org/abs/2104.03340} {arXiv:2104.03340
  [hep-ph]} \BibitemShut {NoStop}%
\bibitem [{\citenamefont {Keung}\ \emph {et~al.}(2021)\citenamefont {Keung},
  \citenamefont {Marfatia},\ and\ \citenamefont {Tseng}}]{Keung:2021rps}%
  \BibitemOpen
  \bibfield  {author} {\bibinfo {author} {\bibfnamefont {W.-Y.}\ \bibnamefont
  {Keung}}, \bibinfo {author} {\bibfnamefont {D.}~\bibnamefont {Marfatia}}, \
  and\ \bibinfo {author} {\bibfnamefont {P.-Y.}\ \bibnamefont {Tseng}},\
  }\href@noop {} {\  (\bibinfo {year} {2021})},\ \Eprint
  {http://arxiv.org/abs/2104.03341} {arXiv:2104.03341 [hep-ph]} \BibitemShut
  {NoStop}%
\bibitem [{\citenamefont {Ferreira}\ \emph {et~al.}(2021)\citenamefont
  {Ferreira}, \citenamefont {Gon\c{c}alves}, \citenamefont {Joaquim},\ and\
  \citenamefont {Sher}}]{Ferreira:2021gke}%
  \BibitemOpen
  \bibfield  {author} {\bibinfo {author} {\bibfnamefont {P.~M.}\ \bibnamefont
  {Ferreira}}, \bibinfo {author} {\bibfnamefont {B.~L.}\ \bibnamefont
  {Gon\c{c}alves}}, \bibinfo {author} {\bibfnamefont {F.~R.}\ \bibnamefont
  {Joaquim}}, \ and\ \bibinfo {author} {\bibfnamefont {M.}~\bibnamefont
  {Sher}},\ }\href@noop {} {\  (\bibinfo {year} {2021})},\ \Eprint
  {http://arxiv.org/abs/2104.03367} {arXiv:2104.03367 [hep-ph]} \BibitemShut
  {NoStop}%
\bibitem [{\citenamefont {Zhang}\ \emph {et~al.}(2021)\citenamefont {Zhang},
  \citenamefont {Liu}, \citenamefont {Yang},\ and\ \citenamefont
  {Feng}}]{Zhang:2021gun}%
  \BibitemOpen
  \bibfield  {author} {\bibinfo {author} {\bibfnamefont {H.-B.}\ \bibnamefont
  {Zhang}}, \bibinfo {author} {\bibfnamefont {C.-X.}\ \bibnamefont {Liu}},
  \bibinfo {author} {\bibfnamefont {J.-L.}\ \bibnamefont {Yang}}, \ and\
  \bibinfo {author} {\bibfnamefont {T.-F.}\ \bibnamefont {Feng}},\ }\href@noop
  {} {\  (\bibinfo {year} {2021})},\ \Eprint {http://arxiv.org/abs/2104.03489}
  {arXiv:2104.03489 [hep-ph]} \BibitemShut {NoStop}%
\bibitem [{\citenamefont {Ahmed}\ \emph {et~al.}(2021)\citenamefont {Ahmed},
  \citenamefont {Khan}, \citenamefont {Li}, \citenamefont {Li}, \citenamefont
  {Raza},\ and\ \citenamefont {Zhang}}]{Ahmed:2021htr}%
  \BibitemOpen
  \bibfield  {author} {\bibinfo {author} {\bibfnamefont {W.}~\bibnamefont
  {Ahmed}}, \bibinfo {author} {\bibfnamefont {I.}~\bibnamefont {Khan}},
  \bibinfo {author} {\bibfnamefont {J.}~\bibnamefont {Li}}, \bibinfo {author}
  {\bibfnamefont {T.}~\bibnamefont {Li}}, \bibinfo {author} {\bibfnamefont
  {S.}~\bibnamefont {Raza}}, \ and\ \bibinfo {author} {\bibfnamefont
  {W.}~\bibnamefont {Zhang}},\ }\href@noop {} {\  (\bibinfo {year} {2021})},\
  \Eprint {http://arxiv.org/abs/2104.03491} {arXiv:2104.03491 [hep-ph]}
  \BibitemShut {NoStop}%
\bibitem [{\citenamefont {Zhou}\ \emph {et~al.}(2021)\citenamefont {Zhou},
  \citenamefont {Bian},\ and\ \citenamefont {Shu}}]{Zhou:2021cfu}%
  \BibitemOpen
  \bibfield  {author} {\bibinfo {author} {\bibfnamefont {R.}~\bibnamefont
  {Zhou}}, \bibinfo {author} {\bibfnamefont {L.}~\bibnamefont {Bian}}, \ and\
  \bibinfo {author} {\bibfnamefont {J.}~\bibnamefont {Shu}},\ }\href@noop {} {\
   (\bibinfo {year} {2021})},\ \Eprint {http://arxiv.org/abs/2104.03519}
  {arXiv:2104.03519 [hep-ph]} \BibitemShut {NoStop}%
\bibitem [{\citenamefont {Yang}\ \emph {et~al.}(2021)\citenamefont {Yang},
  \citenamefont {Zhang}, \citenamefont {Liu}, \citenamefont {Dong},\ and\
  \citenamefont {Feng}}]{Yang:2021duj}%
  \BibitemOpen
  \bibfield  {author} {\bibinfo {author} {\bibfnamefont {J.-L.}\ \bibnamefont
  {Yang}}, \bibinfo {author} {\bibfnamefont {H.-B.}\ \bibnamefont {Zhang}},
  \bibinfo {author} {\bibfnamefont {C.-X.}\ \bibnamefont {Liu}}, \bibinfo
  {author} {\bibfnamefont {X.-X.}\ \bibnamefont {Dong}}, \ and\ \bibinfo
  {author} {\bibfnamefont {T.-F.}\ \bibnamefont {Feng}},\ }\href@noop {} {\
  (\bibinfo {year} {2021})},\ \Eprint {http://arxiv.org/abs/2104.03542}
  {arXiv:2104.03542 [hep-ph]} \BibitemShut {NoStop}%
\bibitem [{\citenamefont {Athron}\ \emph {et~al.}(2021)\citenamefont {Athron},
  \citenamefont {Bal\'azs}, \citenamefont {Jacob}, \citenamefont {Kotlarski},
  \citenamefont {St\"ockinger},\ and\ \citenamefont
  {St\"ockinger-Kim}}]{Athron:2021iuf}%
  \BibitemOpen
  \bibfield  {author} {\bibinfo {author} {\bibfnamefont {P.}~\bibnamefont
  {Athron}}, \bibinfo {author} {\bibfnamefont {C.}~\bibnamefont {Bal\'azs}},
  \bibinfo {author} {\bibfnamefont {D.~H.}\ \bibnamefont {Jacob}}, \bibinfo
  {author} {\bibfnamefont {W.}~\bibnamefont {Kotlarski}}, \bibinfo {author}
  {\bibfnamefont {D.}~\bibnamefont {St\"ockinger}}, \ and\ \bibinfo {author}
  {\bibfnamefont {H.}~\bibnamefont {St\"ockinger-Kim}},\ }\href@noop {} {\
  (\bibinfo {year} {2021})},\ \Eprint {http://arxiv.org/abs/2104.03691}
  {arXiv:2104.03691 [hep-ph]} \BibitemShut {NoStop}%
\bibitem [{\citenamefont {Chen}\ \emph
  {et~al.}(2021{\natexlab{c}})\citenamefont {Chen}, \citenamefont {Wen},
  \citenamefont {Xu},\ and\ \citenamefont {Zhang}}]{Chen:2021vzk}%
  \BibitemOpen
  \bibfield  {author} {\bibinfo {author} {\bibfnamefont {J.}~\bibnamefont
  {Chen}}, \bibinfo {author} {\bibfnamefont {Q.}~\bibnamefont {Wen}}, \bibinfo
  {author} {\bibfnamefont {F.}~\bibnamefont {Xu}}, \ and\ \bibinfo {author}
  {\bibfnamefont {M.}~\bibnamefont {Zhang}},\ }\href@noop {} {\  (\bibinfo
  {year} {2021}{\natexlab{c}})},\ \Eprint {http://arxiv.org/abs/2104.03699}
  {arXiv:2104.03699 [hep-ph]} \BibitemShut {NoStop}%
\bibitem [{\citenamefont {Chun}\ and\ \citenamefont
  {Mondal}(2021)}]{Chun:2021dwx}%
  \BibitemOpen
  \bibfield  {author} {\bibinfo {author} {\bibfnamefont {E.~J.}\ \bibnamefont
  {Chun}}\ and\ \bibinfo {author} {\bibfnamefont {T.}~\bibnamefont {Mondal}},\
  }\href@noop {} {\  (\bibinfo {year} {2021})},\ \Eprint
  {http://arxiv.org/abs/2104.03701} {arXiv:2104.03701 [hep-ph]} \BibitemShut
  {NoStop}%
\bibitem [{\citenamefont {Escribano}\ \emph {et~al.}(2021)\citenamefont
  {Escribano}, \citenamefont {Terol-Calvo},\ and\ \citenamefont
  {Vicente}}]{Escribano:2021css}%
  \BibitemOpen
  \bibfield  {author} {\bibinfo {author} {\bibfnamefont {P.}~\bibnamefont
  {Escribano}}, \bibinfo {author} {\bibfnamefont {J.}~\bibnamefont
  {Terol-Calvo}}, \ and\ \bibinfo {author} {\bibfnamefont {A.}~\bibnamefont
  {Vicente}},\ }\href@noop {} {\  (\bibinfo {year} {2021})},\ \Eprint
  {http://arxiv.org/abs/2104.03705} {arXiv:2104.03705 [hep-ph]} \BibitemShut
  {NoStop}%
\bibitem [{\citenamefont {Aboubrahim}\ \emph {et~al.}(2021)\citenamefont
  {Aboubrahim}, \citenamefont {Klasen},\ and\ \citenamefont
  {Nath}}]{Aboubrahim:2021rwz}%
  \BibitemOpen
  \bibfield  {author} {\bibinfo {author} {\bibfnamefont {A.}~\bibnamefont
  {Aboubrahim}}, \bibinfo {author} {\bibfnamefont {M.}~\bibnamefont {Klasen}},
  \ and\ \bibinfo {author} {\bibfnamefont {P.}~\bibnamefont {Nath}},\
  }\href@noop {} {\  (\bibinfo {year} {2021})},\ \Eprint
  {http://arxiv.org/abs/2104.03839} {arXiv:2104.03839 [hep-ph]} \BibitemShut
  {NoStop}%
\bibitem [{\citenamefont {Bhattacharya}\ \emph {et~al.}(2021)\citenamefont
  {Bhattacharya}, \citenamefont {Datta}, \citenamefont {Marfatia},
  \citenamefont {Nandi},\ and\ \citenamefont {Waite}}]{Bhattacharya:2021ggm}%
  \BibitemOpen
  \bibfield  {author} {\bibinfo {author} {\bibfnamefont {B.}~\bibnamefont
  {Bhattacharya}}, \bibinfo {author} {\bibfnamefont {A.}~\bibnamefont {Datta}},
  \bibinfo {author} {\bibfnamefont {D.}~\bibnamefont {Marfatia}}, \bibinfo
  {author} {\bibfnamefont {S.}~\bibnamefont {Nandi}}, \ and\ \bibinfo {author}
  {\bibfnamefont {J.}~\bibnamefont {Waite}},\ }\href@noop {} {\  (\bibinfo
  {year} {2021})},\ \Eprint {http://arxiv.org/abs/2104.03947} {arXiv:2104.03947
  [hep-ph]} \BibitemShut {NoStop}%
\bibitem [{\citenamefont {Willmann}\ \emph {et~al.}(1999)\citenamefont
  {Willmann} \emph {et~al.}}]{Willmann:1998gd}%
  \BibitemOpen
  \bibfield  {author} {\bibinfo {author} {\bibfnamefont {L.}~\bibnamefont
  {Willmann}} \emph {et~al.},\ }\href {\doibase 10.1103/PhysRevLett.82.49}
  {\bibfield  {journal} {\bibinfo  {journal} {Phys. Rev. Lett.}\ }\textbf
  {\bibinfo {volume} {82}},\ \bibinfo {pages} {49} (\bibinfo {year} {1999})},\
  \Eprint {http://arxiv.org/abs/hep-ex/9807011} {arXiv:hep-ex/9807011}
  \BibitemShut {NoStop}%
\bibitem [{\citenamefont {Tang}(2020)}]{MACE:2020}%
  \BibitemOpen
  \bibfield  {author} {\bibinfo {author} {\bibfnamefont {J.}~\bibnamefont
  {Tang}},\ }\href@noop {} {\enquote {\bibinfo {title} {{Muonium-Antimuonium
  Conversion Experiment (MACE)}},}\ } (\bibinfo {year} {2020}),\ \bibinfo
  {note}
  {\url{https://www.snowmass21.org/docs/files/summaries/RF/SNOWMASS21-RF5_RF0_Jian_Tang-126.pdf}}\BibitemShut
  {NoStop}%
\bibitem [{\citenamefont {Abdallah}\ \emph {et~al.}(2006)\citenamefont
  {Abdallah} \emph {et~al.}}]{Abdallah:2005ph}%
  \BibitemOpen
  \bibfield  {author} {\bibinfo {author} {\bibfnamefont {J.}~\bibnamefont
  {Abdallah}} \emph {et~al.} (\bibinfo {collaboration} {DELPHI}),\ }\href
  {\doibase 10.1140/epjc/s2005-02461-0} {\bibfield  {journal} {\bibinfo
  {journal} {Eur. Phys. J. C}\ }\textbf {\bibinfo {volume} {45}},\ \bibinfo
  {pages} {589} (\bibinfo {year} {2006})},\ \Eprint
  {http://arxiv.org/abs/hep-ex/0512012} {arXiv:hep-ex/0512012} \BibitemShut
  {NoStop}%
\bibitem [{\citenamefont {Dev}\ \emph {et~al.}(2018)\citenamefont {Dev},
  \citenamefont {Mohapatra},\ and\ \citenamefont {Zhang}}]{Dev:2017ftk}%
  \BibitemOpen
  \bibfield  {author} {\bibinfo {author} {\bibfnamefont {P.~S.~B.}\
  \bibnamefont {Dev}}, \bibinfo {author} {\bibfnamefont {R.~N.}\ \bibnamefont
  {Mohapatra}}, \ and\ \bibinfo {author} {\bibfnamefont {Y.}~\bibnamefont
  {Zhang}},\ }\href {\doibase 10.1103/PhysRevLett.120.221804} {\bibfield
  {journal} {\bibinfo  {journal} {Phys. Rev. Lett.}\ }\textbf {\bibinfo
  {volume} {120}},\ \bibinfo {pages} {221804} (\bibinfo {year} {2018})},\
  \Eprint {http://arxiv.org/abs/1711.08430} {arXiv:1711.08430 [hep-ph]}
  \BibitemShut {NoStop}%
\bibitem [{\citenamefont {Li}\ and\ \citenamefont
  {Schmidt}(2019{\natexlab{a}})}]{Li:2018cod}%
  \BibitemOpen
  \bibfield  {author} {\bibinfo {author} {\bibfnamefont {T.}~\bibnamefont
  {Li}}\ and\ \bibinfo {author} {\bibfnamefont {M.~A.}\ \bibnamefont
  {Schmidt}},\ }\href {\doibase 10.1103/PhysRevD.99.055038} {\bibfield
  {journal} {\bibinfo  {journal} {Phys. Rev. D}\ }\textbf {\bibinfo {volume}
  {99}},\ \bibinfo {pages} {055038} (\bibinfo {year} {2019}{\natexlab{a}})},\
  \Eprint {http://arxiv.org/abs/1809.07924} {arXiv:1809.07924 [hep-ph]}
  \BibitemShut {NoStop}%
\bibitem [{\citenamefont {Li}\ and\ \citenamefont
  {Schmidt}(2019{\natexlab{b}})}]{Li:2019xvv}%
  \BibitemOpen
  \bibfield  {author} {\bibinfo {author} {\bibfnamefont {T.}~\bibnamefont
  {Li}}\ and\ \bibinfo {author} {\bibfnamefont {M.~A.}\ \bibnamefont
  {Schmidt}},\ }\href {\doibase 10.1103/PhysRevD.100.115007} {\bibfield
  {journal} {\bibinfo  {journal} {Phys. Rev. D}\ }\textbf {\bibinfo {volume}
  {100}},\ \bibinfo {pages} {115007} (\bibinfo {year} {2019}{\natexlab{b}})},\
  \Eprint {http://arxiv.org/abs/1907.06963} {arXiv:1907.06963 [hep-ph]}
  \BibitemShut {NoStop}%
\bibitem [{\citenamefont {Lindner}\ \emph {et~al.}(2018)\citenamefont
  {Lindner}, \citenamefont {Platscher},\ and\ \citenamefont
  {Queiroz}}]{Lindner:2016bgg}%
  \BibitemOpen
  \bibfield  {author} {\bibinfo {author} {\bibfnamefont {M.}~\bibnamefont
  {Lindner}}, \bibinfo {author} {\bibfnamefont {M.}~\bibnamefont {Platscher}},
  \ and\ \bibinfo {author} {\bibfnamefont {F.~S.}\ \bibnamefont {Queiroz}},\
  }\href {\doibase 10.1016/j.physrep.2017.12.001} {\bibfield  {journal}
  {\bibinfo  {journal} {Phys. Rept.}\ }\textbf {\bibinfo {volume} {731}},\
  \bibinfo {pages} {1} (\bibinfo {year} {2018})},\ \Eprint
  {http://arxiv.org/abs/1610.06587} {arXiv:1610.06587 [hep-ph]} \BibitemShut
  {NoStop}%
\bibitem [{\citenamefont {Calibbi}\ and\ \citenamefont
  {Signorelli}(2018)}]{Calibbi:2017uvl}%
  \BibitemOpen
  \bibfield  {author} {\bibinfo {author} {\bibfnamefont {L.}~\bibnamefont
  {Calibbi}}\ and\ \bibinfo {author} {\bibfnamefont {G.}~\bibnamefont
  {Signorelli}},\ }\href {\doibase 10.1393/ncr/i2018-10144-0} {\bibfield
  {journal} {\bibinfo  {journal} {Riv. Nuovo Cim.}\ }\textbf {\bibinfo {volume}
  {41}},\ \bibinfo {pages} {71} (\bibinfo {year} {2018})},\ \Eprint
  {http://arxiv.org/abs/1709.00294} {arXiv:1709.00294 [hep-ph]} \BibitemShut
  {NoStop}%
\bibitem [{\citenamefont {Tommasini}\ \emph {et~al.}(1995)\citenamefont
  {Tommasini}, \citenamefont {Barenboim}, \citenamefont {Bernabeu},\ and\
  \citenamefont {Jarlskog}}]{Tommasini:1995ii}%
  \BibitemOpen
  \bibfield  {author} {\bibinfo {author} {\bibfnamefont {D.}~\bibnamefont
  {Tommasini}}, \bibinfo {author} {\bibfnamefont {G.}~\bibnamefont
  {Barenboim}}, \bibinfo {author} {\bibfnamefont {J.}~\bibnamefont {Bernabeu}},
  \ and\ \bibinfo {author} {\bibfnamefont {C.}~\bibnamefont {Jarlskog}},\
  }\href {\doibase 10.1016/0550-3213(95)00201-3} {\bibfield  {journal}
  {\bibinfo  {journal} {Nucl. Phys. B}\ }\textbf {\bibinfo {volume} {444}},\
  \bibinfo {pages} {451} (\bibinfo {year} {1995})},\ \Eprint
  {http://arxiv.org/abs/hep-ph/9503228} {arXiv:hep-ph/9503228} \BibitemShut
  {NoStop}%
\bibitem [{\citenamefont {Cai}\ \emph {et~al.}(2017)\citenamefont {Cai},
  \citenamefont {Herrero-Garc\'\i{}a}, \citenamefont {Schmidt}, \citenamefont
  {Vicente},\ and\ \citenamefont {Volkas}}]{Cai:2017jrq}%
  \BibitemOpen
  \bibfield  {author} {\bibinfo {author} {\bibfnamefont {Y.}~\bibnamefont
  {Cai}}, \bibinfo {author} {\bibfnamefont {J.}~\bibnamefont
  {Herrero-Garc\'\i{}a}}, \bibinfo {author} {\bibfnamefont {M.~A.}\
  \bibnamefont {Schmidt}}, \bibinfo {author} {\bibfnamefont {A.}~\bibnamefont
  {Vicente}}, \ and\ \bibinfo {author} {\bibfnamefont {R.~R.}\ \bibnamefont
  {Volkas}},\ }\href {\doibase 10.3389/fphy.2017.00063} {\bibfield  {journal}
  {\bibinfo  {journal} {Front. in Phys.}\ }\textbf {\bibinfo {volume} {5}},\
  \bibinfo {pages} {63} (\bibinfo {year} {2017})},\ \Eprint
  {http://arxiv.org/abs/1706.08524} {arXiv:1706.08524 [hep-ph]} \BibitemShut
  {NoStop}%
\bibitem [{\citenamefont {Branco}\ \emph {et~al.}(2012)\citenamefont {Branco},
  \citenamefont {Ferreira}, \citenamefont {Lavoura}, \citenamefont {Rebelo},
  \citenamefont {Sher},\ and\ \citenamefont {Silva}}]{Branco:2011iw}%
  \BibitemOpen
  \bibfield  {author} {\bibinfo {author} {\bibfnamefont {G.~C.}\ \bibnamefont
  {Branco}}, \bibinfo {author} {\bibfnamefont {P.~M.}\ \bibnamefont
  {Ferreira}}, \bibinfo {author} {\bibfnamefont {L.}~\bibnamefont {Lavoura}},
  \bibinfo {author} {\bibfnamefont {M.~N.}\ \bibnamefont {Rebelo}}, \bibinfo
  {author} {\bibfnamefont {M.}~\bibnamefont {Sher}}, \ and\ \bibinfo {author}
  {\bibfnamefont {J.~P.}\ \bibnamefont {Silva}},\ }\href {\doibase
  10.1016/j.physrep.2012.02.002} {\bibfield  {journal} {\bibinfo  {journal}
  {Phys. Rept.}\ }\textbf {\bibinfo {volume} {516}},\ \bibinfo {pages} {1}
  (\bibinfo {year} {2012})},\ \Eprint {http://arxiv.org/abs/1106.0034}
  {arXiv:1106.0034 [hep-ph]} \BibitemShut {NoStop}%
\bibitem [{\citenamefont {Iguro}\ \emph {et~al.}(2019)\citenamefont {Iguro},
  \citenamefont {Omura},\ and\ \citenamefont {Takeuchi}}]{Iguro:2019sly}%
  \BibitemOpen
  \bibfield  {author} {\bibinfo {author} {\bibfnamefont {S.}~\bibnamefont
  {Iguro}}, \bibinfo {author} {\bibfnamefont {Y.}~\bibnamefont {Omura}}, \ and\
  \bibinfo {author} {\bibfnamefont {M.}~\bibnamefont {Takeuchi}},\ }\href
  {\doibase 10.1007/JHEP11(2019)130} {\bibfield  {journal} {\bibinfo  {journal}
  {JHEP}\ }\textbf {\bibinfo {volume} {11}},\ \bibinfo {pages} {130} (\bibinfo
  {year} {2019})},\ \Eprint {http://arxiv.org/abs/1907.09845} {arXiv:1907.09845
  [hep-ph]} \BibitemShut {NoStop}%
\bibitem [{\citenamefont {Raidal}\ \emph {et~al.}(2008)\citenamefont {Raidal}
  \emph {et~al.}}]{Raidal:2008jk}%
  \BibitemOpen
  \bibfield  {author} {\bibinfo {author} {\bibfnamefont {M.}~\bibnamefont
  {Raidal}} \emph {et~al.},\ }\href {\doibase 10.1140/epjc/s10052-008-0715-2}
  {\bibfield  {journal} {\bibinfo  {journal} {Eur. Phys. J. C}\ }\textbf
  {\bibinfo {volume} {57}},\ \bibinfo {pages} {13} (\bibinfo {year} {2008})},\
  \Eprint {http://arxiv.org/abs/0801.1826} {arXiv:0801.1826 [hep-ph]}
  \BibitemShut {NoStop}%
\bibitem [{\citenamefont {Han}\ \emph {et~al.}(2020)\citenamefont {Han},
  \citenamefont {L\'opez-Ib\'a\~nez}, \citenamefont {Melis}, \citenamefont
  {Vives}, \citenamefont {Wu},\ and\ \citenamefont {Yang}}]{Han:2020exx}%
  \BibitemOpen
  \bibfield  {author} {\bibinfo {author} {\bibfnamefont {C.}~\bibnamefont
  {Han}}, \bibinfo {author} {\bibfnamefont {M.~L.}\ \bibnamefont
  {L\'opez-Ib\'a\~nez}}, \bibinfo {author} {\bibfnamefont {A.}~\bibnamefont
  {Melis}}, \bibinfo {author} {\bibfnamefont {O.}~\bibnamefont {Vives}},
  \bibinfo {author} {\bibfnamefont {L.}~\bibnamefont {Wu}}, \ and\ \bibinfo
  {author} {\bibfnamefont {J.~M.}\ \bibnamefont {Yang}},\ }\href {\doibase
  10.1007/JHEP05(2020)102} {\bibfield  {journal} {\bibinfo  {journal} {JHEP}\
  }\textbf {\bibinfo {volume} {05}},\ \bibinfo {pages} {102} (\bibinfo {year}
  {2020})},\ \Eprint {http://arxiv.org/abs/2003.06187} {arXiv:2003.06187
  [hep-ph]} \BibitemShut {NoStop}%
\bibitem [{\citenamefont {Cuypers}\ and\ \citenamefont
  {Davidson}(1998)}]{Cuypers:1996ia}%
  \BibitemOpen
  \bibfield  {author} {\bibinfo {author} {\bibfnamefont {F.}~\bibnamefont
  {Cuypers}}\ and\ \bibinfo {author} {\bibfnamefont {S.}~\bibnamefont
  {Davidson}},\ }\href {\doibase 10.1007/s100520050157} {\bibfield  {journal}
  {\bibinfo  {journal} {Eur. Phys. J. C}\ }\textbf {\bibinfo {volume} {2}},\
  \bibinfo {pages} {503} (\bibinfo {year} {1998})},\ \Eprint
  {http://arxiv.org/abs/hep-ph/9609487} {arXiv:hep-ph/9609487} \BibitemShut
  {NoStop}%
\bibitem [{\citenamefont {Lavoura}(2003)}]{Lavoura:2003xp}%
  \BibitemOpen
  \bibfield  {author} {\bibinfo {author} {\bibfnamefont {L.}~\bibnamefont
  {Lavoura}},\ }\href {\doibase 10.1140/epjc/s2003-01212-7} {\bibfield
  {journal} {\bibinfo  {journal} {Eur. Phys. J. C}\ }\textbf {\bibinfo {volume}
  {29}},\ \bibinfo {pages} {191} (\bibinfo {year} {2003})},\ \Eprint
  {http://arxiv.org/abs/hep-ph/0302221} {arXiv:hep-ph/0302221} \BibitemShut
  {NoStop}%
\bibitem [{\citenamefont {Hanneke}\ \emph {et~al.}(2008)\citenamefont
  {Hanneke}, \citenamefont {Fogwell},\ and\ \citenamefont
  {Gabrielse}}]{Hanneke:2008tm}%
  \BibitemOpen
  \bibfield  {author} {\bibinfo {author} {\bibfnamefont {D.}~\bibnamefont
  {Hanneke}}, \bibinfo {author} {\bibfnamefont {S.}~\bibnamefont {Fogwell}}, \
  and\ \bibinfo {author} {\bibfnamefont {G.}~\bibnamefont {Gabrielse}},\ }\href
  {\doibase 10.1103/PhysRevLett.100.120801} {\bibfield  {journal} {\bibinfo
  {journal} {Phys. Rev. Lett.}\ }\textbf {\bibinfo {volume} {100}},\ \bibinfo
  {pages} {120801} (\bibinfo {year} {2008})},\ \Eprint
  {http://arxiv.org/abs/0801.1134} {arXiv:0801.1134 [physics.atom-ph]}
  \BibitemShut {NoStop}%
\bibitem [{\citenamefont {Hanneke}\ \emph {et~al.}(2011)\citenamefont
  {Hanneke}, \citenamefont {Hoogerheide},\ and\ \citenamefont
  {Gabrielse}}]{Hanneke:2010au}%
  \BibitemOpen
  \bibfield  {author} {\bibinfo {author} {\bibfnamefont {D.}~\bibnamefont
  {Hanneke}}, \bibinfo {author} {\bibfnamefont {S.~F.}\ \bibnamefont
  {Hoogerheide}}, \ and\ \bibinfo {author} {\bibfnamefont {G.}~\bibnamefont
  {Gabrielse}},\ }\href {\doibase 10.1103/PhysRevA.83.052122} {\bibfield
  {journal} {\bibinfo  {journal} {Phys. Rev. A}\ }\textbf {\bibinfo {volume}
  {83}},\ \bibinfo {pages} {052122} (\bibinfo {year} {2011})},\ \Eprint
  {http://arxiv.org/abs/1009.4831} {arXiv:1009.4831 [physics.atom-ph]}
  \BibitemShut {NoStop}%
\bibitem [{\citenamefont {Parker}\ \emph {et~al.}(2018)\citenamefont {Parker},
  \citenamefont {Yu}, \citenamefont {Zhong}, \citenamefont {Estey},\ and\
  \citenamefont {M\"uller}}]{Parker:2018vye}%
  \BibitemOpen
  \bibfield  {author} {\bibinfo {author} {\bibfnamefont {R.~H.}\ \bibnamefont
  {Parker}}, \bibinfo {author} {\bibfnamefont {C.}~\bibnamefont {Yu}}, \bibinfo
  {author} {\bibfnamefont {W.}~\bibnamefont {Zhong}}, \bibinfo {author}
  {\bibfnamefont {B.}~\bibnamefont {Estey}}, \ and\ \bibinfo {author}
  {\bibfnamefont {H.}~\bibnamefont {M\"uller}},\ }\href {\doibase
  10.1126/science.aap7706} {\bibfield  {journal} {\bibinfo  {journal}
  {Science}\ }\textbf {\bibinfo {volume} {360}},\ \bibinfo {pages} {191}
  (\bibinfo {year} {2018})},\ \Eprint {http://arxiv.org/abs/1812.04130}
  {arXiv:1812.04130 [physics.atom-ph]} \BibitemShut {NoStop}%
\bibitem [{\citenamefont {Morel}\ \emph {et~al.}(2020)\citenamefont {Morel},
  \citenamefont {Yao}, \citenamefont {Clad\'e},\ and\ \citenamefont
  {Guellati-Kh\'elifa}}]{Morel:2020dww}%
  \BibitemOpen
  \bibfield  {author} {\bibinfo {author} {\bibfnamefont {L.}~\bibnamefont
  {Morel}}, \bibinfo {author} {\bibfnamefont {Z.}~\bibnamefont {Yao}}, \bibinfo
  {author} {\bibfnamefont {P.}~\bibnamefont {Clad\'e}}, \ and\ \bibinfo
  {author} {\bibfnamefont {S.}~\bibnamefont {Guellati-Kh\'elifa}},\ }\href
  {\doibase 10.1038/s41586-020-2964-7} {\bibfield  {journal} {\bibinfo
  {journal} {Nature}\ }\textbf {\bibinfo {volume} {588}},\ \bibinfo {pages}
  {61} (\bibinfo {year} {2020})}\BibitemShut {NoStop}%
\bibitem [{\citenamefont {Han}\ \emph {et~al.}(2021{\natexlab{b}})\citenamefont
  {Han}, \citenamefont {Huang}, \citenamefont {Tang},\ and\ \citenamefont
  {Zhang}}]{Han:2021nod}%
  \BibitemOpen
  \bibfield  {author} {\bibinfo {author} {\bibfnamefont {C.}~\bibnamefont
  {Han}}, \bibinfo {author} {\bibfnamefont {D.}~\bibnamefont {Huang}}, \bibinfo
  {author} {\bibfnamefont {J.}~\bibnamefont {Tang}}, \ and\ \bibinfo {author}
  {\bibfnamefont {Y.}~\bibnamefont {Zhang}},\ }\href {\doibase
  10.1103/PhysRevD.103.055023} {\bibfield  {journal} {\bibinfo  {journal}
  {Phys. Rev. D}\ }\textbf {\bibinfo {volume} {103}},\ \bibinfo {pages}
  {055023} (\bibinfo {year} {2021}{\natexlab{b}})},\ \Eprint
  {http://arxiv.org/abs/2102.00758} {arXiv:2102.00758 [hep-ph]} \BibitemShut
  {NoStop}%
\bibitem [{\citenamefont {Belfatto}\ \emph {et~al.}(2020)\citenamefont
  {Belfatto}, \citenamefont {Beradze},\ and\ \citenamefont
  {Berezhiani}}]{Belfatto:2019swo}%
  \BibitemOpen
  \bibfield  {author} {\bibinfo {author} {\bibfnamefont {B.}~\bibnamefont
  {Belfatto}}, \bibinfo {author} {\bibfnamefont {R.}~\bibnamefont {Beradze}}, \
  and\ \bibinfo {author} {\bibfnamefont {Z.}~\bibnamefont {Berezhiani}},\
  }\href {\doibase 10.1140/epjc/s10052-020-7691-6} {\bibfield  {journal}
  {\bibinfo  {journal} {Eur. Phys. J. C}\ }\textbf {\bibinfo {volume} {80}},\
  \bibinfo {pages} {149} (\bibinfo {year} {2020})},\ \Eprint
  {http://arxiv.org/abs/1906.02714} {arXiv:1906.02714 [hep-ph]} \BibitemShut
  {NoStop}%
\bibitem [{\citenamefont {Grossman}\ \emph {et~al.}(2020)\citenamefont
  {Grossman}, \citenamefont {Passemar},\ and\ \citenamefont
  {Schacht}}]{Grossman:2019bzp}%
  \BibitemOpen
  \bibfield  {author} {\bibinfo {author} {\bibfnamefont {Y.}~\bibnamefont
  {Grossman}}, \bibinfo {author} {\bibfnamefont {E.}~\bibnamefont {Passemar}},
  \ and\ \bibinfo {author} {\bibfnamefont {S.}~\bibnamefont {Schacht}},\ }\href
  {\doibase 10.1007/JHEP07(2020)068} {\bibfield  {journal} {\bibinfo  {journal}
  {JHEP}\ }\textbf {\bibinfo {volume} {07}},\ \bibinfo {pages} {068} (\bibinfo
  {year} {2020})},\ \Eprint {http://arxiv.org/abs/1911.07821} {arXiv:1911.07821
  [hep-ph]} \BibitemShut {NoStop}%
\bibitem [{\citenamefont {Coutinho}\ \emph {et~al.}(2020)\citenamefont
  {Coutinho}, \citenamefont {Crivellin},\ and\ \citenamefont
  {Manzari}}]{Coutinho:2019aiy}%
  \BibitemOpen
  \bibfield  {author} {\bibinfo {author} {\bibfnamefont {A.~M.}\ \bibnamefont
  {Coutinho}}, \bibinfo {author} {\bibfnamefont {A.}~\bibnamefont {Crivellin}},
  \ and\ \bibinfo {author} {\bibfnamefont {C.~A.}\ \bibnamefont {Manzari}},\
  }\href {\doibase 10.1103/PhysRevLett.125.071802} {\bibfield  {journal}
  {\bibinfo  {journal} {Phys. Rev. Lett.}\ }\textbf {\bibinfo {volume} {125}},\
  \bibinfo {pages} {071802} (\bibinfo {year} {2020})},\ \Eprint
  {http://arxiv.org/abs/1912.08823} {arXiv:1912.08823 [hep-ph]} \BibitemShut
  {NoStop}%
\bibitem [{\citenamefont {Feinberg}\ and\ \citenamefont
  {Weinberg}(1961{\natexlab{a}})}]{Feinberg:1961zz}%
  \BibitemOpen
  \bibfield  {author} {\bibinfo {author} {\bibfnamefont {G.}~\bibnamefont
  {Feinberg}}\ and\ \bibinfo {author} {\bibfnamefont {S.}~\bibnamefont
  {Weinberg}},\ }\href {\doibase 10.1103/PhysRevLett.6.381} {\bibfield
  {journal} {\bibinfo  {journal} {Phys. Rev. Lett.}\ }\textbf {\bibinfo
  {volume} {6}},\ \bibinfo {pages} {381} (\bibinfo {year}
  {1961}{\natexlab{a}})}\BibitemShut {NoStop}%
\bibitem [{\citenamefont {Feinberg}\ and\ \citenamefont
  {Weinberg}(1961{\natexlab{b}})}]{Feinberg:1961zza}%
  \BibitemOpen
  \bibfield  {author} {\bibinfo {author} {\bibfnamefont {G.}~\bibnamefont
  {Feinberg}}\ and\ \bibinfo {author} {\bibfnamefont {S.}~\bibnamefont
  {Weinberg}},\ }\href {\doibase 10.1103/PhysRev.123.1439} {\bibfield
  {journal} {\bibinfo  {journal} {Phys. Rev.}\ }\textbf {\bibinfo {volume}
  {123}},\ \bibinfo {pages} {1439} (\bibinfo {year}
  {1961}{\natexlab{b}})}\BibitemShut {NoStop}%
\bibitem [{\citenamefont {Conlin}\ and\ \citenamefont
  {Petrov}(2020)}]{Conlin:2020veq}%
  \BibitemOpen
  \bibfield  {author} {\bibinfo {author} {\bibfnamefont {R.}~\bibnamefont
  {Conlin}}\ and\ \bibinfo {author} {\bibfnamefont {A.~A.}\ \bibnamefont
  {Petrov}},\ }\href {\doibase 10.1103/PhysRevD.102.095001} {\bibfield
  {journal} {\bibinfo  {journal} {Phys. Rev. D}\ }\textbf {\bibinfo {volume}
  {102}},\ \bibinfo {pages} {095001} (\bibinfo {year} {2020})},\ \Eprint
  {http://arxiv.org/abs/2005.10276} {arXiv:2005.10276 [hep-ph]} \BibitemShut
  {NoStop}%
\bibitem [{\citenamefont {Mariam}\ \emph {et~al.}(1982)\citenamefont {Mariam}
  \emph {et~al.}}]{Mariam:1982bq}%
  \BibitemOpen
  \bibfield  {author} {\bibinfo {author} {\bibfnamefont {F.~G.}\ \bibnamefont
  {Mariam}} \emph {et~al.},\ }\href {\doibase 10.1103/PhysRevLett.49.993}
  {\bibfield  {journal} {\bibinfo  {journal} {Phys. Rev. Lett.}\ }\textbf
  {\bibinfo {volume} {49}},\ \bibinfo {pages} {993} (\bibinfo {year}
  {1982})}\BibitemShut {NoStop}%
\bibitem [{\citenamefont {Klempt}\ \emph {et~al.}(1982)\citenamefont {Klempt},
  \citenamefont {Schulze}, \citenamefont {Wolf}, \citenamefont {Camani},
  \citenamefont {Gygax}, \citenamefont {Ruegg}, \citenamefont {Schenck},\ and\
  \citenamefont {Schilling}}]{Klempt:1982ge}%
  \BibitemOpen
  \bibfield  {author} {\bibinfo {author} {\bibfnamefont {E.}~\bibnamefont
  {Klempt}}, \bibinfo {author} {\bibfnamefont {R.}~\bibnamefont {Schulze}},
  \bibinfo {author} {\bibfnamefont {H.}~\bibnamefont {Wolf}}, \bibinfo {author}
  {\bibfnamefont {M.}~\bibnamefont {Camani}}, \bibinfo {author} {\bibfnamefont
  {F.~N.}\ \bibnamefont {Gygax}}, \bibinfo {author} {\bibfnamefont
  {W.}~\bibnamefont {Ruegg}}, \bibinfo {author} {\bibfnamefont
  {A.}~\bibnamefont {Schenck}}, \ and\ \bibinfo {author} {\bibfnamefont
  {H.}~\bibnamefont {Schilling}},\ }\href {\doibase 10.1103/PhysRevD.25.652}
  {\bibfield  {journal} {\bibinfo  {journal} {Phys. Rev. D}\ }\textbf {\bibinfo
  {volume} {25}},\ \bibinfo {pages} {652} (\bibinfo {year} {1982})}\BibitemShut
  {NoStop}%
\bibitem [{\citenamefont {Alloul}\ \emph {et~al.}(2014)\citenamefont {Alloul},
  \citenamefont {Christensen}, \citenamefont {Degrande}, \citenamefont {Duhr},\
  and\ \citenamefont {Fuks}}]{Alloul:2013bka}%
  \BibitemOpen
  \bibfield  {author} {\bibinfo {author} {\bibfnamefont {A.}~\bibnamefont
  {Alloul}}, \bibinfo {author} {\bibfnamefont {N.~D.}\ \bibnamefont
  {Christensen}}, \bibinfo {author} {\bibfnamefont {C.}~\bibnamefont
  {Degrande}}, \bibinfo {author} {\bibfnamefont {C.}~\bibnamefont {Duhr}}, \
  and\ \bibinfo {author} {\bibfnamefont {B.}~\bibnamefont {Fuks}},\ }\href
  {\doibase 10.1016/j.cpc.2014.04.012} {\bibfield  {journal} {\bibinfo
  {journal} {Comput. Phys. Commun.}\ }\textbf {\bibinfo {volume} {185}},\
  \bibinfo {pages} {2250} (\bibinfo {year} {2014})},\ \Eprint
  {http://arxiv.org/abs/1310.1921} {arXiv:1310.1921 [hep-ph]} \BibitemShut
  {NoStop}%
\bibitem [{\citenamefont {Alwall}\ \emph {et~al.}(2014)\citenamefont {Alwall},
  \citenamefont {Frederix}, \citenamefont {Frixione}, \citenamefont {Hirschi},
  \citenamefont {Maltoni}, \citenamefont {Mattelaer}, \citenamefont {Shao},
  \citenamefont {Stelzer}, \citenamefont {Torrielli},\ and\ \citenamefont
  {Zaro}}]{Alwall:2014hca}%
  \BibitemOpen
  \bibfield  {author} {\bibinfo {author} {\bibfnamefont {J.}~\bibnamefont
  {Alwall}}, \bibinfo {author} {\bibfnamefont {R.}~\bibnamefont {Frederix}},
  \bibinfo {author} {\bibfnamefont {S.}~\bibnamefont {Frixione}}, \bibinfo
  {author} {\bibfnamefont {V.}~\bibnamefont {Hirschi}}, \bibinfo {author}
  {\bibfnamefont {F.}~\bibnamefont {Maltoni}}, \bibinfo {author} {\bibfnamefont
  {O.}~\bibnamefont {Mattelaer}}, \bibinfo {author} {\bibfnamefont {H.~S.}\
  \bibnamefont {Shao}}, \bibinfo {author} {\bibfnamefont {T.}~\bibnamefont
  {Stelzer}}, \bibinfo {author} {\bibfnamefont {P.}~\bibnamefont {Torrielli}},
  \ and\ \bibinfo {author} {\bibfnamefont {M.}~\bibnamefont {Zaro}},\ }\href
  {\doibase 10.1007/JHEP07(2014)079} {\bibfield  {journal} {\bibinfo  {journal}
  {JHEP}\ }\textbf {\bibinfo {volume} {07}},\ \bibinfo {pages} {079} (\bibinfo
  {year} {2014})},\ \Eprint {http://arxiv.org/abs/1405.0301} {arXiv:1405.0301
  [hep-ph]} \BibitemShut {NoStop}%
\bibitem [{\citenamefont {Sj\"ostrand}\ \emph {et~al.}(2015)\citenamefont
  {Sj\"ostrand}, \citenamefont {Ask}, \citenamefont {Christiansen},
  \citenamefont {Corke}, \citenamefont {Desai}, \citenamefont {Ilten},
  \citenamefont {Mrenna}, \citenamefont {Prestel}, \citenamefont {Rasmussen},\
  and\ \citenamefont {Skands}}]{Sjostrand:2014zea}%
  \BibitemOpen
  \bibfield  {author} {\bibinfo {author} {\bibfnamefont {T.}~\bibnamefont
  {Sj\"ostrand}}, \bibinfo {author} {\bibfnamefont {S.}~\bibnamefont {Ask}},
  \bibinfo {author} {\bibfnamefont {J.~R.}\ \bibnamefont {Christiansen}},
  \bibinfo {author} {\bibfnamefont {R.}~\bibnamefont {Corke}}, \bibinfo
  {author} {\bibfnamefont {N.}~\bibnamefont {Desai}}, \bibinfo {author}
  {\bibfnamefont {P.}~\bibnamefont {Ilten}}, \bibinfo {author} {\bibfnamefont
  {S.}~\bibnamefont {Mrenna}}, \bibinfo {author} {\bibfnamefont
  {S.}~\bibnamefont {Prestel}}, \bibinfo {author} {\bibfnamefont {C.~O.}\
  \bibnamefont {Rasmussen}}, \ and\ \bibinfo {author} {\bibfnamefont {P.~Z.}\
  \bibnamefont {Skands}},\ }\href {\doibase 10.1016/j.cpc.2015.01.024}
  {\bibfield  {journal} {\bibinfo  {journal} {Comput. Phys. Commun.}\ }\textbf
  {\bibinfo {volume} {191}},\ \bibinfo {pages} {159} (\bibinfo {year}
  {2015})},\ \Eprint {http://arxiv.org/abs/1410.3012} {arXiv:1410.3012
  [hep-ph]} \BibitemShut {NoStop}%
\bibitem [{\citenamefont {de~Favereau}\ \emph {et~al.}(2014)\citenamefont
  {de~Favereau}, \citenamefont {Delaere}, \citenamefont {Demin}, \citenamefont
  {Giammanco}, \citenamefont {Lema\^\i{}tre}, \citenamefont {Mertens},\ and\
  \citenamefont {Selvaggi}}]{deFavereau:2013fsa}%
  \BibitemOpen
  \bibfield  {author} {\bibinfo {author} {\bibfnamefont {J.}~\bibnamefont
  {de~Favereau}}, \bibinfo {author} {\bibfnamefont {C.}~\bibnamefont
  {Delaere}}, \bibinfo {author} {\bibfnamefont {P.}~\bibnamefont {Demin}},
  \bibinfo {author} {\bibfnamefont {A.}~\bibnamefont {Giammanco}}, \bibinfo
  {author} {\bibfnamefont {V.}~\bibnamefont {Lema\^\i{}tre}}, \bibinfo {author}
  {\bibfnamefont {A.}~\bibnamefont {Mertens}}, \ and\ \bibinfo {author}
  {\bibfnamefont {M.}~\bibnamefont {Selvaggi}} (\bibinfo {collaboration}
  {DELPHES 3}),\ }\href {\doibase 10.1007/JHEP02(2014)057} {\bibfield
  {journal} {\bibinfo  {journal} {JHEP}\ }\textbf {\bibinfo {volume} {02}},\
  \bibinfo {pages} {057} (\bibinfo {year} {2014})},\ \Eprint
  {http://arxiv.org/abs/1307.6346} {arXiv:1307.6346 [hep-ex]} \BibitemShut
  {NoStop}%
\bibitem [{\citenamefont {Zyla}\ \emph {et~al.}(2020)\citenamefont {Zyla} \emph
  {et~al.}}]{Zyla:2020zbs}%
  \BibitemOpen
  \bibfield  {author} {\bibinfo {author} {\bibfnamefont {P.~A.}\ \bibnamefont
  {Zyla}} \emph {et~al.} (\bibinfo {collaboration} {Particle Data Group}),\
  }\href {\doibase 10.1093/ptep/ptaa104} {\bibfield  {journal} {\bibinfo
  {journal} {PTEP}\ }\textbf {\bibinfo {volume} {2020}},\ \bibinfo {pages}
  {083C01} (\bibinfo {year} {2020})}\BibitemShut {NoStop}%
\bibitem [{\citenamefont {Iguro}\ \emph {et~al.}(2020)\citenamefont {Iguro},
  \citenamefont {Omura},\ and\ \citenamefont {Takeuchi}}]{Iguro:2020rby}%
  \BibitemOpen
  \bibfield  {author} {\bibinfo {author} {\bibfnamefont {S.}~\bibnamefont
  {Iguro}}, \bibinfo {author} {\bibfnamefont {Y.}~\bibnamefont {Omura}}, \ and\
  \bibinfo {author} {\bibfnamefont {M.}~\bibnamefont {Takeuchi}},\ }\href
  {\doibase 10.1007/JHEP09(2020)144} {\bibfield  {journal} {\bibinfo  {journal}
  {JHEP}\ }\textbf {\bibinfo {volume} {09}},\ \bibinfo {pages} {144} (\bibinfo
  {year} {2020})},\ \Eprint {http://arxiv.org/abs/2002.12728} {arXiv:2002.12728
  [hep-ph]} \BibitemShut {NoStop}%
\bibitem [{\citenamefont {Endo}\ \emph {et~al.}(2020)\citenamefont {Endo},
  \citenamefont {Iguro},\ and\ \citenamefont {Kitahara}}]{Endo:2020mev}%
  \BibitemOpen
  \bibfield  {author} {\bibinfo {author} {\bibfnamefont {M.}~\bibnamefont
  {Endo}}, \bibinfo {author} {\bibfnamefont {S.}~\bibnamefont {Iguro}}, \ and\
  \bibinfo {author} {\bibfnamefont {T.}~\bibnamefont {Kitahara}},\ }\href
  {\doibase 10.1007/JHEP06(2020)040} {\bibfield  {journal} {\bibinfo  {journal}
  {JHEP}\ }\textbf {\bibinfo {volume} {06}},\ \bibinfo {pages} {040} (\bibinfo
  {year} {2020})},\ \Eprint {http://arxiv.org/abs/2002.05948} {arXiv:2002.05948
  [hep-ph]} \BibitemShut {NoStop}%
\bibitem [{\citenamefont {the FCC-ee~design study}(2017)}]{FCCee:2017}%
  \BibitemOpen
  \bibfield  {author} {\bibinfo {author} {\bibnamefont {the FCC-ee~design
  study}},\ }\href@noop {} {}\bibinfo {howpublished}
  {\url{http://tlep.web.cern.ch/content/machine-parameters}} (\bibinfo {year}
  {2017})\BibitemShut {NoStop}%
\bibitem [{\citenamefont {Charles}\ \emph {et~al.}(2018)\citenamefont {Charles}
  \emph {et~al.}}]{Charles:2018vfv}%
  \BibitemOpen
  \bibfield  {author} {\bibinfo {author} {\bibfnamefont {T.~K.}\ \bibnamefont
  {Charles}} \emph {et~al.} (\bibinfo {collaboration} {CLICdp, CLIC}),\ }\href
  {\doibase 10.23731/CYRM-2018-002} {\  (\bibinfo {year} {2018}),\
  10.23731/CYRM-2018-002},\ \Eprint {http://arxiv.org/abs/1812.06018}
  {arXiv:1812.06018 [physics.acc-ph]} \BibitemShut {NoStop}%
\bibitem [{\citenamefont {Aaboud}\ \emph {et~al.}(2019)\citenamefont {Aaboud}
  \emph {et~al.}}]{Aaboud:2019lxo}%
  \BibitemOpen
  \bibfield  {author} {\bibinfo {author} {\bibfnamefont {M.}~\bibnamefont
  {Aaboud}} \emph {et~al.} (\bibinfo {collaboration} {ATLAS}),\ }\href
  {\doibase 10.1007/JHEP04(2019)048} {\bibfield  {journal} {\bibinfo  {journal}
  {JHEP}\ }\textbf {\bibinfo {volume} {04}},\ \bibinfo {pages} {048} (\bibinfo
  {year} {2019})},\ \Eprint {http://arxiv.org/abs/1902.05892} {arXiv:1902.05892
  [hep-ex]} \BibitemShut {NoStop}%
\bibitem [{\citenamefont {Delahaye}\ \emph {et~al.}(2019)\citenamefont
  {Delahaye}, \citenamefont {Diemoz}, \citenamefont {Long}, \citenamefont
  {Mansouli\'e}, \citenamefont {Pastrone}, \citenamefont {Rivkin},
  \citenamefont {Schulte}, \citenamefont {Skrinsky},\ and\ \citenamefont
  {Wulzer}}]{Delahaye:2019omf}%
  \BibitemOpen
  \bibfield  {author} {\bibinfo {author} {\bibfnamefont {J.~P.}\ \bibnamefont
  {Delahaye}}, \bibinfo {author} {\bibfnamefont {M.}~\bibnamefont {Diemoz}},
  \bibinfo {author} {\bibfnamefont {K.}~\bibnamefont {Long}}, \bibinfo {author}
  {\bibfnamefont {B.}~\bibnamefont {Mansouli\'e}}, \bibinfo {author}
  {\bibfnamefont {N.}~\bibnamefont {Pastrone}}, \bibinfo {author}
  {\bibfnamefont {L.}~\bibnamefont {Rivkin}}, \bibinfo {author} {\bibfnamefont
  {D.}~\bibnamefont {Schulte}}, \bibinfo {author} {\bibfnamefont
  {A.}~\bibnamefont {Skrinsky}}, \ and\ \bibinfo {author} {\bibfnamefont
  {A.}~\bibnamefont {Wulzer}},\ }\href@noop {} {\  (\bibinfo {year} {2019})},\
  \Eprint {http://arxiv.org/abs/1901.06150} {arXiv:1901.06150 [physics.acc-ph]}
  \BibitemShut {NoStop}%
\bibitem [{\citenamefont {Han}\ \emph {et~al.}(2021{\natexlab{c}})\citenamefont
  {Han}, \citenamefont {Ma},\ and\ \citenamefont {Xie}}]{Han:2020uid}%
  \BibitemOpen
  \bibfield  {author} {\bibinfo {author} {\bibfnamefont {T.}~\bibnamefont
  {Han}}, \bibinfo {author} {\bibfnamefont {Y.}~\bibnamefont {Ma}}, \ and\
  \bibinfo {author} {\bibfnamefont {K.}~\bibnamefont {Xie}},\ }\href {\doibase
  10.1103/PhysRevD.103.L031301} {\bibfield  {journal} {\bibinfo  {journal}
  {Phys. Rev. D}\ }\textbf {\bibinfo {volume} {103}},\ \bibinfo {pages}
  {L031301} (\bibinfo {year} {2021}{\natexlab{c}})},\ \Eprint
  {http://arxiv.org/abs/2007.14300} {arXiv:2007.14300 [hep-ph]} \BibitemShut
  {NoStop}%
\bibitem [{\citenamefont {Long}\ \emph {et~al.}(2021)\citenamefont {Long},
  \citenamefont {Lucchesi}, \citenamefont {Palmer}, \citenamefont {Pastrone},
  \citenamefont {Schulte},\ and\ \citenamefont {Shiltsev}}]{Long:2020wfp}%
  \BibitemOpen
  \bibfield  {author} {\bibinfo {author} {\bibfnamefont {K.}~\bibnamefont
  {Long}}, \bibinfo {author} {\bibfnamefont {D.}~\bibnamefont {Lucchesi}},
  \bibinfo {author} {\bibfnamefont {M.}~\bibnamefont {Palmer}}, \bibinfo
  {author} {\bibfnamefont {N.}~\bibnamefont {Pastrone}}, \bibinfo {author}
  {\bibfnamefont {D.}~\bibnamefont {Schulte}}, \ and\ \bibinfo {author}
  {\bibfnamefont {V.}~\bibnamefont {Shiltsev}},\ }\href {\doibase
  10.1038/s41567-020-01130-x} {\bibfield  {journal} {\bibinfo  {journal}
  {Nature Phys.}\ }\textbf {\bibinfo {volume} {17}},\ \bibinfo {pages} {289}
  (\bibinfo {year} {2021})},\ \Eprint {http://arxiv.org/abs/2007.15684}
  {arXiv:2007.15684 [physics.acc-ph]} \BibitemShut {NoStop}%
\bibitem [{\citenamefont {Al~Ali}\ \emph {et~al.}(2021)\citenamefont {Al~Ali}
  \emph {et~al.}}]{AlAli:2021let}%
  \BibitemOpen
  \bibfield  {author} {\bibinfo {author} {\bibfnamefont {H.}~\bibnamefont
  {Al~Ali}} \emph {et~al.},\ }\href@noop {} {\  (\bibinfo {year} {2021})},\
  \Eprint {http://arxiv.org/abs/2103.14043} {arXiv:2103.14043 [hep-ph]}
  \BibitemShut {NoStop}%
\bibitem [{\citenamefont {von Weizsacker}(1934)}]{vonWeizsacker:1934nji}%
  \BibitemOpen
  \bibfield  {author} {\bibinfo {author} {\bibfnamefont {C.~F.}\ \bibnamefont
  {von Weizsacker}},\ }\href {\doibase 10.1007/BF01333110} {\bibfield
  {journal} {\bibinfo  {journal} {Z. Phys.}\ }\textbf {\bibinfo {volume}
  {88}},\ \bibinfo {pages} {612} (\bibinfo {year} {1934})}\BibitemShut
  {NoStop}%
\bibitem [{\citenamefont {Williams}(1934)}]{Williams:1934ad}%
  \BibitemOpen
  \bibfield  {author} {\bibinfo {author} {\bibfnamefont {E.~J.}\ \bibnamefont
  {Williams}},\ }\href {\doibase 10.1103/PhysRev.45.729} {\bibfield  {journal}
  {\bibinfo  {journal} {Phys. Rev.}\ }\textbf {\bibinfo {volume} {45}},\
  \bibinfo {pages} {729} (\bibinfo {year} {1934})}\BibitemShut {NoStop}%
\bibitem [{\citenamefont {Chankowski}\ \emph {et~al.}(1999)\citenamefont
  {Chankowski}, \citenamefont {Krawczyk},\ and\ \citenamefont
  {Zochowski}}]{Chankowski:1999ta}%
  \BibitemOpen
  \bibfield  {author} {\bibinfo {author} {\bibfnamefont {P.~H.}\ \bibnamefont
  {Chankowski}}, \bibinfo {author} {\bibfnamefont {M.}~\bibnamefont
  {Krawczyk}}, \ and\ \bibinfo {author} {\bibfnamefont {J.}~\bibnamefont
  {Zochowski}},\ }\href {\doibase 10.1007/s100520050662} {\bibfield  {journal}
  {\bibinfo  {journal} {Eur. Phys. J. C}\ }\textbf {\bibinfo {volume} {11}},\
  \bibinfo {pages} {661} (\bibinfo {year} {1999})},\ \Eprint
  {http://arxiv.org/abs/hep-ph/9905436} {arXiv:hep-ph/9905436} \BibitemShut
  {NoStop}%
\bibitem [{\citenamefont {Chankowski}\ \emph {et~al.}(2000)\citenamefont
  {Chankowski}, \citenamefont {Farris}, \citenamefont {Grzadkowski},
  \citenamefont {Gunion}, \citenamefont {Kalinowski},\ and\ \citenamefont
  {Krawczyk}}]{Chankowski:2000an}%
  \BibitemOpen
  \bibfield  {author} {\bibinfo {author} {\bibfnamefont {P.~H.}\ \bibnamefont
  {Chankowski}}, \bibinfo {author} {\bibfnamefont {T.}~\bibnamefont {Farris}},
  \bibinfo {author} {\bibfnamefont {B.}~\bibnamefont {Grzadkowski}}, \bibinfo
  {author} {\bibfnamefont {J.~F.}\ \bibnamefont {Gunion}}, \bibinfo {author}
  {\bibfnamefont {J.}~\bibnamefont {Kalinowski}}, \ and\ \bibinfo {author}
  {\bibfnamefont {M.}~\bibnamefont {Krawczyk}},\ }\href {\doibase
  10.1016/S0370-2693(00)01293-4} {\bibfield  {journal} {\bibinfo  {journal}
  {Phys. Lett. B}\ }\textbf {\bibinfo {volume} {496}},\ \bibinfo {pages} {195}
  (\bibinfo {year} {2000})},\ \Eprint {http://arxiv.org/abs/hep-ph/0009271}
  {arXiv:hep-ph/0009271} \BibitemShut {NoStop}%
\bibitem [{\citenamefont {Barbieri}\ \emph {et~al.}(2007)\citenamefont
  {Barbieri}, \citenamefont {Hall}, \citenamefont {Nomura},\ and\ \citenamefont
  {Rychkov}}]{Barbieri:2006bg}%
  \BibitemOpen
  \bibfield  {author} {\bibinfo {author} {\bibfnamefont {R.}~\bibnamefont
  {Barbieri}}, \bibinfo {author} {\bibfnamefont {L.~J.}\ \bibnamefont {Hall}},
  \bibinfo {author} {\bibfnamefont {Y.}~\bibnamefont {Nomura}}, \ and\ \bibinfo
  {author} {\bibfnamefont {V.~S.}\ \bibnamefont {Rychkov}},\ }\href {\doibase
  10.1103/PhysRevD.75.035007} {\bibfield  {journal} {\bibinfo  {journal} {Phys.
  Rev. D}\ }\textbf {\bibinfo {volume} {75}},\ \bibinfo {pages} {035007}
  (\bibinfo {year} {2007})},\ \Eprint {http://arxiv.org/abs/hep-ph/0607332}
  {arXiv:hep-ph/0607332} \BibitemShut {NoStop}%
\end{thebibliography}%

\end{document}